\documentclass[12pt]{book}

\usepackage[numbers]{natbib}
\bibpunct{[}{]}{;}{n}{,}{,}
\usepackage{amsmath,amssymb,amsmath}
\usepackage{fullpage}
\usepackage{graphicx}
\usepackage{bm}
\usepackage{url}
\usepackage[pdftex,bookmarks,colorlinks=true,citecolor=blue]{hyperref}
\newcommand{\nodeset}{\mathcal{N}}     % the entire set of nodes
\newcommand{\edgeset}{\mathcal{E}}     % the entire set of edges
\newcommand{\relationset}{\mathcal{Y}}  % the entire set of relations in a graph (usually repr by adj matr)
\newcommand{\node}{N}       % general notation for a node or the number of nodes
\renewcommand{\deg}{\delta}      % node degree
\newcommand{\pexp}{\gamma}  % power-law exponent
\newcommand{\cmeventspace}{{\cal Y}} %Event space for a continuous markov process model (CMPM)
\newcommand{\cmtimeint}{{\cal T}}  %interval of time for a CMPM
\newcommand{\X}{{Y}}       % the random variable denoting network configuration
\newcommand{\x}{y}       % a value of the network conf r.v.
\newcommand{\xv}{{\bf y}}  % set of network edge configuration values (bold low-case)
\newcommand{\Latent}{Z}     % latent position random variable
\newcommand{\latent}{z}     % latent position value
\newcommand\Bet{\textrm{Beta}}

\newcommand{\pone}{{$p_1$}}
\newcommand{\ptwo}{{$p_2$}}
\newcommand{\ER}{{Erd\"{o}s-R\'{e}nyi-Gilbert}}
\newcommand{\EG}{{exchangeable graph}}

\newcommand{\natnum}{{\mathbb N}} % set of natural numbers
\newcommand{\argmax}{\textrm{argmax}}

\newcommand{\Gaussian}{\mathcal N}  % Gaussian distribution
\newcommand{\mult}{\textrm{mult}}   % Multinomial
\newcommand{\Bern}{\textrm{Bern}}   % Bernoulli
\newcommand{\Poi}{\textrm{Poi}}     % Poisson
\newcommand{\unif}{\textrm{unif}}
\newcommand{\hypercube}{\textrm{hypercube}}

\newcommand{\bv}{\begin{array}}
\newcommand{\ev}{\end{array}}
\newcommand{\bit}{\begin{itemize}}
\newcommand{\eit}{\end{itemize}}
\newcommand{\ben}{\begin{enumerate}}
\newcommand{\een}{\end{enumerate}}
\newcommand{\beq}{\begin{equation}}
\newcommand{\eeq}{\end{equation}}
\newcommand{\bvq}{\begin{eqnarray*}}
\newcommand{\evq}{\end{eqnarray*}}
\newcommand{\bvqa}{\begin{eqnarray}}
\newcommand{\evqa}{\end{eqnarray}}
\newcommand{\ptoq}{p \rightarrow q}
\newcommand{\pfromq}{p \leftarrow q}

\newcommand{\comment}[1]{}

\begin{document}

\title{\bf A Survey of Statistical Network Models}
\author{
Anna Goldenberg\\ University of Toronto %\texttt{anna.goldenberg@utoronto.ca}
\and
Alice X. Zheng\\ Microsoft Research %\texttt{alicez@microsoft.com}
\and
Stephen E. Fienberg\\ Carnegie Mellon University %\texttt{fienberg@stat.cmu.edu}
\and
Edoardo M. Airoldi\\ Harvard University %\texttt{airoldi@fas.harvard.edu}
\vspace{10cm}}
\date{December 2009}

%Center for Cellular and Biomolecular Research, University of Toronto, Toronto, Ont. M5S 3G4, Canada
%
%Microsoft Research, One Microsoft Way, Redmond, WA 98052, USA
%
%Department of Statistics and Machine Learning Department, Carnegie Mellon University, Pittsburgh, PA 15213-3890, USA
%
%Department of Statistics and FAS Center for Systems Biology, Harvard University, 1 Oxford Street, Cambridge, MA 02138, USA

\maketitle
\cleardoublepage \pagenumbering{roman}
\tableofcontents
\clearpage
\setcounter{page}{0}
\pagenumbering{arabic}

\chapter*{Preface}
\addcontentsline{toc}{chapter}{Preface}%

Networks are ubiquitous in science and have become a focal
  point for discussion in everyday life.  Formal statistical models
  for the analysis of network data have emerged as a major topic of
  interest in diverse areas of study, and most of these involve a form
  of graphical representation.  Probability models on graphs date back
  to 1959.  Along with empirical studies in social psychology and
  sociology from the 1960s, these early works generated an active
  ``network community'' and a substantial literature in the 1970s.
  This effort moved into the statistical literature in the late 1970s
  and 1980s, and the past decade has seen a burgeoning network
  literature in statistical physics and computer science.  The growth
  of the World Wide Web and the emergence of online ``networking
  communities" such as {\it Facebook}, {\it MySpace}, and {\it
    LinkedIn}, and a host of more specialized professional network
  communities has intensified interest in the study of networks and
  network data.\\

Our goal in this review is to provide the reader with an entry point
to this burgeoning literature.  We begin with an overview of the
historical development of statistical network modeling and then we
introduce a number of examples that have been studied in the network
literature.  Our subsequent discussion focuses on a number of
prominent static and dynamic network models and their
interconnections.  We emphasize formal model descriptions, and pay
special attention to the interpretation of parameters and their
estimation. We end with a description of some open problems and
challenges for machine learning and statistics.

%%%%%%%%%%%%%%%%%%%%%%%%%%%%%%%%%%%%%%%%%%%%%%%%%%%%%%%%%%%%%%%%%%%%%%%%%%%%%%
%%%%%%%%%%%%%%%%%%%%%%%%%%%%%%%%%%%%%%%%%%%%%%%%%%%%%%%%%%%%%%%%%%%%%%%%%%%%%%
%%%%%%%%%%%%%%%%%%%%%%%%%%%%%%%%%%%%%%%%%%%%%%%%%%%%%%%%%%%%%%%%%%%%%%%%%%%%%%
%%%%%%%%%%%%%%%%%%%%%%%%%%%%%%%%%%%%%%%%%%%%%%%%%%%%%%%%%%%%%%%%%%%%%%%%%%%%%%
%%%%%%%%%%%%%%%%%%%%%%%%%%%%%%%%%%%%%%%%%%%%%%%%%%%%%%%%%%%%%%%%%%%%%%%%%%%%%%
%%%%%%%%%%%%%%%%%%%%%%%%%%%%%%%%%%%%%%%%%%%%%%%%%%%%%%%%%%%%%%%%%%%%%%%%%%%%%%
%%%%%%%%%%%%%%%%%%%%%%%%%%%%%%%%%%%%%%%%%%%%%%%%%%%%%%%%%%%%%%%%%%%%%%%%%%%%%%
%%%%%%%%%%%%%%%%%%%%%%%%%%%%%%%%%%%%%%%%%%%%%%%%%%%%%%%%%%%%%%%%%%%%%%%%%%%%%%
%%%%%%%%%%%%%%%%%%%%%%%%%%%%%%%%%%%%%%%%%%%%%%%%%%%%%%%%%%%%%%%%%%%%%%%%%%%%%%
%%%%%%%%%%%%%%%%%%%%%%%%%%%%%%%%%%%%%%%%%%%%%%%%%%%%%%%%%%%%%%%%%%%%%%%%%%%%%%

\chapter{Introduction}
%\addcontentsline{toc}{unnumchapter}{Introduction}%
%\markboth{Introduction}{Introduction}

Many scientific fields involve the study of networks in
some form.  Networks have been used to analyze interpersonal social
relationships, communication networks, academic paper coauthorships
and citations, protein interaction patterns, and much more.  Popular
books on networks and their analysis began to appear a decade ago,
\citep[see, e.g.,][]{Bara:2002,Buch:2002,Watt:1999,Watt:2003,Chri:Fowl:2009}
and online ``networking communities" such as {\it Facebook}, {\it
  MySpace}, and {\it LinkedIn} are an even more recent phenomenon.

In this work, we survey selective aspects of the literature on
statistical modeling and analysis of networks in social sciences,
computer science, physics, and biology.  Given the volume of books,
papers, and conference proceedings published on the subject in these
different fields, a single comprehensive survey would be impossible.
Our goal is far more modest.  We attempt to chart the progress of
statistical modeling of network data over the past seventy years and
to outline succinctly the major schools of thought and approaches to
network modeling and to describe some of their interconnections.  We
also attempt to identify major statistical gaps in these modeling
efforts.  From this overview one might then synthesize and deduce
promising future research directions. Kolaczyk~\cite{Kola:2009} provides 
a complementary statistical overview.

The existing set of statistical network models may be organized along
several major axes.  For this article, we choose the axis of static
vs.\ dynamic models.  Static network models concentrate on explaining
the observed set of links based on a single snapshot of the network,
whereas dynamic network models are often concerned with the mechanisms
that govern changes in the network over time.  Most early examples of
networks were single static snapshots.  Hence static network models
have been the main focus of research for many years.  However, with
the emergence of online networks, more data is available for dynamic
analysis, and in recent years there has been growing interest in
dynamic modeling.

In the remainder of this chapter we provide a brief historical
overview of network modeling approaches.  In subsequent chapters we
introduce some examples studied in the network literature and give a
more detailed comparative description of select modeling approaches.

\section{Overview of Modeling Approaches}

Almost all of the ``statistically'' oriented literature on the
analysis of networks derives from a handful of seminal papers.  In
social psychology and sociology there is the early work of
\citet{Simm:1950} at the turn of the last century and
\citet{More:1934} in the 1930s as well as the empirical studies of
Stanley Milgram~\citep{Milg:1967,Trav:Milg:1969} in the 1960s; in
mathematics/probability there is the Erd\"{o}s-R\'{e}nyi paper on
random graph models \citep{Erdo:Reny:1960}.  There are other papers
that dealt with these topics contemporaneously or even earlier.  But
these are the ones that appear to have had lasting impact.

 \citet{More:1934} invented the sociogram --- a diagram of points and
 lines used to represent relations among persons, a precursor to the
 graph representation for networks. Luce and others developed a
 mathematical structure to go with Moreno's sociograms using incidence
 matrices and graphs (see,
 e.g.,~\cite{Luce:Perr:1949,Luce:1950,Luce:1973,Luce:Macy:Tagi:1955,Radn:Trit:1954,Spil:1966,Alba:1973}),
 but the structure they explored was essentially deterministic.
 Milgram gave the name to what is now referred to as the "Small World"
phenomenon --- short paths of connections linking most people in
social spheres --- and his experiments had provocative results: the
shortest path between any two people for completed chains has a median length of around 6;
however,  the majority of chains initiated in his experiments were never completed!  (His studies provided
the title for the play and movie {\it Six Degrees of Separation}, ignoring the compleity of his results due to the censoring.)
 \citet{Whit:1970} and \citet{Fien:Lee:1975} gave a formal
 Markov-chain like model and analysis of the Milgram experimental
 data, including information on the uncompleted chains.  
 Milgram's data were gathered in batches of transmission, and
 thus these models can be thought of as representing early examples of
 generative descriptions of dynamic network evolution.  Recently,
 \citet{Dodd:Muha:Watt:2003} studied a global ``replication" variation
 on the Milgram study in which more than 60,000 e-mail users attempted
 to reach one of 18 target persons in 13 countries by forwarding
 messages to acquaintances. Only 384 of 24,163 chains reached their
 targets but they estimate the median length for completions to be 7, by
 assuming that attrition occurs at random.

The social science network research community that arose in the 1970s
was built upon these earlier efforts, in particular the {\ER} model.
Research on the {\ER} model (along with works by Katz et
al.~\cite{Katz:1951,Katz:Wils:1956,Katz:Powe:1957}) engendered the
field of random graph theory.  In their papers, Erd\"{o}s and
R\'{e}nyi worked with fixed number of vertices, $N$, and number of
edges, $E$, and studied the properties of this model as $E$ increases.
Gilbert studied a related two-parameter version of the model, with $N$
as the number of vertices and $p$ the fixed probability for choosing
edges.  Although their descriptions might at first appear to be static
in nature, we could think in terms of adding edges sequentially and
thus turn the model into a dynamic one.  In this alternative binomial
version of the {\ER} model, the key to asymptotic behavior is the
value $\lambda=pN$.  There is a ``phase change'' associated with the
value of $\lambda=1$, at which point we shift from seeing many small
connected components in the form of trees to the emergence of a single
``giant connected component.''  Probabilists such as \citet{Pitt:1990}
imported ideas and results from stochastic processes into the random
graph literature. % \citet{Aldo:1997} made the link between stochastic processes and random graphs, giving rise to a fundamental distribution in population genetics.

\citet{Holl:Lein:1981}'s {\pone} model extended the {\ER} model to
allow for differential attraction (popularity) and expansiveness, as
well as an additional effect due to reciprocation.  The {\pone} model
was log-linear in form, which allowed for easy computation of maximum
likelihood estimates using a contingency table formulation of the
model \citep{Fien:Wass:1981,Fien:Wass:1981a}.  It also allowed for
various generalizations to multidimensional network structures
\citep{Fien:Meye:Wass:1985} and stochastic blockmodels.  This approach
to modeling network data quickly evolved into the class of $p^*$ or
exponential random graph models (ERGM) originating in the work of
\citet{Fran:Stra:1986} and \citet{Stra:Iked:1990}.  A trio of papers
demonstrating procedures for using ERGMs
\citep{Wass:Patt:1996,Patt:Wass:1999,Robi:Patt:Wass:1999} led to the
wide-spread use of ERGMs in a descriptive form for cross sectional
network structures or cumulative links for networks---what we refer to
here as static models.  Full maximum likelihood approaches for ERGMs
appeared in the work of Snijders and Handcock and their collaborators,
some of which we describe in \autoref{chapter:static}.

Most of the early examples of networks in the social science
literature were relatively small (in terms of the number of nodes) and
involved the study of the network at a fixed point in time or
cumulatively over time.  Only a few studies (e.g., Sampson's 1968 data
on novice monks in the monastery \citep{Samp:1968}) collected,
reported, and analyzed network data at multiple points in time so that
one could truly study the evolution of the network, i.e., network
dynamics.  The focus on relatively small networks reflected the
state-of-art of computation but it was sufficient to trigger the
discussion of how one might assess the fit of a network model.  Should
one focus on ``small sample" properties and exact distributions given
some form of minimal sufficient statistic, as one often did in other
areas of statistics, or should one look at asymptotic properties, 
where there is a sequence of networks of increasing size?
Even if we have ``repeated cross-sections" of the network, if the
network is truly evolving in continuous time we need to ask how to
ensure that the continuous time parameters are estimable.  We return
to many of these question in subsequent chapters.

In the late 1990s, physicists began to work on network models and
study their properties in a form similar to the macro-level
descriptions of statistical physics. Barab\'asi, Newman, and Watts,
among others, produced what we can think of as variations on the {\ER}
model which either controlled the growth of the network or allowed for
differential probabilities for edge addition and/or deletion.  These
variations were intended to produce phenomena such as ``hubs," ``local
clustering," and ``triadic closures."  The resulting models gave us
fixed degree distribution limits in the form of power laws ---
variations on preferential attachment models (``the rich get richer'')
that date back to \citet{Yule:1925} and \citet{Simo:1955} (see also
\citep{Mitz:2004}) --- as well as what became known as ``small world"
models.  The small-world phenomenon, which harks back to Milgram's
1960s studies, usually refers to two distinct properties: (1) small
average distance and (2) the ``clustering" effect, where two nodes with
a common neighbor are more likely to be adjacent. Many of these
authors claim that these properties are ubiquitous in realistic
networks.  To model networks with the small-world phenomenon, it is
natural to utilize randomly generated graphs with a power law degree
distribution, where the fraction of nodes with degree $k$ is
proportional to $k^{-a}$ for some positive exponent $a$.  Many of the
most relevant papers are included in an edited collection by
\citet{Newm:Bara:Watt:2006}.  More recently this style of statistical
physics models have been used to detect community structure in networks, e.g., see
\citet{Girv:Newm:2002} and \citet{Back:Hutt:Klei:Lan:2006}, a
phenomenon which has its counterpart description in the social science
network modeling literature.

The probabilistic literature on random graph models from the 1990s
made the link with epidemics and other evolving stochastic phenomena.
Picking up on this idea, \citet{Watt:Stro:1998} and others used epidemic
models to capture general characteristics of the evolution of these
new variations on random networks.  \citet{Durr:2006} has provided us
with a book-length treatment on the topic with a number of interesting
variations on the theme.  The appeal of stochastic processes as
descriptions of dynamic network models comes from being able to
exploit the extensive literature already developed, including the
existence and the form of stationary distributions and other model
features or properties.  \citet{Chun:Lu:2006} provide a complementary
treatment of these models and their probabilistic properties.

One of the principal problems with this diverse network literature that we see is
that, with some notable exceptions, the statistical tools for
estimation and assessing the fit of ``statistical physics" or
stochastic process models is lacking.  Consequently, no attention is paid to the fact that real data may often be biased and noisy. What authors in the network
literature have often relied upon is the extraction of key features of
the related graphical network representation, e.g., the use of power
laws to represent degree distributions or measures of centrality and
clustering, without any indication that they are either necessary or
sufficient as descriptors for the actual network data.  Moreover,
these summary quantities can often be highly misleading as the
critique by
\citet{Stou:Malm:Amar:2005,Stou:Malm:Amar:2008} of
methods used by \citet{Bara:2005} and \citet{Vazq:Oliv:Dezs:Goh:2006}
suggest.  Barab\'asi claimed that the dynamics of a number of human
activities are scale-free, i.e., he specifically reported that the
probability distribution of time intervals between consecutive e-mails
sent by a single user and time delays for e-mail replies follow a
power-law with exponent $-1$, and he proposed a priority-queuing
process as an explanation of the bursty nature of human
activity. \citet{Stou:Malm:Amar:2008} demonstrated that the
reported power-law distribution was solely an artifact of the
analysis of the empirical data and used Bayes factors to show that the
proposed model is not representative of e-mail communication patterns.
See a related discussion of the poor fit of power laws in
\citet{Clau:Shal:Newm:2007}. There are several works, however, that try to address model fitting and model comparison.  For example, the work of \citet{Will:Mart:2000} showed how a simple two-parameter model predicted ``key structural properties of the most complex and comprehensive food webs in the primary literature". Another good example is the work of \citet{Midd:Ziv:Wigg:2004} where the authors used network motif counts as input to a discriminative systematic classification for deciding which configuration model the actual observed network came from; they looked at power law, small-world, duplication-mutation and duplication-mutation-complementation and other models (seven in total) and concluded that the duplication-mutation-complementation model described the protein-protein interaction data in {\it Drosophila melanogaster} species best. 

Machine learning approaches emerged in several forms over the past
decade with the empirical studies of \citet{Falo:Falo:Falo:1999} and
\citet{Klei:2000a,Klei:2000,Klei:2001}, who introduced
a model for which the underlying graph is a grid---the graphs
generated do not have a power law degree distribution, and each vertex
has the same expected degree. The strict requirement that the
underlying graph be a cycle or grid renders the model inapplicable to
webgraphs or biological networks. \citet{Durr:2006} treats variations
on this model as well.  More recently, a number of authors have looked
to combine the stochastic blockmodel ideas from the 1980s with latent
space models, model-based clustering
\citep{Hand:Raft:Tant:2007} or mixed-membership models
\citep{Airo:Blei:Fien:Xing:2008}, to provide generative models that
scale in reasonable ways to substantial-sized networks.  The class of
mixed membership models resembles a form of soft clustering
\citep{Eros:Fien:Laff:2004} and includes the latent
Dirichlet allocation model \citep{Blei:Ng:Jord:2003} from machine
learning as a special case.  This class of models offers much promise
for the kinds of network dynamical processes we discuss here.

\section{What This Survey Does Not Cover}

This survey focuses primarily on statistical network models and their
applications.  As a consequence there are a number of topics that we
touch upon only briefly or essentially not at all, such as

\begin{itemize}

\item {\it Probability theory associated with random graph models.}
  The probabilistic literature on random graph models is now truly
  extensive and the bulk of the theorems and proofs, while interesting
  in their own right, are largely unconnected with the present
  exposition.  For excellent introductions to this literature, see
  \citet{Chun:Lu:2006} and \citet{Durr:2006}. For related results
  on the mathematics of graph theory, see \citet{Boll:2001}.

\item {\it Efficient computation on networks.}  There is a substantial
  computer science literature dealing with efficient calculation of
  quantities associated with network structures, such as shortest
  paths, network diameter, and other measures of connectivity,
  centrality, clustering, etc.  The edited volume by
  \citet{Bran:Erle:2005} contains good overviews of a number of
  these topics as well as other computational issues associated with
  the study of graphs.

\item {\it Use of the network as a tool for sampling.}  Adaptive
  sampling strategies modify the sampling probabilities of selection
  based on observed values in a network structure.  This strategy is
beneficial when searching for rare or clustered populations.  \citet{Thom:Sebe:1996} and
  \citet{Thom:2006} discuss adaptive sampling in detail.  There is
  also related work on target sampling~\cite{Thom:2006a} and
  respondent-driven sampling~\cite{Salg:Heck:2004,Volz:Heck:2008}.
  
\item {\it Neural networks.} Neural networks originated as simple
  models for connections in the brain but have more recently been used
  as a computational tool for pattern recognition (e.g.,
  \citet{Bish:1995}), machine learning (e.g., \citet{Neal:1996}),
  and models of cognition (e.g., \citet{Roge:McCl:2004}).

\item {\it Networks and economic theory.} A relatively new area of
  study is the link between network problems, economic theory, and
  game theory.  Some useful entrees to this literature are
  \citet{Even:Kear:2006}, \citet{Goya:2007},
  \citet{Kear:Suri:Mont:2006}, and \citet{Jack:2008}, whose book
  contains an excellent semi-technical introduction to network concepts
  and structures.

\item {\it Relational networks.} This is a very popular area in
  machine learning.  It uses probabilistic graphical models to
  represent uncertainty in the data. The types of ``networks" in this
  area, such as Bayes nets, dependency diagrams, etc., have a
  different meaning than the networks we consider in this review. The
  main difference is that the networks in our work are considered to
  ``be given" or arising directly from properties of the network under
  study, rather than being representative of the uncertainty of the
  relationships between nodes and node attributes. There is a
  multitude of literature on relational networks, e.g., see
  \citet{Frie:Geto:Koll:Pfef:1999}, \citet{Geto:Frie:Koll:Task:2002},
  \citet{Jens:2003,Nevi:Jens:Frie:Hay:2003a}, and \citet{Geto:Task:2007}.

\item {\it Bi-partite graphs.} These are graphs that represent measurement on two populations of objects, such as individuals and features. The graphs in this context are seldom the best representation of the data, with exception perhaps of binary measurements or when the true populations have comparable sizes. Recent work on exchangeable Rasch matrices is related to to this topic and potentially relevant for network analysis.   \citet{Laur:2003,Laur:2008,Bass:Cose:Mand:2007}   suggest applications to bipartite graphs.

\item{\it Agent-based modeling.} Building on older ideas such as
  cellular automata, agent-based modeling attempts to simulate the
  simultaneous operations of multiple agents, in an effort to
  re-create and predict the actions of complex phenomena.  Because the
  interest is often on the interaction among the agents, this domain of
  research has been linked with network ideas.  With the recent
  advances in high-performance computing, simulations of large-scale
  social systems have become an active area of research, e.g.,
  see~\cite{Bona:2002}. In particular, there is a strong interest
  in areas that revolve around national security and the military,
  with studies on the effects of catastrophic events and biological
  warfare, as well as computational explorations of possible recovery
  strategies \citep{Carl:Remi:2004,Carl:Frid:Casm:Yahj:2006}. These
  works are the contemporary counterparts of more classical work at
  the interface between artificial intelligence and the social
  sciences \citep{Carl:1990,Carl:Newe:1994,Carl:2002}.

\end{itemize}

%%%%%%%%%%%%%%%%%%%%%%%%%%%%%%%%%%%%%%%%%%%%%%%%%%%%%%%%%%%%%%%%%%%%%%%%%%%%%%
%%%%%%%%%%%%%%%%%%%%%%%%%%%%%%%%%%%%%%%%%%%%%%%%%%%%%%%%%%%%%%%%%%%%%%%%%%%%%%
%%%%%%%%%%%%%%%%%%%%%%%%%%%%%%%%%%%%%%%%%%%%%%%%%%%%%%%%%%%%%%%%%%%%%%%%%%%%%%
%%%%%%%%%%%%%%%%%%%%%%%%%%%%%%%%%%%%%%%%%%%%%%%%%%%%%%%%%%%%%%%%%%%%%%%%%%%%%%
%%%%%%%%%%%%%%%%%%%%%%%%%%%%%%%%%%%%%%%%%%%%%%%%%%%%%%%%%%%%%%%%%%%%%%%%%%%%%%
%%%%%%%%%%%%%%%%%%%%%%%%%%%%%%%%%%%%%%%%%%%%%%%%%%%%%%%%%%%%%%%%%%%%%%%%%%%%%%
%%%%%%%%%%%%%%%%%%%%%%%%%%%%%%%%%%%%%%%%%%%%%%%%%%%%%%%%%%%%%%%%%%%%%%%%%%%%%%
%%%%%%%%%%%%%%%%%%%%%%%%%%%%%%%%%%%%%%%%%%%%%%%%%%%%%%%%%%%%%%%%%%%%%%%%%%%%%%
%%%%%%%%%%%%%%%%%%%%%%%%%%%%%%%%%%%%%%%%%%%%%%%%%%%%%%%%%%%%%%%%%%%%%%%%%%%%%%
%%%%%%%%%%%%%%%%%%%%%%%%%%%%%%%%%%%%%%%%%%%%%%%%%%%%%%%%%%%%%%%%%%%%%%%%%%%%%%
% WAS: \input{examples-ema-10-15-08.tex}

\chapter{Motivation and Dataset Examples}
\label{chap:examples_nets}

\section{Motivations for Network Analysis}

Why do we analyze networks?  The motivation behind network analysis is
as diverse as the origin of network problems within differing academic
fields.  Before we delve into details of the ``how'' of statistical
network modeling, we start with some examples of the ``why.''  This
chapter also includes descriptions of popular datasets for interested
readers who may wish to exercise their modeling muscles.

Social scientists are often interested in questions of interpretation
such as the meanings of edges in a social network~\citep{Krac:1999}.
Do they arise out of friendliness, strategic alliance, obligation, or
something else?  When the meaning of edges are known, the object is
often to characterize the structure of those relations (e.g., whether
friendships or strategic alliances are hierarchical or transitive).  A
large volume of statistically-oriented social science literature is
dedicated to modeling the mechanisms and relations of network
properties and testing hypotheses about network structure, see, e.g.,~\citep{Snij:Patt:Robi:Hand:2006}.

Physicists, on the other hand, tend to be interested in understanding
parsimonious mechanisms for network formation
\citep{Bara:Jeon:Neda:Rava:2002,Newm:Watt:Stro:2002}. For example, a
common modeling goal is to explain how a given network comes to have
its particular degree distribution or diameter at time $t$. \comment{Despite
the seemingly dynamic nature of the proposed models, they study what
we refer to in this review as static networks. Research in this area
typically focuses on global network properties such as degree
distribution, without addressing the semantics of individual edges or
characterizing the fit of the model associated with the mechanism. We review some of these models in the static network (\autoref{sec:static}) and others in the dynamic network section (\autoref{sec:dynamic})alluding to the dynamic nature of the underlying mechanisms.}

Several network analysis concepts have found niches in computational
biology.  For example, work on protein function classification can be
thought of as finding hidden groups in the protein-protein interaction
network \citep{Airo:Blei:Xing:Fien:2005,Airo:Blei:Fien:Xing:2007} to gain better understanding of underlying biological processes. Label propagation (node similarity) in networks can be harnessed to help with functional gene annotation \citep{Most:Ray:Ward:Grou:2008}.
Graph alignment can be used to locate subgraphs that are common among
species, thus advancing our understanding of evolution
\citep{Flan:Nova:Srin:McAd:2006}. Motif finding, or more generally the search for
subgraph patterns, also has many applications \citep{Alon:2007}. Combining networks from heterogeneous data sources helps to improve the accuracy of predicted genetic interactions \citep{Wong:Zhan:Tong:Li:2003}. Heterogeneity of network data sources in biology introduces a lot of noise into the global network structure, especially when networks created for different purposes (such as protein co-regulation and gene co-expression) are combined. \citep{Morr:Frey:Paig:2003} addresses network de-noising via degree-based structure priors on graphs.  For a review of biological applications of networks, please see \citep{Zhu:Gers:Snyd:2007}.

The task of finding hidden groups is also relevant in analyzing
communication networks, e.g., in detecting possible latent terrorist
cells~\citep{Baum:Gold:Magd:Wall:2004}.  The related task of
discovering the ``roles" of individual nodes is useful for identity
disambiguation \citep{Bhat:2006} and for business organization
analysis \citep{McCa:Corr:Wang:2005}.  These applications often take
the machine learning approach of graph partitioning, a topic
previously known in social science and statistics literature as
blockmodeling \citep{Lorr:Whit:1971,Dore:Bata:Ferl:2004a}. A related
question is \emph{functional} clustering, where the goal is not to
statistically cluster the network, but to discover members of dynamic
communities with similar functions based on existing network connectivity
\citep{Girv:Newm:2002,Newm:2004,Newm:2006a,Shal:Camp:Klin:2007}.

In the machine learning community, networks are often used to predict missing information, which can be edge related, e.g., predicting missing links
in the network \citep{OMad:Smyt:Adam:2005,Clau:Moor:Newm:2008,Libe:Klei:2003}, or attribute related, e.g.,
predicting how likely a movie is to be a box office hit
\citep{Jens:2003}.  Other applications include locating the crucial
missing link in a business or a terrorist network, or calculating the
probability that a customer will purchase a new product, given the
pattern of purchases of his friends~\citep{Hill:Prov:Voli:2006}. The latter question can more generally be stated as predicting individual's preferences given the preferences of her ``friends''. This research direction has evolved into an area of its own under the name of \emph{recommender systems}, which has recently received a lot of media attention due to the competition by the largest online movie rental company Netflix. The company has awarded a prize of one million dollars to a team of researchers that were able to predict customer ratings of movies with higher than $10\%$ accuracy than their own in-house system \citep{Swas:2009}.

The concept of information propagation also finds many applications in
the network domain, such as virus propagation in computer networks
\citep{Wang:Chak:Wang:Falo:2003}, HIV infection networks
\citep{Morr:Kret:1995,Jone:Hand:2003,Jone:Hand:2003a}, viral marketing
\citep{Domi:2005} and more generally gossiping \citep{Kemp:Klei:Tard:2005}. Here some work focuses on finding network configurations optimal for routing, while other research assumes that the network structure is given and focus on suitable models for disease or information spread.

\section{Sample Datasets}

A plethora of data sets are available for network analysis, and more
are emerging every year.  We provide a quick guided tour of the most
popular datasets and applications in each field.

In his ground-breaking paper, \citet{Milg:1967} experimented with
the construction of interpersonal social networks.  His result that
the median length of completed chains was approximately 6 led to the
pop-culture coining of the phrase ``six degrees of separation.''
Subjects of subsequent studies ranged from social interactions of
monks \citep{Samp:1968}, to hierarchies of elephants
\citep{McCo:Moss:Dura:Bake:2001,Vanc:Arch:Moss:2009}, to sexual
relationships between adults of Colorado \citep{Klov:Pott:Wood:Muth:1994}, to
friendships amongst elementary school students
\citep{Harr:Flor:Tabo:Bear:2003,Udry:2003}. 

While a lot of biological applications focus on the study of protein-protein
interaction networks
\citep{Gavi:Bosc:Krau:Gran:2002,Gavi:Aloy:Gran:Krau:2006,Krog:Cagn:Yu:Zhon:2006,Regu:Brei:Bouc:Brei:2006,Yu:Brau:Yild:Lemm:2008}, metabolic
networks \citep{Huss:Holm:2007}, functional and co-expression gene similarity networks
and gene regulatory networks \citep{Frie:2004,Walc:Mugl:Wigg:2009}, computer science
applications revolve around e-mail \citep{McCa:Corr:Wang:2005}, the
internet
\citep{Falo:Falo:Falo:1999,Chen:Chan:Govi:Jami:2002,Holm:Karl:Forr:2008},
the web \citep{Hube:Adam:1999,Albe:Jeon:Bara:1999}, academic paper
co-authorship \citep{Gold:Moor:2005} and citation networks
\citep{Mann:Mimn:McCa:2006,Mimn:McCa:2007}.  Citation networks have a long history of modeling in different areas of research starting with the seminal paper of \citet{DeSo:1965} and more recently in physics \citep{Leic:Clar:Shed:Newm:2007}. 
With the recent rise of
online networks, computer science and social science researchers are
also starting to examine blogger networks such as {\it LiveJournal}, 
social networks found on {\it Friendster}, {\it Facebook}, {\it
  Orkut}, and dating networks such as {\it Match.com}.  
  
  Terrorist
networks (often simulated) and telecommunication networks have come
under similar scrutiny, especially since the events of September 11,
2001 (e.g.,
see~\citep{Kreb:2002,Reid:Qin:Chun:Xu:2004,Reid:Chen:2005,Chen:Reid:Sina:Silk:2008}).
There has also been work on ecological networks such as foodwebs
\citep{Will:Mart:2000,Alle:Alon:Pasc:2008}, neuronal networks
\citep{Lee:Stev:2007}, network epidemiology \citep{Volz:Meye:2009},
economic trading networks
\citep{Gled:2002}, transportation networks (roads, railways,
airplanes; e.g.,~\citep{Gast:Newm:2006}), resource
distribution networks, mobile phone networks
\citep{Eagl:Pent:Laze:2009} and many others. 

%Depending upon the amount of available information, network data are often categorized as one-mode, two-mode, or relational (multi-modal). To give a concrete example, in a one-mode network, nodes may represent authors of academic papers and links co-authorships. A corresponding two-mode network would be a bi-partite graph of nodes representing papers and authors, and links specifying authorship.  A relational network is a somewhat overloaded term. In computer science it may describe any network where either the nodes or the links or both are heterogeneous. For example, a relational network representation of author-paper data would contain two components: a data graph containing the two-mode network described above, and a model graph that explicitly represents the type of each node (author or paper), attributes of each type (author's affiliation, research area, paper's title, etc.), and relationships between the attributes (a paper's title might depend on the authors' research areas).  Thus a one-mode network only encodes whether a pair of researchers have co-authored one or more papers, a two-mode network encodes also the identity of the papers, and a relational network contains even more information on attributes of the authors and papers and possible variety of dependencies among them.

Several network data repositories are available on public websites and as part of packages. For example, UCINet\footnote{\url{http://www.analytictech.com/ucinet/}}  includes a lot of well known smaller scale datasets such as the Davis Southern Club Women dataset \cite{Davi:Gard:Gard:Wall:1941}, Zachary's karate club dataset \citep{Zach:1977}, and Sampson's monk data \citep{Samp:1968} described below. Pajek\footnote{\url{http://vlado.fmf.uni-lj.si/pub/networks/data/}}  contains a larger set of small and large networks from domains such as biology, linguistics, and food-web. Additional datasets in a variety of domains include power grid networks, US politics, cellular and protein networks and others\footnote{\url{http://www-personal.umich.edu/~mejn/netdata/}\\\url{http://cdg.columbia.edu/cdg/datasets}\\\url{http://www.nd.edu/~networks/resources.htm}}. A collection of large and very large directed and undirected networks in the areas of communication, citation, internet and others are available as part of Stanford Network Analysis Package (SNAP)\footnote{\url{http://snap.stanford.edu/data/}}.

We now introduce six examples of networks studied in the literature,
describing the data in reasonable detail and including graphs
depicting the networks wherever feasible.  For each network example we
articulate specific questions of interest.

\subsection{Sampson's ``Monastery'' Study}
\label{sec:sampson}

\begin{figure}[t]
\begin{center}
 {\resizebox{3.2in}{2.8in}{\includegraphics{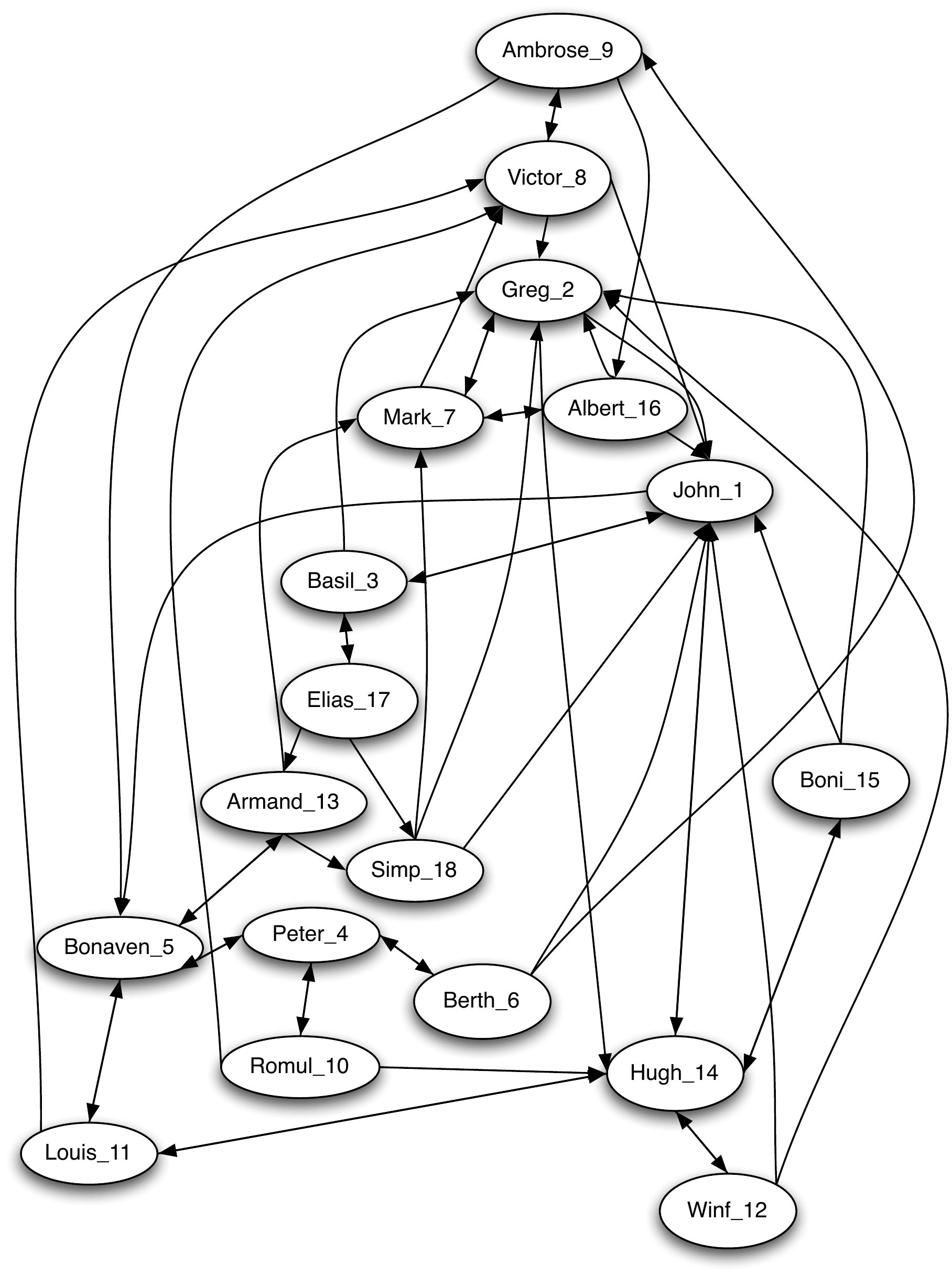}}}
\end{center}
\caption{Network derived from ``whom do you like'' sociometric relations collected by Sampson.}
\label{fig:sampson_whomdolike}
\end{figure}

 A classic example of a social network is the one derived from the
 survey administered by  Samspon and published in his doctoral
 dissertation \citep{Samp:1968}. \autoref{fig:sampson_whomdolike}
 displays the network derived from the ``whom do you like''
 sociometric relations in this dataset.
 Sampson spent several months in a monastery in New England, where a
 number of novices were preparing to join a monastic order. Sampson's
 original analysis was rooted in direct anthropological
 observations. He strongly suggested the existence of tight factions
 among the novices: the loyal opposition (whose members joined the
 monastery first), the young turks (whose members joined later on),
 the outcasts (who were not accepted in either of the two main
 factions), and the waverers (who did not take sides). The events that
 took place during Sampson's stay at the monastery supported his
 observations. For instance, John and Gregory, two members of the young turks,  
 were expelled over religious differences, and other members resigned shortly after these events. About a year after
 leaving the monastery, Sampson surveyed all of the novices, and asked
 them to rank the other novices in terms of four sociometric
 relations: like/dislike, esteem, personal influence, and alignment
 with the monastic credo, retrospectively, at four different epochs
 spanning his stay at the monastery.

 The presence of a well defined social structure within the monastery
 (the factions) that can be inferred from responses to the survey, as
 well as the social dynamics of subtle ideological conflicts that led
 to the dissolution of the monastic order, have much intrigued both
 statisticians and social scientists for the past four
 decades. Researchers typically consider the faction labels assigned
 by Sampson to the novices as the anthropological ground truth in
 their analysis. For example analyses, we refer to
 \cite{Fien:Meye:Wass:1985,Hand:Raft:Tant:2007,Davi:Carl:2008,Airo:Blei:Fien:Xing:2008}.

\subsection{The Enron Email Corpus}
\label{sec:enron}

The Enron email corpus has been widely studied in recent machine
learning network literature.  Enron Corporation was an energy and
trading company specializing in the marketing of electricity and gas.
In 2000 it was the seventh largest company in the United States with
reported revenues of over \$100 billion.  On December 2, 2001, Enron
filed for bankruptcy.
\begin{figure}[t!]
\begin{center}
{\resizebox{4.6in}{2.2in}{\includegraphics{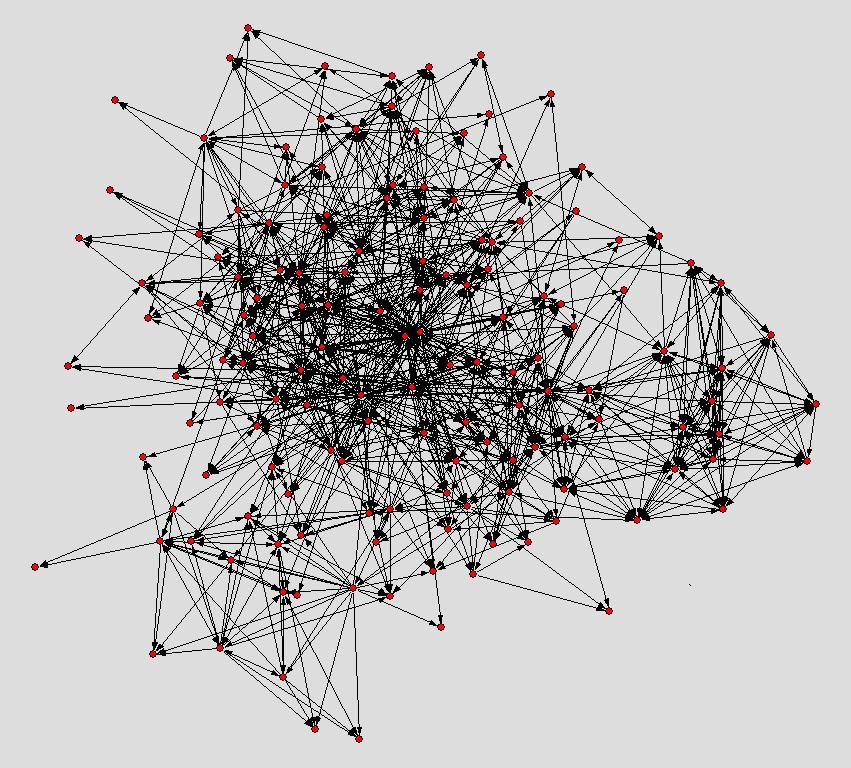}}}
\end{center}
\caption{E-mail exchange data among 151 Enron executives, using a
threshold of a minimum of 5 messages for each
link. Source:~\cite{Huh:Fien:2008}.}
\label{fig:enron1}
\end{figure}
\begin{figure}[b!]
\begin{center}
{\resizebox{4.6in}{2.2in}{\includegraphics{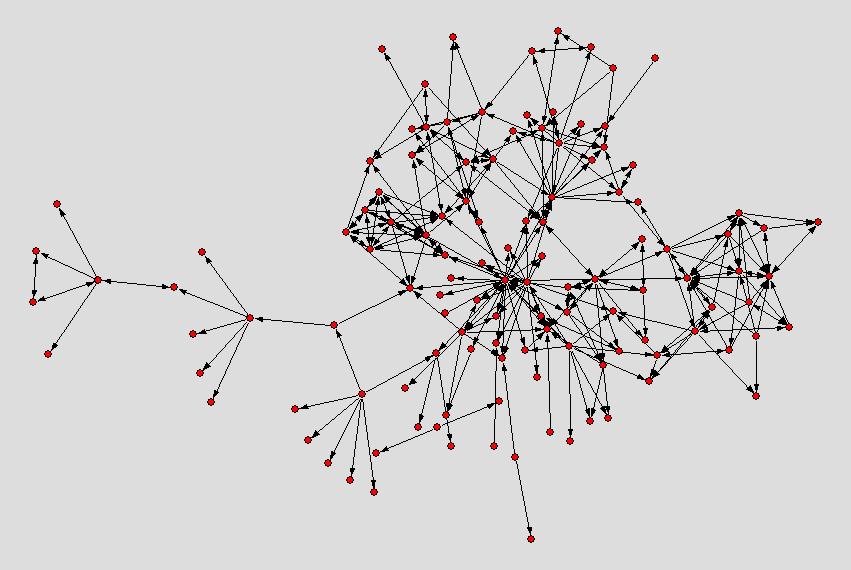}}}
\end{center}
\caption{E-mail exchange data among 151 Enron executives, using a
threshold of a minimum of 30 messages for each
link. Source:~\cite{Huh:Fien:2008}.}
\label{fig:enron2}
\end{figure}
The sudden collapse cast suspicions over its management and prompted
federal investigations.  Thirty-four Enron officials were prosecuted
and top Enron executives and associates were subsequently found to be
guilty of accounting fraud.  During the investigation, the courts
subpoenaed extensive email logs from most of Enron's employees, and
the Federal Energy Regulatory Commission (FERC) published the database
online.\footnote{\url{http://www.ferc.gov/industries/electric/indus-act/wec/enron/info-release.asp}}
Subsequently, researchers in the CALO (Cognitive Assistant that Learns
and Organizes) project corrected integrity problems in the
dataset.\footnote{\url{http://www.cs.cmu.edu/~enron/}} The original
FERC dataset contains 619,446 email messages (about 92\% of Enron's
staff emails), and the cleaned-up CALO dataset contains 200,399
messages from 158 users.  Another version of the data consists of the
contents of the mail folders of the top 151 executives, containing
about 225,000 messages covering a period from 1997 to
2004.\footnote{\url{http://www.isi.edu/~adibi/Enron/Enron.htm}}
\autoref{fig:enron1} and \autoref{fig:enron2} give network snapshots
of the e-mail traffic among these 151 executives with thresholds of 5
and 30 messages, respectively.

Research activity on the Enron dataset range from document
classification to social-network analysis to visualization.  A
collection of papers working with the Enron corpus were gathered
together in a special 2005 issue of {\it Computational \& Mathematical
  Organization Theory}, see \cite{Carl:Skil:2005}.

\subsection{The Protein Interaction Network in Budding Yeast}

\begin{figure}[ht!]
 \centering
  \includegraphics[width=0.8\textwidth]{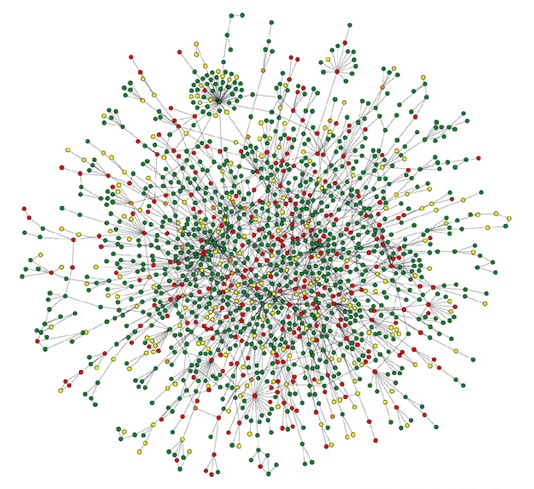}
\caption{A popular image of the protein interaction network in {\em Saccharomyces cerevisiae}, also known as the budding yeast. The figure is reproduced with permission. Source:~\cite{Bara:Oltv:2004}.}
\label{fig:yeast_protein_network}
\end{figure}

 The budding yeast is a unicellular organism that has become a
 de-facto model organism for the study of molecular and cellular
 biology \citep{Bots:Cher:Cher:1997}.  There are about 6,000 proteins
 in the budding yeast, which interact in a number of ways
 \citep{Cher:Ball:Weng:Juvi:1997}. For instance, proteins bind
 together to form protein complexes, the physical units that carry out
 most functions in the cell \citep{Krog:Cagn:Yu:Zhon:2006}. In recent
 years, a large amount of resources has been directed to collect
 experimental evidence of physical proteins binding, in an effort to
 infer and catalogue protein complexes and their multifaceted
 functional roles
 \citep[e.g.][]{Fiel:Song:1989,Ito:Tash:Muta:Ozaw:2000,Uetz:Giot:Cagn:Mans:2000,Gavi:Bosc:Krau:Gran:2002,Ho:Gruh:Heil:Bade:2002}.
 Currently, there are four main sources of interactions between pairs
 of proteins that target proteins localized in different cellular
 compartments with variable degrees of success: (i) literature curated
 interactions \citep{Regu:Brei:Bouc:Brei:2006}, (ii) yeast two-hybrid
 (Y2H) interaction assays \citep{Yu:Brau:Yild:Lemm:2008}, (iii)
 protein fragment complementation (PCA) interaction assays
 \citep{Tara:Mess:Land:Radi:2008}, and (iv) tandem affinity
 purification (TAP) interaction assays
 \citep{Gavi:Aloy:Gran:Krau:2006,Krog:Cagn:Yu:Zhon:2006}.  These
 collections include a total of about 12,292 protein interactions
 \citep{Jens:Bork:2008}, although the number of such interactions is
 estimated to be between 18,000 \citep{Yu:Brau:Yild:Lemm:2008} and
 30,000
 \citep{VonM:Krau:Snel:Corn:2002}. \autoref{fig:yeast_protein_network}
 shows a popular image of the interaction network among proteins in
 the budding yeast, produced as part of an analysis by
 \citet{Bara:Oltv:2004}.

 Statistical methods have been developed for analyzing many aspects of
 this large protein interaction network, including de-noising
 \citep{Bern:Vaug:Hart:2007,Airo:Blei:Fien:Xing:2007}, function
 prediction \citep{Nabi:Jim:Agar:Chaz:2005}, and identification of
 binding motifs \citep{Bank:Nabi:Pete:Sing:2008}.
 
\subsection{The Add Health Adolescent Relationship and HIV Transmission Study}
The National Longitudinal Study of Adolescent Health (Add Health) is a
study of adolescents in the United States drawn from a representative
sample of middle, junior high, and highschools.  The study focused on
patterns of friendship, sexual relationships, as well as disease
transmissions.  To date, four waves of surveys have been collected
over the course of fifteen years.

Wave I surveys occurred between 1994 to 1995 and included 90,118
students from 145 schools across the country.  Each student completed
an in-school questionnaire on his or her family background, school
life and activities, friendships, and health status.  Administrators
from participating schools also completed questionnaires about student
demography and school curriculum and services.  In addition, 20,745
students were chosen for an in-home interview that included more
sensitive topics such as sexual behavior.  For 16 selected schools
(two large and fourteen small), Add Health attempted to administer the
in-home survey to all enrolled students.  This saturated sample
distinguishes itself from the ego-centric and snowball samples
collected from past studies; it allows for the construction of
relationship networks with more accurate global characteristics. The fully observed friendship networks in all the schools are also a valuable resource and an important contribution of this work.

\begin{figure}[t]
\begin{center}
 \includegraphics[width=0.8\textwidth]{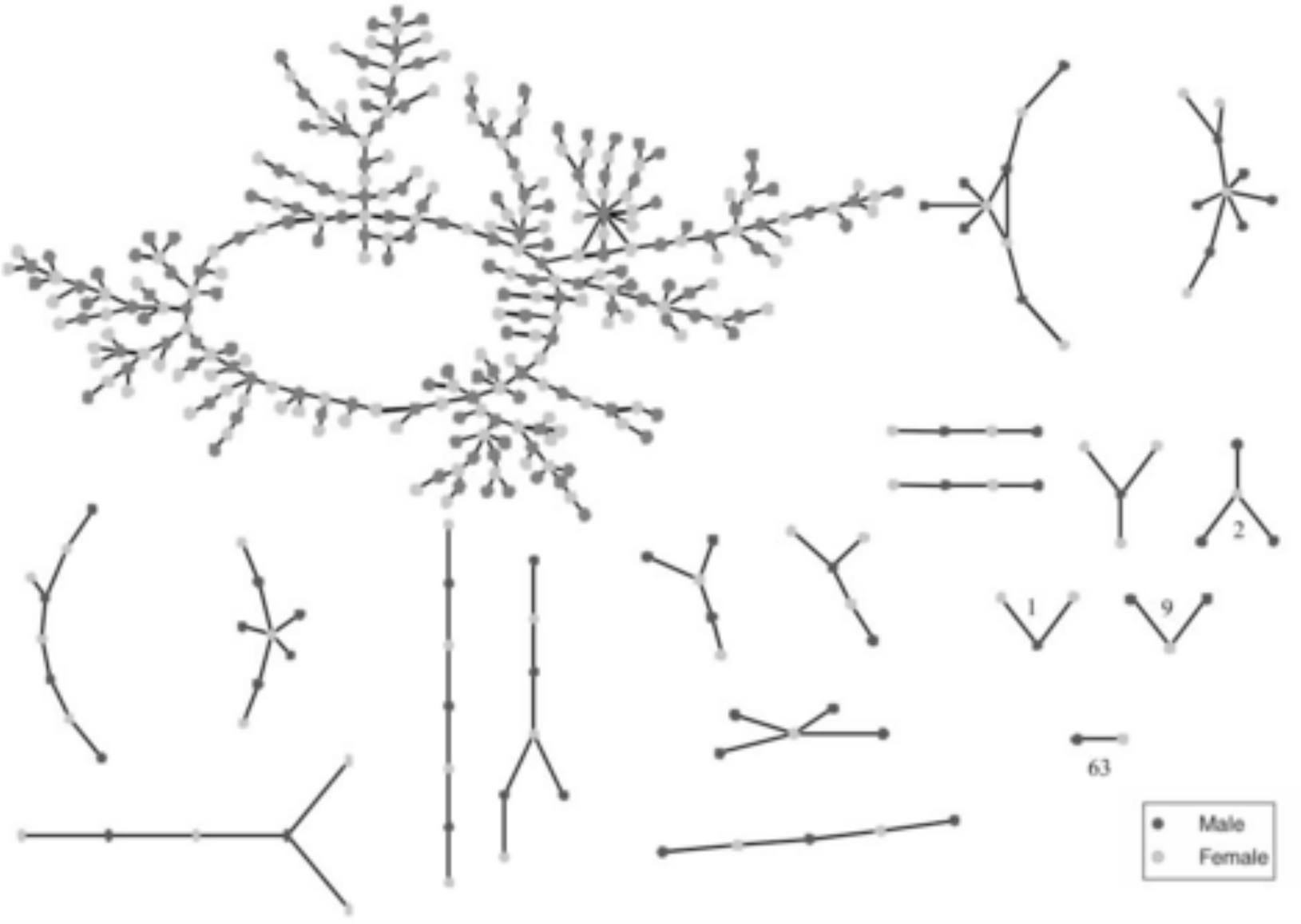}
 \caption{The Add Health sexual relationships network of US highschool adolescents. This figure is reproduced with permission. Source: \citet{Bear:Mood:Stov:2004}}
\label{fig:add-health-bearman}
\end{center}
\end{figure}

Wave II data collection occurred 18-months after Wave I in 1996 and
followed up on the in-home interviews.  The dataset covered 14,738
adolescents and 128 school administrators.  Based on the data
collected from Wave I and II, \citet{Bear:Mood:Stov:2004} constructed the
timed sequence of relationship networks amongst students from the two
large schools with saturated sampling.  The resulting sexual
relationship network bears strong resemblance to a spanning tree as opposed to previously hypothesized core or inverse-core structures\footnote{A core is a group of inter-connected individuals who sit at the center of the graph and interact with individuals on the periphery.  An inverse core is a group of central individuals who are connected to those on the periphery but not to each other.}  (See
\autoref{fig:add-health-bearman}.)

Wave III interviews were conducted in 2001 and 2002 with topics
including marriage, childbearing, and sexually transmitted diseases.
Of the original Wave I in-home respondents, 15,170 were interviewed
again for Wave III.  Of these, 13,184 participants provided oral fluid
specimens for HIV testing.  \citet{Morr:Hand:Mill:Ford:2006} studied
the prevalence of HIV infections among young adults based on data
collected in Wave III.

Wave IV interviews were conducted in 2007 and 2008 with the original
Wave I respondents, who are now dispersed across the nation in all 50
states.  Of the original respondents, 92.5\% were located and 80.3\%
were interviewed.  The interview included a comprehensive survey of
the social, emotional, spiritual, and physical aspects of health.
Physical measurements, biospecimen, and geographical data were also
collected.

For detailed information about the data, as well as access to the
public-domain and restricted-access datasets, see
\url{http://www.cpc.unc.edu/projects/addhealth}.

\subsection{The Framingham ``Obesity'' Study}

One of the most famous and important epidemiological studies was
initiated in Framingham, Massachusetts, a suburb of Boston, in 1948
with an originally enrolled cohort of 5209 people.  In 1971
investigators initiated an ``offspring'' cohort study which enrolled
most of the children of the original cohort and their spouses.
Participants completed a questionnaire and underwent physical
examinations (including measurements of height and weight) in
three-year periods beginning 1973, 1981, 1985, 1989, 1992, 1997,
1999. Christakis and Fowler~\cite{Chri:Fowl:2007} derive body
mass index information on a total of 12,067 individuals who appeared in any of the Framingham Heart cohorts (one ``close friend'' for each cohort
member).\footnote{A body-mass index value (weight in kg. divided by
  the square of the height in meters) of 30 or more was taken to
  indicate obesity.}  There were 38,611 observed family and social
ties (edges) to the core 5,124 cohort members.

\begin{figure}[ht!]
\begin{center}
\includegraphics*[width=\textwidth]{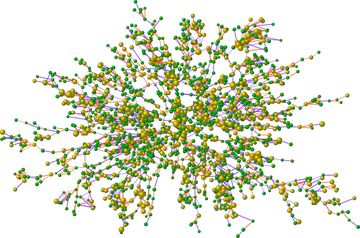}
\end{center}
\caption{Obesity network from Framingham offspring cohort data.  Each node represents one person in the dataset (a total of 2200 in this picture). Circles with red borders denote women, with blue borders {--} men. The size of each circle is proportional to the body-mass index. The color inside the circle denotes obesity status - yellow is obese (body-mass index $\ge30$, green is non-obese. The colors of ties between nodes indicate relationships - purple denotes a friendship or marital tie and orange is a familial tie. This figure is reproduced with permission. Source:~\citep{Chri:Fowl:2007}.}
\label{obesitynetwork}
\end{figure}

Through a series of network snapshots and statistical analyses,
Christakis and Fowler described the evolution of the ``clustering" of
obesity in this social network.  In particular they claim to have
examined whether the data conformed to ``small-world," ``scale-free,"
and ``hierarchical" types of of random graph network models.
\autoref{obesitynetwork} depicts data on the largest connected
subcomponent (the so-called giant component) for the network in 2000,
which consists of 2200 individuals. Other analyses in their paper
explore attributions of the individuals via longitudinal
logistic-regression models with lagged effects.  Subsequently, they
have published similar papers focused on the dynamics of smoking
behavior over time~\citep{Chri:Fowl:2008} and on
happiness~\citep{Chri:Fowl:2008a}, both using the structure of
Framingham ``offspring'' cohort.

This work has come under criticism by others.  For example Cohen-Cole
and Fletcher note that there are plausible alternative explanations to
the network structure based on contextual
factors~\citep{Cohe:Flet:2008}, and in a separate paper
demonstrate that the same methodology detects ``implausible" social
network effects for such medical conditions as acne and headaches as
well as for physical height~\citep{Cohe:Flet:2008a}. The authors answer to these criticisms can be found in \citep{Fowl:Chri:2008}. The question of the magnitude and significance of social network effects is still a subject of an ongoing debate. 

\subsection{The NIPS Paper Co-Authorship Dataset}
\label{sec:nips-data}
The NIPS dataset contains information on publications that appeared in
the {\it Neural Information Processing Systems} (NIPS) conference
proceedings, volumes 1 through 12, corresponding to years
1987-1999---the pre-electronic submission era.  The original
collection contained scanned full papers made available by Yann
LeCunn.  Sam Roweis subsequently processed the data to glean
information such as title, authorship information, and word counts per
document. In total, there are 2,037 authors and 1,740 papers with an
average of 2.29 authors per paper and 1.96 papers per author. The NIPS database is available from Sam Roweis'
website\footnote{\url{http://www.cs.toronto.edu/~roweis/data.html}} in
raw and {\it MATLAB} formats along with a detailed description and
information on its construction.

\begin{figure}[!ht]
\begin{center}
{\resizebox{5in}{2.2in}{\includegraphics{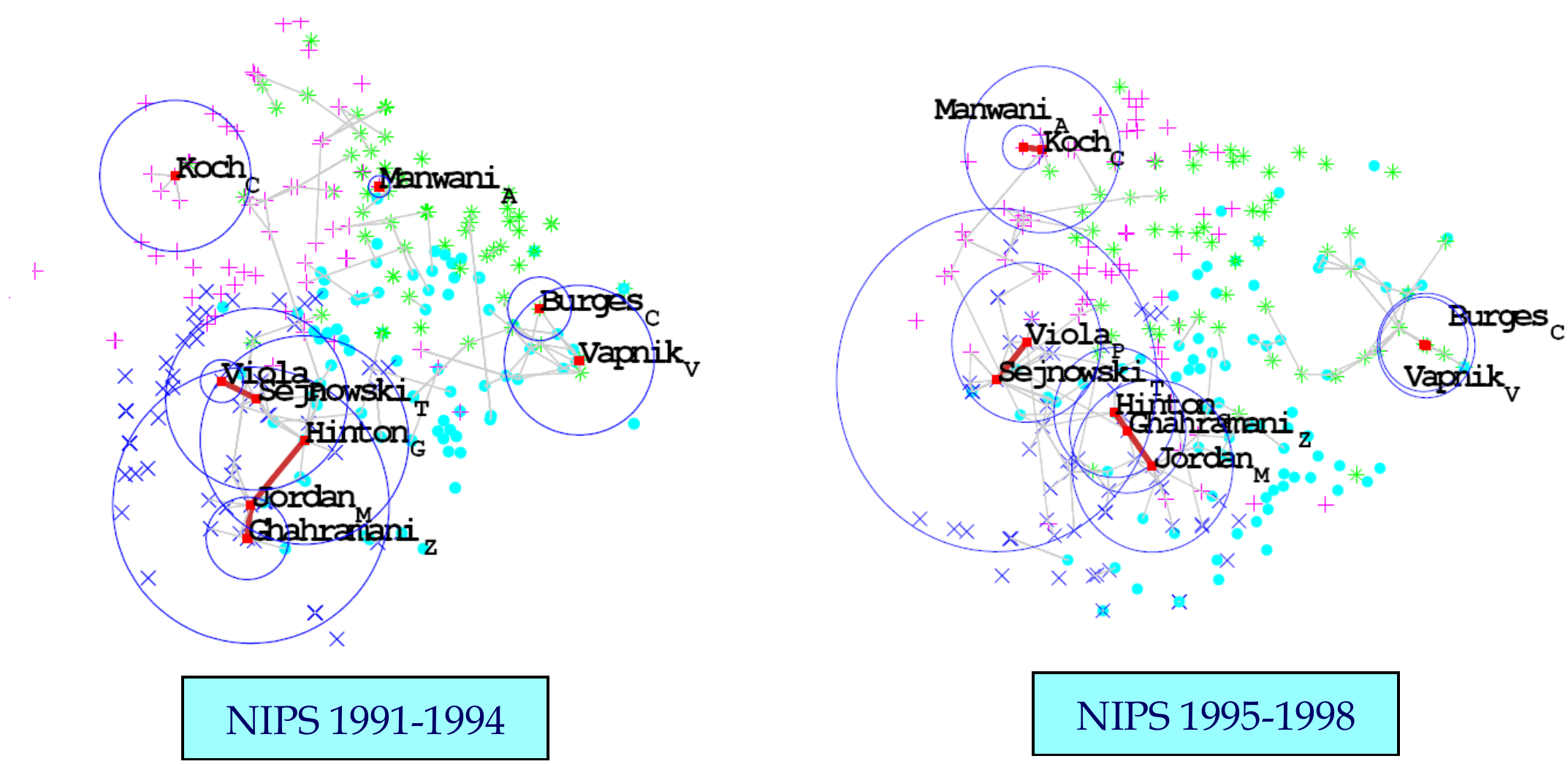}}}
\end{center}
\caption{NIPS paper co-authorship data. Each point represents an
  author.  Two authors are linked by an edge if they have co-authored
  at least one paper at NIPS.  Left: 1991-1994.  Right:
  1995-1998. Each graph contains all the links for the selected
  period. Several well known people in the Machine Learning field are
  highlighted. The size of the circles around selected individuals
  depend on their number of collaborations.  Colors are meant to
  facilitate visualization. This figure is reproduced with permission. Source:~\cite{Sark:Moor:2005}.}
\label{fig:nips}
\end{figure}

Various authors have used the NIPS data  to analyze author-to-author connectivity in
static \citep{Gold:Moor:2004} as well as dynamic settings
\citep{Sark:Moor:2005a}. \citet{Li:McCa:2006} modeled the text of the documents
and \citet{Sark:Sidd:Gord:2007} analyzed the two-mode network
(author-word-author) in a dynamic context.  In \autoref{fig:nips} we
reproduce a graphic illustration of the inferred dynamic evolution of
the network from \citep{Sark:Moor:2005}.  

%%%%%%%%%%%%%%%%%%%%%%%%%%%%%%%%%%%%%%%%%%%%%%%%%%%%%%%%%%%%%%%%%%%%%%%%%%%%%%
%%%%%%%%%%%%%%%%%%%%%%%%%%%%%%%%%%%%%%%%%%%%%%%%%%%%%%%%%%%%%%%%%%%%%%%%%%%%%%
%%%%%%%%%%%%%%%%%%%%%%%%%%%%%%%%%%%%%%%%%%%%%%%%%%%%%%%%%%%%%%%%%%%%%%%%%%%%%%
%%%%%%%%%%%%%%%%%%%%%%%%%%%%%%%%%%%%%%%%%%%%%%%%%%%%%%%%%%%%%%%%%%%%%%%%%%%%%%
%%%%%%%%%%%%%%%%%%%%%%%%%%%%%%%%%%%%%%%%%%%%%%%%%%%%%%%%%%%%%%%%%%%%%%%%%%%%%%
%%%%%%%%%%%%%%%%%%%%%%%%%%%%%%%%%%%%%%%%%%%%%%%%%%%%%%%%%%%%%%%%%%%%%%%%%%%%%%
%%%%%%%%%%%%%%%%%%%%%%%%%%%%%%%%%%%%%%%%%%%%%%%%%%%%%%%%%%%%%%%%%%%%%%%%%%%%%%
%%%%%%%%%%%%%%%%%%%%%%%%%%%%%%%%%%%%%%%%%%%%%%%%%%%%%%%%%%%%%%%%%%%%%%%%%%%%%%
%%%%%%%%%%%%%%%%%%%%%%%%%%%%%%%%%%%%%%%%%%%%%%%%%%%%%%%%%%%%%%%%%%%%%%%%%%%%%%
%%%%%%%%%%%%%%%%%%%%%%%%%%%%%%%%%%%%%%%%%%%%%%%%%%%%%%%%%%%%%%%%%%%%%%%%%%%%%%
% WAS: \input{staticmodels-ema-10-15-08.tex}

\chapter{Static Network Models}
\label{chapter:static}

A number of basic network models are essentially static in nature.
The statistical activities associated with them focus on certain local
and global network statistics and the extent to which they capture the
main elements of actual realized networks. In this chapter, we briefly
summarize two lines of research. The first originates in the
mathematics community with the {\ER} model and led to two types of
generalizations: (i) the ``statistical physics'' generalizations that
led to power laws for degree distributions---the so-called scale-free
graphs, and (ii) the {\EG} models that introduce weak dependences
among the edges in a controlled fashion, which ultimately lead to a
range of more structured connectivity patterns and enable model
comparison strategies rooted in information theory. A second line of
research originated in the statistics and social sciences communities
in response to a need for models of social networks. The {\pone} model
of Holland and Leinhardt, which in some sense generalizes the {\ER}
model, and the more general descriptive family of exponential random
graph models effectively initiate this line of modeling. Some of these
models also have a {\it generative} interpretation that allows us to
think about their use in a dynamic, evolutionary setting.  We define
and discuss popular dynamic interpretations of the data generating
process, including the generative interpretation, in
\autoref{chap:dym}.

\section{Basic Notation and Terminology}

 In theoretical computer science, a graph or network $G$ is often
 defined in terms of nodes and edges, $G \equiv G(\nodeset,\edgeset)$,
 where $\nodeset$ is a set of nodes and $\edgeset$ a set of edges, and
 $N = |\nodeset|, E = |\edgeset|$.  In the statistical literature, $G$
 is often defined in terms of the nodes and the corresponding
 measurements on pairs of nodes, $G \equiv
 G(\nodeset,\relationset)$. $\relationset$ is usually represented as a
 square matrix of size $N \times N$. For instance, $\relationset$ may
 be represented as an adjacency matrix $Y$ with binary elements in a
 setting where we are only concerned with encoding presence or absence
 of edges between pairs of nodes.  For undirected relations the
 adjacency matrix is symmetric.
 
Henceforth we will work with graphs mostly defined in terms of its
set of $N$ nodes and its binary adjacency matrix $Y$ containing
$\sum_{ij} Y_{ij} = E$ directed edges.  Nodes in the network may
represent individuals, organizations, or some other kind of unit of
study.  Edges correspond to types of links, relationships, or
interactions between the units, and they may be directed, as in the
Holland-Leinhardt model, or undirected, as in the {\ER} model. 

A note about terminology: in computer science, graphs contain nodes and
edges; in social sciences, the corresponding terminology is  usually
actors and ties.  We largely follow the computer science terminology in this 
review.

\section{The {\ER} Random Graph Model}
\label{sec:er-model}

The mathematical biology literature of the 1950s contains a number of
papers using what we now know as the network model $G(N,p)$, which for
a network of $N$ nodes sets the probability of an edge between each
pair of nodes equal to $p$, independently of the other edges, e.g.,
see \citet{Solo:Rapo:1951} who discuss this model as a description of
a neural network.  But the formal properties of simple random graph
network models are usually traced back to \citet{Gilb:1959}, who
examined $G(N,p)$, and to \citet{Erdo:Reny:1959}.  The {\ER} random
graph model, $G(N,E)$, describes an undirected graph involving $N$
nodes and a fixed number of edges, $E$, chosen randomly from the
$N\choose 2$ possible edges in the graph; an equivalent interpretation
is that all ${N\choose 2}\choose E$ graphs are equally
likely.\footnote{Both versions are often referred to as
Erd\"os-R\'enyi models in the current literature.}  The $G(N,p)$ model
has a binomial likelihood where the probability of $E$ edges is
\[
 \ell(G(N,p) \hbox{ has $E$ edges}\mid p) = p^E (1-p)^{{N\choose 2}-E}, 
\]
or, equivalently, in terms of the $N\times N$ binary adjacency matrix $Y$
\[
 \textstyle
 \ell(Y\mid p) = \prod_{i\neq j} ~ p^{Y_{ij}} (1-p)^{1-Y_{ij}}.
\]
The likelihood of the $G(N,E)$ model is a hypergeometric distribution
 and this induces a uniform distribution over the sample space of
 possible graphs.  The $G(N,p)$ model specifies the probability of
 every edge, $p$, and controls the expected number of edges, $p\cdot
 {N\choose 2}$. The $G(N,E)$ model specifies the number of edges, $E$,
 and implies the expected ``marginal" probability of every edge,
 $E/{N\choose 2}$.  The $G(N, p)$ model is more commonly found in
 modern literature on random graph theory, in part because the
 independence of edges simplifies analysis \cite[see,
 e.g.,][]{Chun:Lu:2006,Durr:2006}.

 \citet{Erdo:Reny:1960} went on to describe in detail the behavior of
 $G(N,E)$ as $p=E/{N\choose 2}$ increases from 0 to 1.  In the
 binomial version the key to asymptotic behavior is the value of
 $\lambda = pN$. One of the important Erd\"os-R\'enyi results is that
 there is a phase change at $\lambda = 1$, where a giant connected
 component emerges while the other components remain relatively small
 and mostly in the form of
 trees~\citep[see][]{Chun:Lu:2006,Durr:2006}. More formally,
\begin{itemize}
\item[P1.] If $\lambda < 1$, then a graph in $G(N,p)$ will have no connected components of size larger than $O(\log N)$, a.s. as $N\rightarrow\infty$.
\item[P2.] If $\lambda = 1$, then a graph in $G(N,p)$ will have a largest component whose size is of $O(N^{2/3})$, a.s. as $n\rightarrow\infty$.
\item[P3.] If $\lambda$ tends to a constant $c > 1$, then a graph in
  $G(N, p)$ will have a unique ``giant" component
  containing a positive fraction of the nodes, a.s. as $N\rightarrow\infty$. No other component will contain more than $O(\log N)$ nodes, a.s. as $N\rightarrow\infty$.
\end{itemize}

A summary of a proof using branching processes is given in the
appendix of this chapter.  Some of the proof concepts will be useful
for discussion of {\EG} models in \autoref{sec:exchangeableGraph}.

The {\ER} model has spawned an enormous number of mathematical papers
that study and generalize it, e.g., see \citep{Boll:2001}.  But few of
them are especially relevant for the actual statistical analysis of
network data. In essence, the model dictates that every node in a
graph has approximately the same number of neighbors. Empirically
there are few observed networks with such simple structure, but we
still need formal tools for deciding on how poor a fit the model
provides for a given observed network, and what kinds of generalized
network models appear to be more appropriate.  This has led to two
separate literatures, one of which has focused on formal statistical
properties associated with estimating parameters of network
models---the {\pone} and exponential random graph models described
below---and a second that identifies selected predicted features of
models and empirically checks observed networks for those features.
The latter is largely associated with papers emanating from
statistical physics and computer science, several of which are
described in detail in \autoref{chap:dym}.
 
\section{The Exchangeable Graph Model}
\label{sec:exchangeableGraph}

% The {\ER} model and the extensions discussed above focus exclusively on the connectivity within a graph, while ignoring the possible influence of node attributes. In the examples of \autoref{chap:examples_nets}, consider, for instance, the influence that attributes such as location, religious credo and gender may have on the formation of social ties, or the effect of the presence or absence of complementary docking sites on the physical binding between two proteins. 

The exchangeable graph model provides the simplest possible extension
of the original random graph model by introducing a weak form of
dependence among the probability of sampling edges (i.e.,
exchangeability) that is due to {\em non-observable} node attributes,
in the form of node-specific binary strings.
This extension helps focus the analysis, whether empirical or
theoretical, on the interplay between connectivity of a graph and its
node-specific sources of variability \citep{Airo:2006,Airo:2009a}.

 Consider the following data generating process for an exchangeable graph model, which generates binary observations on pairs of nodes.
\begin{enumerate}
 \item[1.]  Sample node-specific $K$-bit binary strings for each node $n \in \mathcal{N}$
 \vspace{3pt}
  \item[]~~$\vec b_n \sim \unif~(\hbox{vertex set of $K$-hypercube})$, 
  \vspace{3pt}
 \item[2.] Sample directed edges for all node pairs $n,m \in \mathcal{N}\times\mathcal{N}$
 \vspace{3pt}
  \item[]~~$Y_{nm} \sim \Bern\bigm(q(\vec b_n, \vec b_m)\bigm)$,
\end{enumerate}
where $\vec b_{1:N}$ are $K$-bit binary strings\footnote{Note that the
space of $K$-bit binary strings can be mapped one-to-one to the vertex
set of the $K$-hypercube, i.e., the unit hypercube in $K$
dimensions.}, and $q$ maps pairs of binary strings into the $[0,1]$
interval.
This generation process induces weakly dependent edges. The edges are
conditionally independent given the binary string representations of
the incident nodes. They are {\em exchangeable} in the sense of De
Finetti \citep{DeFi:1990}.

From a statistical perspective, the exchangeable graph model we survey
here \citep{Airo:2006,Airo:2009a} provides perhaps the simplest step-up
in complexity from the random graph model
\citep{Erdo:Reny:1959,Gilb:1959}. In the data generation process, the
bit strings are equally probable but the induced probabilities of
observing edges are different. A class of random graphs with such a
property has been recently rediscovered and further explored in the
mathematics literature, where the class of such graphs is referred to
as {\em inhomogeneous} random graphs \citep{Boll:Jans:Rior:2007}.
An alternative and arguably more interesting set of specifications can
be obtained by imposing dependence among the bits at each node. This
can be accomplished by sampling sets of dependent probabilities from a
family of distributions on the unit hypercube, $\vec p_n \in [0,1]^K$,
and then sampling the bits independently given these dependent
probabilities.
\begin{enumerate}
 \item[1.]  Sample node-specific $K$-bit binary strings for each node
 $n \in \mathcal{N}$
 \vspace{3pt}
  \item[]~~$\vec p_n \sim \hypercube~(\vec\mu, \sigma,\alpha)$, where
  $\sigma>(K-1)\cdot\alpha>0,$
  \item[]~~$b_{nk} \sim \Bern~(p_{nk})$, for $k=1,\dots,K$ 
  \vspace{3pt}
 \item[2.] Sample directed edges for all node pairs $n,m \in \mathcal{N}\times\mathcal{N}$
 \vspace{3pt}
  \item[]~~$Y_{nm} \sim \Bern\bigm(q(\vec b_n, \vec b_m)\bigm)$,
\end{enumerate}
In the $\hypercube$ distribution\footnote{The hypercube distribution can be obtained using a hierarchical construction as follows. Sample $\vec u \sim$ Normal$_{k}(\vec\mu,\Sigma)$, where $u\in\mathbb{R}^k$ and $\Sigma_{ii}=\alpha, \Sigma_{ij}=\beta$ for $i\neq j$. Then define $p_i=(1 + e^{-u_i})^{-1}$ for $i=1\dots k$. The resulting density for $\vec p$, where $\vec p \in [0,1]^k$ is
\[
f_P(\vec p \mid \vec\mu,\alpha,\beta) = \frac{|2\pi \Sigma|^{-\frac{1}{2}}}{\prod_{j=1}^d p_j (1-p_j)} \exp\left(-\frac{1}{2} \left(\log(\vec p / (1 - \vec p)) - \vec\mu\right)^\prime \Sigma^{-1} \left(\log(\vec p / (1 - \vec p)) - \vec\mu\right)\right).
\]
For more details see \cite{Airo:2009}.}, $\vec\mu,\sigma,\alpha$ control the
frequency, variability and correlation of the bits within a string,
respectively; and $q$ maps binary pairs of strings into the unit
interval.

\begin{figure}[ht!]
 \centering
 \includegraphics[width=.9\textwidth]{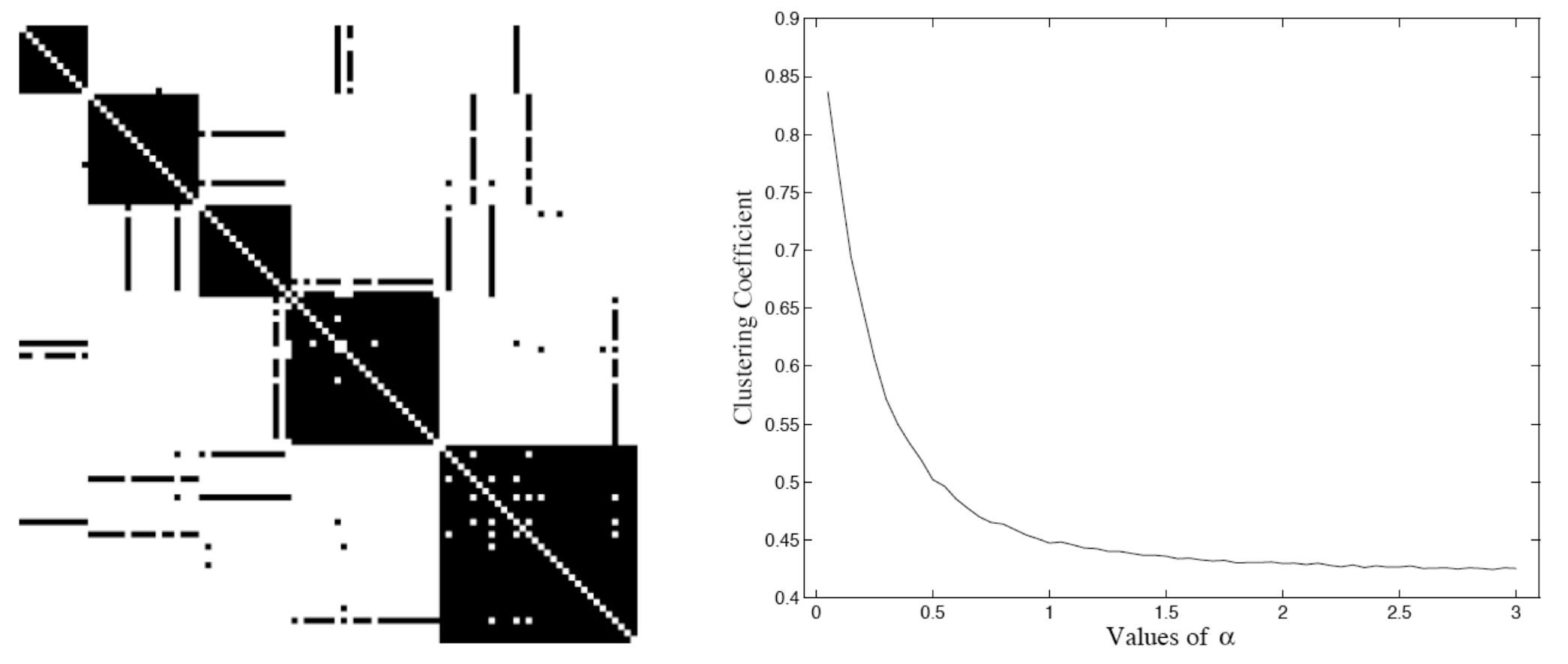}
 \caption{{\em Left panel.} An example adjacency matrix that correspond to a fully connected component among 100 nodes. {\em Right panel.} The clustering coefficient as a function of $\alpha$ on a sequence of graphs with 100 nodes. Here $\sigma=12$, and $\log(\mu_i)=\frac{1}{K}$ for every $i=1\dots K$.}
\label{fig:giant-component}
\end{figure}

In the exchangeable graph model, the number of bits, $K$, captures the
complexity of the graph. For instance, for $K<N$ the model provides a
compression of the graph. For directed graphs the function $q$ is
asymmetric in the arguments. The sparsity of the bit strings is
controlled by the parameter $\alpha>0$.  A larger value of $\alpha$
leads to larger negative correlation among the bits and thereby a
sparser network.
 In such an exchangeable graph model there are two main sources of
 variability: (i) the probability of an edge decreases with the number
 of bits $K$, as more complexity reduces the chances of an edge, and
 (ii) the probability of an edge increases with $1/\alpha$, as
 concentrating density in the corners of the unit $K$-hypercube
 improves the chances of an edge. While this model does not quite fit
 the definition of non-homogeneous models of
 \citet{Boll:Jans:Rior:2007}, it is tractable enough to allow the
 analysis of the giant component in $(K,\alpha)$ space, by leveraging
 the branching process strategy developed by \citet{Durr:2006} (see
 the appendix at the end of the chapter).  As in Durrett's analysis,
 the giant component emerges because a number of smaller components
 must intersect with high probability. In exchangeable graph models
 however, the giant component has a peculiar structure; connected
 components are themselves connected to form the giant component as
 soon as bit strings that match on two bits appear with high
 probability. Figure \ref{fig:giant-component} provides a graphical
 illustration of this intuition.
Nodes that {\em bridge} two connected components are evident in the
left panel. Note that there are no nodes that bridge three components,
as bit strings that match on three bits is an unlikely event in a
graph with 100 nodes.

Given a graph, we can infer the corresponding set of binary strings
from data. The likelihood that correspond to an exchangeable graph
model is simple to write,
\[
 \ell(Y|\theta) = \int d~\vec b_{1:N} \bigm( \prod_{n,m} ~\Pr~(Y_{n,m} | \vec b_n, \vec b_m, q) \prod_n \Pr~(\vec b_n| \theta)\bigm),
\]
where $\theta=(\vec\mu,\sigma,\alpha)$ or an appropriate set of
parameters. We can apply standard inference techniques
\citep{Airo:2007,Airo:Blei:Fien:Xing:2008}.
Fitting an exchangeable graph model allows us to assess the complexity
of an observed graph, leveraging notions from information theory. For
instance, we can use the minimum description length (MDL) principle to
decide how many bits we need to explain the observed connectivity
patterns with high probability. We can also quantify how much {\em
information} is retained at different bit-lengths, and plot the
corresponding information profile for $K<N$ and an entropy histogram
for any given value of $K$.

% Mapping nodes in a graph onto the space of binary strings via the exchangeable graph model, for instance, enables algorithmic comparisons between different statistical models of pairwise measurements, and provides assessments of the statistical significance associated with the observed overlap between two cliques in a graph.
 
The exchangeable graph model allows for algorithmic comparison of any set
of statistical models that are proposed to summarize an observed
graph. As an illustration, consider an observed graph $G$ and two
alternative models $A$ and $B$. Rather than comparing how well models
$A$ and $B$ recover the degree distribution of $G$ or other graph
statistics, and independently of whether it makes sense to
directly compare the two likelihoods of $A$ and $B$ (in fact, these
models need not have a likelihood), we can proceed as follows.
\begin{enumerate}
 \item Given a graph $G$, fit models $A(\Theta_a)$ and $B(\Theta_b)$
   to obtain an estimate of their parameters ${\Theta_a}^{Est}$ and ${\Theta_b}^{Est}$ respectively.
 \item Sample $M$ graphs at random from the support of
   $A({\Theta_a}^{Est})$ and $B({\Theta_b}^{Est})$.
 \item Compute the distributions of summary statistics based on notion
   from information theory, such as information profile and entropy
   histogram, corresponding to the $2M$ graphs sampled from $A$ and
   $B$.
 \item Compare models in terms of the distribution on the statistics
   above, such as the complexity of the two models' supports and their
   similarity to the complexity of $G$.
\end{enumerate}
 
The exchangeable graph model also allows for evaluation of the
distribution of the number of bit strings with $I$ matching bits, for
any integer $I<K$. In theory this distribution leads to expectations
on the number of nodes that bridge $I$ communities, where the members
of each community have only one out of $I$ matching bits. In practice,
we may want to specify $K$ in advance so that each bit corresponds to
a well defined property. For instance, in applications to biology,
nodes may correspond to proteins and the $K$ bits encode
presence or absence of specific protein domains. The distribution on the
number of $I$ matchings leads to p-values that summarize how
unexpected it is to observe binding events among a set of proteins
that share a certain combination of domains.

Overall, the exchangeable graph model introduces weak dependences
among the edges of a random graph in a controlled fashion, which
ultimately lead to a range of more structured connectivity patterns
and enable model comparison strategies rooted in notions from
information theory. The focus here is not on modeling per se. In fact,
the model is kept as simple as possible. Rather, the focus is on
modeling as a means to establish a technical link between graph
connectivity and node attributes. This technical link is useful to
address some of the issues listed in Chapter \ref{ch:issues}. For more
details see \cite{Airo:2009a}.

There exist other complex graph models in the network analysis
literature that induce exchangeable or partially exchangeable edges.
We will discuss latent space models
\citep{Hoff:Raft:Hand:2002,Hand:Raft:Tant:2007} and stochastic
blockmodels
\citep{Nowi:Snij:2001,Airo:Blei:Xing:Fien:2005,Airo:Blei:Fien:Xing:2008}
as examples.  These models can all be traced back to an original
analysis of multivariate sociometric relations, measurements of relations represented as vectors rather than scalars, that was developed a few
decades ago \citep{Fien:Meye:Wass:1985}.  The difference in these
models and the exchangeable graph model lies in the interpretation of
the latent variables and in the goal of the analysis.  Latent space
models interprets the latent variables as latent positions in a social
space, and blockmodels interpret the latent membership vectors in
terms of functional association or community membership.  In the
exchangeable graph model, the latent binary strings do not carry
semantic meaning, rather they are mathematical artifacts that help to
represent a graph and induce an expressive parametric family of
distributions~\citep{Aldo:1985,Kall:2005,Airo:2009a}.  Most
importantly, the exchangeable graph model is meant to be a tool to
{\em represent and explore} the space of connectivity patterns in a
smooth, principled semi-parametric fashion. In this regard,
exchangeable graph models differ substantially from latent space
models or stochastic blockmodels.

\section{The {\pone} Model for Social Networks}
\label{sec:pone}

A conceptually separate thread of research developed in parallel in
the statistics and social sciences literature, starting with the
introduction of the {\pone} model.  Consider a directed graph on the
set of $n$ nodes.  Holland and Leinhardt's {\pone} model focuses on
dyadic pairings and keeps track of whether node $i$ links to
$j$, $j$ to $i$, neither, or both.  It contains the following 
parameters:
\begin{itemize}
\item $\theta$: a base rate for edge propagation,
\item $\alpha _i$ (expansiveness): the effect of an outgoing edge from $i$,
\item $\beta_j$ (popularity): the effect of an incoming edge into $j$,
\item $\rho_{ij}$ (reciprocation/mutuality): the added effect of reciprocated
  edges. 
\end{itemize}
Let $P(0,0)$ be the probability for the absence of an edge between $i$
and $j$, $P_{ij}(1,0)$ the probability of $i$ linking to $j$ (``1"
indicates the outgoing node of the edge), $P_{ij}(1,1)$ the
probability of $i$ linking to $j$ {\emph{and}} $j$ linking to $i$.
The {\pone} model posits the following probabilities (see
\cite{Holl:Lein:1981}):
\begin{align}
    \log P_{ij}(0,0) &= \lambda _{ij}, \\
    \log P_{ij}(1,0) &= \lambda _{ij} + \alpha_i + \beta_j + \theta, \\
    \log P_{ij}(0,1) &= \lambda _{ij} + \alpha_j + \beta_i + \theta, \\
    \log P_{ij}(1,1) &= \lambda _{ij} + \alpha_i + \beta_j + \alpha_j + \beta_i + 2\theta + \rho_{ij}.
\end{align}
In this representation of {\pone}, $\lambda_{ij}$ is a normalizing
constant to ensure that the probabilities for each dyad $(i,j)$ add to
1.  For our present purposes, assume that the dyad is in one and
only one of the four possible states. The reciprocation effect,
$\rho_{ij}$, implies that the odds of observing a mutual dyad, with an
edge from node $i$ to node $j$ and one from $j$ to $i$, is enhanced by
a factor of $\exp({\rho_{ij}})$ over and above what we would
expect if the edges occured independently of one another.

The problem with this general {\pone} representation is that there is
a lack of identification of the reciprocation parameters. The
following special cases of {\pone} are identifiable and of special
interest:
\begin{enumerate}
\item $\alpha_i =0$, $\beta_j =0$, and $\rho_{ij}=0$.  This is
  basically an {\ER} model for directed graphs: each directed edge has
  the same probability of appearance.
\item $\rho_{ij}=0$, {\it no reciprocal effect}.  This model
  effectively focuses solely on the degree distributions into and out
  of nodes.
\item $\rho_{ij}=\rho$, {\it constant reciprocation}.  This was the
  version of {\pone} studied in depth by Holland and Leinhardt using
  maximum likelihood estimation.
%    \item $\rho_{ij}=\rho_i+\rho_j$, edge-dependent.
\item $\rho_{ij}=\rho + \rho_i+\rho_j$, {\it edge-dependent
  reciprocation}.
  \citet{Fien:Wass:1981,Fien:Wass:1981a} described
  this model and how to find maximum likelihood estimate for the
  parameters.
\end{enumerate}
In the constant reciprocation setting, the elevated probability of
reciprocal edges does not depend on the dyad, whereas edge-dependent
reciprocation dictates multiplicative increases of the reciprocation
probability based on node-specific parameters.

The likelihood function for the {\pone} model is clearly in
exponential family form.  For the constant reciprocation version,
we have\\
\begin{equation}
\log {Pr}_{p_1}(y) \propto  y_{++} \theta + \sum_i y_{i+} \alpha_i +\sum_j y_{+j} \beta_j + \sum_{ij} y_{ij}y_{ji} \rho ,
\end{equation}
where a ``+'' denotes summing over the corresponding subscript.  The
minimal sufficient statistics (MSSs) are $y_{i+}$, $y_{+j}$, and
$\sum_{ij} y_{ij}y_{ji}$.  Then using the usual exponential family
theory we know that the likelihood equations are found by setting the
MSSs equal to their expectations (cf.~\cite{Wain:Jord:2008}).  Holland
and Leinhardt gave an explicit iterative algorithm for solving these
equations with the added constraints that the probabilities for each
dyad add to 1.

A major problem with the {\pone} and related models, recognized by
Holland and Leinhardt, is the lack of standard asymptotics to assist
in the development of goodness-of-fit procedures for the model.  Since
the number of {$\{\alpha_i\}$} and {$\{\beta_j\}$} increase directly
with the number of nodes, we have no consistency results for the
maximum likelihood estimates, and no simple way to test for $\rho=0$,
for example.  A few ad hoc fixes have been suggested in literature,
the most direct of which deals with the problem by setting subsets of
the {$\{\alpha_i\}$} and {$\{\beta_j\}$} equal to one another (see the
discussion of blockmodels below) or by considering them as arising from
common prior distributions (see, e.g., \cite{Wang:Wong:1987}).
\citet{Fien:Petr:Rina:2009} recently suggested the use of tools
from algebraic statistics to find Markov basis generators for the
model and the conditional distribution of the data given the MSSs.

Fienberg and Wasserman proposed a slightly different dyad-based data
representation for the {\pone} model.  Conceptually, the dyad
considers the two directed measurements together: $\{D_{ij} =
(y_{ij},y_{ji})\}$. In their work, they define
\begin{equation*}
x_{ijkl} = 
\begin{cases}
  1 \quad\mbox{if}\quad  D(y_{ij},y_{ji})=(k,l),\\  
  0 \quad\mbox{otherwise},
\end{cases}
\end{equation*}
where $k$ and $l$ take the values of 1 or 0.  This representation
converts the dyad $\{D_{ij} = (y_{ij},y_{ji})\}$ into a $2 \times 2$
table with exactly one entry of 1 and the rest 0.  Now if we collect
the data for the $n(n-1)/2$ dyads together, they form an $n \times n
\times 2 \times 2$ incomplete contingency table with ``structural"
zeros down the diagonal of the $n \times n$ marginal (i.e., no self
loops), and ``duplicate" data for each dyad above and below the
diagonal. In this redundant 4-way table, the model of no second-order
interaction corresponds to {\pone} with constant reciprocation, and
the standard iterative proportional fitting algorithm\footnote{For
details on IPF for contingency tables, see
~\cite{Bish:Fien:Holl:1975,Fien:1980}} can be used to compute the
maximum likelihood estimates.  \citet{Fien:Meye:Wass:1985} show that
same type of contingency table representation also works for the
correlated {\pone} model for multiple relations, and Meyer~
\cite{Meye:1982} provides a technical statistical rational for these
contingency table representations.

Holland and Leinhardt analyzed Sampson's monk dataset
(c.f.~\autoref{sec:sampson} and \cite{Samp:1968}) using the {\pone}
model.  \citet{Fien:Meye:Wass:1985} analyzed an 8-relation
version of the Sampson data (4 positive and 4 negative) using their
multiple-relation generalizations of {\pone}, but focusing on an
aggregation of the 18 monks into the three blocks identified
in~\cite{Whit:Boor:Brei:1976}: a top-esteemed block of 7 monks
with an unambivalently positive attitude towards itself, in conflict
with a more ambivalent block of 7, and a block of 4 outcasts and
waiverers.  

\section{{\ptwo} Models for Social Networks and Their Bayesian Relatives}
\label{sec:ptwo}

In the statistical literature, the notion of {\it fixed effects}
typically refers to a set of unknown constant quantities, each of
which is used to partly explain the variability of the observations
corresponding to a unit of analysis, e.g., an individual or a pair of
individuals. This contrasts the notion of {\it random effects}, which
refers to a set of unknown variable quantities that serve a similar
purpose and are drawn from the same underlying distribution.

The {\pone} model treats expansiveness, $\{\alpha _i\}$, and
popularity, $\{\beta_j\}$, as fixed effects associated with unique
nodes in the network.  Often it makes more sense to think about the
ensemble of expansiveness and/or popularity effects as a sample drawn
from some underlying distribution, and then estimate the parameters of
that distribution.  This type of random effects network model
has been developed in a series of papers by Snijders and his
collaborators and they refer to it as the {\ptwo} network model, e.g.,
see~\citet{VanD:Snij:Zijl:2004}.  It is reasonably
straightforward to take any of the multivariate variations on {\pone}
and generate a family of multi-level models with mixtures of fixed and
random effects in the spirit of {\ptwo}, e.g., see
\citet{Zijl:VanD:Snij:2006}.

Bayesian extensions of frequentist approaches often involve positing a
statistical model for fixed effects, thus converting them into random
effects. The principal distinction between the {\ptwo} models and
Bayesian extensions of {\pone} is that, in the latter, the other
unknown constant quantities, $\lambda,\theta,\rho$, may be also
converted into random effects. Furthermore, there may be additional
levels to the multilevel hierarchy in these models, and there are
prior distributions on the parameters at the highest level of the
hierarchy (cf.~\citet{Gill:Swar:2004,Wang:Wong:1987}).  It should come
as no surprise that authors using the Bayesian approach have worked
with Monte Carlo Markov chain (MCMC) methods as have those using
versions of {\ptwo}.

MCMC implementations of {\ptwo} models in STOCNET\footnote{STOCNET is
  a freestanding Software package for the statistical analysis of
  social networks, available at
  \url{http://stat.gamma.rug.nl/stocnet/}.} are well-suited for
  networks with a relatively large number of nodes, e.g.,
  \citet{Zijl:VanD:Snij:2006} study network data from 20 Dutch high
  schools with a total of 1,232 pupils.

\section{Exponential Random Graph Models}
\label{sec:ergm}

% {\em STEVE'S OUTLINE. This section will begin with the fundamental paper of \cite{Fran:Stra:1986} and the idea of pseudo-likelihood estimation in \cite{Stra:Iked:1990} that appears to circumvent the estimation conundrum of using the original apparatus except on very simple problems.  This then led to the trio of papers co-authored by Wasserman, which explained how to implement the strategy in three different forms. Next we have the development of ``full information" MLEs using MCMC by Handcock, Snijders and their students and colleagues.  But these models are no longer have simple exponential family structure and there are degeneracies and near-degeneracies as found in recent papers by Handcock and Fienberg, Rinaldo, and Zhou. END.}

Under the assumption that two possible edges are dependent only if
they share a common node,\footnote{This is the definition of Markov
  property for spatial processes on a lattice in \cite{Besa:1974}.}
\citet{Fran:Stra:1986} proved the following characterization for
the probability distribution of undirected Markov graphs: 
\beq
\Pr_\theta \, \{Y=y\} = \exp \Bigm( \sum_{k=1}^{n-1} \theta_k S_k(y) +
\tau T(y) + \psi(\theta,\tau) \Bigm) \quad y \in \mathcal{Y}, 
\eeq
where $\theta := \{\theta_k\}$ and $\tau$ are parameters,
$\psi(\theta,\tau)$ is the normalizing constant, and the statistics $S_k$
and $T$ are counts of specific structures such as edges, triangles, and
$k$-stars:
\[
 \bv{lcll}
  \hbox{number of edges:} \qquad & S_1(y) & = & \sum_{1\leq i\leq j\leq n} y_{ij}, \\
  &&&\\
  \hbox{number of $k$-stars ($k \geq 2$):} \qquad & S_k(y) & = & \sum_{1\leq i\leq n} \binom{y_{i+}}{k}, \\
  &&&\\
  \hbox{number of triangles:} \qquad & T(y)   & = & \sum_{1\leq i\leq j\leq h\leq n} y_{ij}\,y_{ih}\,y_{jh}.
 \ev
\]
Note that there is a dependence structure to the parameters of this
model, with edges being contained in 2-stars, and 2-stars being
contained in both triangles and three-stars.  Certain variations of
this ERGM model that involve directed edges are natural
generalizations of the {\pone} model.  Alternative parameterizations
that go beyond Markov graph models have been recently proposed, e.g.,
see
\citep{Snij:Patt:Robi:Hand:2006,Wass:Robi:Stei:2007,Bank:Carl:1994}.

\citet{Fran:Stra:1986} worked mainly with the three parameter model
where $\theta_3, \dots, \theta_{n-1}=0$.  They proposed a
pseudo-likelihood parameter estimation method~\citep{Stra:Iked:1990}
that maximizes
\beq 
\ell(\theta) = \sum_{i<j} \log \Bigm( \Pr_\theta
\,\{ Y_{ij} = y_{ij} \,|\, Y_{uv}=y_{uv} \hbox{ for all } u<v, (u,v)
\neq (i,j) \}\Bigm). \nonumber 
\eeq
\citet{Wass:Patt:1996} proposed the current formulation of these {\it
  Exponential Random Graph Models} (ERGM), also referred to as $p^*$
models, as a generalization of the Markov graphs of Frank and Strauss.
For both directed and undirected graphs, they maintain a similar
characterization of the probabilities where the statistics $S_k$ and
$T$ are replaced by arbitrary statistics $U$.  This leads to
likelihood functions of the form 
\beq 
\Pr_\theta \, \{ Y = y \} =
\exp \Bigm( \theta^\top u(y) - \psi(\theta) \Bigm).  
\eeq

The statistics $u(y)$ are counts of graph structures.  Although they
are not independent---they count overlapping sets of edges---they are
assumed independent in the pseudo-likelihood.  Ignoring these
correlations is a bad idea; it causes extreme sensitivity of the
predicted number of edges to small changes in the value of certain
parameters \citep{VanD:Gile:Hand:2009}.  \citet{Park:Newm:2004a}
formally characterized sensitivity issues.
\citet{Snij:Patt:Robi:Hand:2006} recently proposed a variant of these
models where the major problem of double-counting is mitigated but not
overcome.  \citet{Hunt:Hand:2006} estimate likelihood ratios for
nearby $\{\theta_i\}$ using a MCMC procedure related to the work of
\citet{Geye:Thom:1992}.  Their estimation procedure can be used for
models based on distributions in the curved exponential family.
 
\citet{Robi:Snij:Wang:Hand:2007} describe problems associated with the
estimation of parameters in many ERGMs, involving near degeneracies of
the likelihood function and thus of methods used to estimate
parameters using maximum likelihood. For example, for a certain
combination of ERGM statistics, the likelihood function may have
multiple, clearly distinct modes, and there are very few network
configurations---often radically different from each other---that have
non-zero probabilities. This is a topic of current theoretical and
empirical investigation rooted in the theory of discrete exponential
families \citep{Hand:2003,Rina:Fien:Zhou:2009}.  For a discussion of
mixing times of MCMC methods for ERGMs and the relevance to
convergence and degeneracies, see~\cite{Bham:Bres:Sly:2008}.

 There are two carefully constructed packages of routines that are
 available for analyzing network data using ERGMs: {\it statnet}\footnote{A
   package written for the R statistical environment described at
   \url{http://csde.washington.edu/statnet/}.  See also the
   documentation in
   \citep{Hand:Hunt:Butt:Good:2008,Hunt:Hand:Butt:Good:2008,Morr:Hand:Hunt:2008,Good:Hand:Hunt:Butt:2008}.}
 and {\it SIENA}\footnote{Simulation Investigation for Empirical Network
   Analysis---a freestanding package available at
   \url{http://stat.gamma.rug.nl/snijders/siena.html}.}.  These packages focus on the use of MCMC methods for estimating the parameters in ERGMs.  
 
\paragraph{Remark.}
 It is possible to express the current formulation of exponential
 random graphs using the formalism of undirected graphical models and
 the Hammersley-Clifford theorem \citep{Clif:1990,Besa:1974}.  We can
 write the likelihood of an arbitrary undirected graph as \beq \Pr
 (\xv | \bm{\theta}) = \frac{\prod_{c\in\mathcal{C}} \psi(\xv_c |
 \bm{\theta}_c)}{z}, \eeq where $\xv_c$ denotes the nodes in clique
 $c$, $\bm{\theta}_c$ denotes the corresponding set of parameters,
 $\psi$ are non-normalized potentials over the cliques, and $z = \sum
 \prod_{c\in\mathcal{C}} \psi(\xv_c | \bm{\theta}_c)$ is the
 normalization constant.  If the likelihood is in the exponential
 family, then the log potentials are linear in $\theta_c$ and
 ``features'' $u(\xv_c)$, and we can write: 
\bvq 
\Pr(\xv | \bm{\theta}) &=& \exp \Bigm \{ \sum_{c\in\mathcal{C}} \log \psi(\xv_c
 | \bm{\theta}_c) - \log z \Bigm \} \\ 
&=& \exp \Bigm \{ \sum_{c\in\mathcal{C}} \bm{\theta}_c^\top u(\xv_c) - \log z \Bigm \} \\
&=& \exp \Bigm \{ \bm{\theta}^\top u(\xv) - \log z \Bigm \}.  
\evq
 Within the exponential family, the advantage is that computing
 derivatives and likelihood and deriving the corresponding EM
 algorithm are feasible, although possibly computationally expensive,
 by using variational approximation strategies and Monte Carlo
 methods. A lot of methodology on the subject has been developed in
 the area of machine learning. There, undirected graphs appear
 primarily in the context of relational learning and imaging. For an
 in-depth discussion on exact and approximation methods and for
 references see \cite{Ravi:2007,Wain:Jord:2008}.

\section{Random Graph Models with Fixed Degree Distribution}

The {\ER} random graph model is fully symmetric and the expected
degree (the number of edges associated with a node) is the same for
all nodes in the graph, following a binomial distribution.  A number
of natural extensions of the {\ER} model result in varying node 
degrees.  For example,
\begin{itemize}
\item the preferential attachment model~\citep{Bara:Albe:1999}
 captures the formation of hubs in a graph (see
 \autoref{sec:dym-randomgraph});
\item the one-parameter ``small-world" model~\citep{Watt:Stro:1998}
  interpolates between an ordered finite-dimensional lattice and
  an {\ER} random graph in order to produce local clustering and
  triadic closures (see \autoref{sec:dym-smallworld}).
\end{itemize}
Albert and Barab\'asi~\citep{Albe:Bara:2002} describe a number of
variants on these themes.  Many of the investigators exploring the use
of such models often focus on the empirical degree distribution,
claiming for example that it follows a power-law in many real world
networks
(cf.~\citep{Bara:Albe:1999,Newm:2004,Chun:Lu:2006,Durr:2006}).  The
papers utilizing these ``statistical physics" style models often talk
about fixed-degree distributions \cite[e.g.,][]{Park:Newm:2004}, and
they either fix the degree-distribution parameters or compute
distributions that are conditional on some function of the degree
distributions or sequences, such as their expectations
(cf.~\citep{Newm:Watt:Stro:2002,Chun:Lu:Vu:2003}). Software is
available to sample from the space of random graphs with a given
degree distribution based on Monte Carlo Markov chain methods
\citep{Blit:Diac:2006,Hand:Hunt:Butt:Good:2008}.

% We mention this aspect of the literature on random graph models largely as a segue into descriptions of statistical models for graphs that address the issue of empirical degree distributions or sequences more directly, e.g., as the minimal sufficient statistics associate with collections of parameters in the models.

There would appear to be a direct link between these ideas and the
representation of degree distributions in the family of {\pone}
models.  In the latter, the $\alpha_i$ and $\beta_i$ parameters
represent the out-degree and in-degree for the $i$th node, and the
corresponding sufficient statistics are the empirical values for
these.  In the statistical literature there is a long tradition of
looking at distributions conditional on minimal sufficient statistics,
and for network models such a notion was investigated as early as 1975
by Holland and Leinhardt, who looked at the version of {\pone} with
$\rho =0$, conditioned on the empirical in-degree and out-degree for
all nodes in the network \citep{Holl:Lein:1975}.  This allows for the
calculation of an exact distribution that is independent of the $\{
\alpha_i \}$ and $\{ \beta_i \}$ by enumerating all possible adjacency
matrices in the reference set with the observed in-degrees and
out-degrees.  There is the expectation that such an approach could
lead to a uniformly most powerful test for $\rho=0$, but there is no
theory to support this expectation as of yet.  McDonald, Smith and
Forster~\citep{McDo:Smit:Fors:2007} suggest an iterative approach for
such calculations using a Metropolis-Hastings algorithm to generate
from the conditional distribution of the triad census given the
indegrees, the out-degrees and the number of mutual dyads.  In a pair
of papers \citep{Snij:VanD:2002,Snij:Patt:Robi:Hand:2006}, Snijders
and colleagues explore such conditioning for maximum likelihood
estimation for exponential random graph models, largely as a mechanism
for avoiding the degeneracies and near degeneracies observed when
unconditional maximum likelihood is used, cf.~\autoref{sec:ergm} and
\citep{Robi:Snij:Wang:Hand:2007}.  \citet{Snij:2003} does something
similar for dynamic models for graphs.  Roberts~\cite{Robe:2000}
suggests an algorithm for the conditional distribution of the {\pone}
model where $\rho_{ij}=\rho$ given the full set of minimal sufficient
statistics, but \citet{McDo:Smit:Fors:2007} offer a counterexample and
suggest an alteration of their algorithm to generate the proper exact
distribution.  Generating such exact distributions is a very tricky
matter in discrete exponential families because of the need to utilize
appropriate Markov bases, either explicitly as in Diaconis and
Sturmfels~\citep{Diac:Stur:1998} or implicitly. It is unclear whether
the proposals in this literature are in fact reaching all possible
tables associated with the distribution.

\citet{Blit:Diac:2006} explore different efficient
mechanisms for generating random graphs with fixed degree sequence and
explicitly make the link between the ``statistical physics" and
``sociological" literatures, whereas the earlier papers by
\citet{Newm:2004} and \citet{Park:Newm:2004} reference exponential
random graphs but only approach the notion of fixed degree
distributions from a statistical physics perspective, focusing on
characteristics of network ensembles rather that maximum likelihood
estimation and assessment of goodness-of-fit.

\section{Blockmodels, Stochastic Blockmodels and Community Discovery}
\label{sec:mmsb}

A problem which has been a focus of attention for at least 40 years in
the network literature has been the search for an ``optimal partition"
of the nodes into groups or blocks.  In the sociometric literature
this was known as blockmodeling. A formalization of networks in terms
of non-stochastic blocks goes back at least as far as
\citet{Lorr:Whit:1971}. Their paper and the discussion of structural
equivalence gave rise to innumerable papers in mathematical sociology,
(see, e.g., \citep{Burt:1980}) and algorithmic search strategies for
determining blocks (see, e.g.,
\citep{Arab:Boor:Levi:1978,Dore:Bata:Ferl:2004,Dore:Bata:Ferl:2004a}).
By embedding these ideas within a framework of random graphs,
\citet{Holl:Lask:Lein:1983} explained how a special version of {\pone}
could be used to describe a random graph model with predefined
blocks. (See also the related discussion in \citep{Fien:Meye:Wass:1985}
and \citep{Wang:Wong:1987}.)

A true stochastic blockmodel approach, however, involves the {\em
discovery} of the block structure as part of the model search strategy
\citep{Wass:Ande:1987a}, and the first attempts at doing this within
the framework of {\pone} and its exponential family generalizations
was due to Nowicki and Snijders, who focused on technical issues such
as non-identifiability in a restricted version of the blockmodel
\citep{Snij:Nowi:1997,Nowi:Snij:2001,Nunk:Sawi:2005,Copi:Jack:Kirm:2009}. A
comprehensive statistical treatment of these models was recently
developed for analyzing protein interaction data
\citep{Airo:Blei:Xing:Fien:2005,Airo:Blei:Fien:Xing:2007} and then
further developed in the context of social network data
\citep{Airo:Blei:Fien:Xing:2008}. \citet{Hand:Raft:Tant:2007} approach
this stochastic blockmodeling problem through a combination of latent
space models and traditional clustering.  We decribe some of this work
in more detail below.

More recently in the statistical physics and computer science
literatures the problem has gone under the label of detection of
community structure, e.g.,
see~\cite{Girv:Newm:2002,Newm:2004,Clau:2005,Newm:2006,
Shal:Camp:Klin:2007,Mish:Schr:Stan:Tarj:2009}.  This literature is now
voluminous and seemingly unconnected to the statistical blockmodel
work.

The basic idea, in both the model-based and algorithmic approaches as
well as the community detection literature, is
that nodes that are heavily interconnected should form a block or
community.  The nodes are reordered to display the blocks down the
diagonal of the adjacency matrix representing the network.  Moreover,
the connections between nodes in different blocks appear in much
sparser off-diagonal blocks. In model-based approaches, the partition
of the nodes maximizes a statistical criterion linked to the model,
e.g., a likelihood function, whereas most algorithmic solutions
maximize ad hoc criteria related to the ``density" of links
within and between blocks.

More formally, a blockmodel is a model of network data that relies on
the intuitive notion of {\em structural equivalence}: two nodes are
defined to be structurally equivalent if their connectivity with
similar nodes is similar---this is a ``soft'' definition.\footnote{The
term {\em stochastic equivalence} is often used in place of structural
equivalence, e.g., see~\citep{Wass:Faus:1994}.} Following up this
idea, we can imagine collapsing structurally equivalent nodes together
to form a super-node, or a block in the language of
blockmodels. Keeping the notion of a block in mind we can now revisit
and sharpen the definition of structurally equivalent nodes: given $N$
nodes and $K$ blocks, let $Y_{N\times N}$ be the adjacency matrix of
the graph $G(\nodeset,\relationset)$, then two nodes $a$ and $b$ are
structurally equivalent, and thus belong to the same block $h$, if
their connectivity patterns $\mathcal{C}_a$ and $\mathcal{C}_b$ with
nodes in other blocks are similar. The equivalence between
connectivity patterns of nodes $a$ and $b$ can be formally stated as
follows:
\[
 \mathcal{C}_a \equiv \bigm\{ Y(a,i\in h_k): \forall h_k\neq h \bigm\}~\approx~\mathcal{C}_b,
\]
where the index $i$ runs over the nodes other than $a,b$, the index
$k$ runs over the blocks other than $h$, $h_k$ is the set of nodes
in block $k$, and $\approx$ quantifies similarity according to a suitable distance
metric. This definition relies on a pre-specified partitioning of the
$N$ nodes into $K$ blocks.
A blockmodel is useful, for instance, in the analysis of social
relations where blocks may correspond to social factions,
% see~\autoref{fig:example_analysis1}, 
as well as in the analysis of protein interactions where blocks may
correspond to stable protein complexes.
% e.g., see~\autoref{fig:example_analysis2}.

%\begin{figure}[b!]
% \centering
%  \includegraphics[width=0.7\textwidth]{figure_airoldi.pdf} 
%\caption{The estimated blockmodel from whom-do-you-like relations
%  among the 18 monks in Sampson's data offers a succinct and
%  suggestive summary of the social dynamics within the
%  monastery. \citep[Source:][]{Airo:Blei:Fien:Xing:2008}. }
%\label{fig:example_analysis1}
%\end{figure}

%\begin{figure}[t!]
% \centering
%  \includegraphics[width=0.9\textwidth]{figure_krogan.pdf}
%\caption{A blockmodel summarizing interactions among stable protein
%  complexes in budding yeast, formed by groups of tightly interacting
%  proteins. \citep[Source:][]{Krog:Cagn:Yu:Zhon:2006}.}
%\label{fig:example_analysis2}
%\end{figure}

\begin{figure}[t!]
  \includegraphics[width=0.45\textwidth]{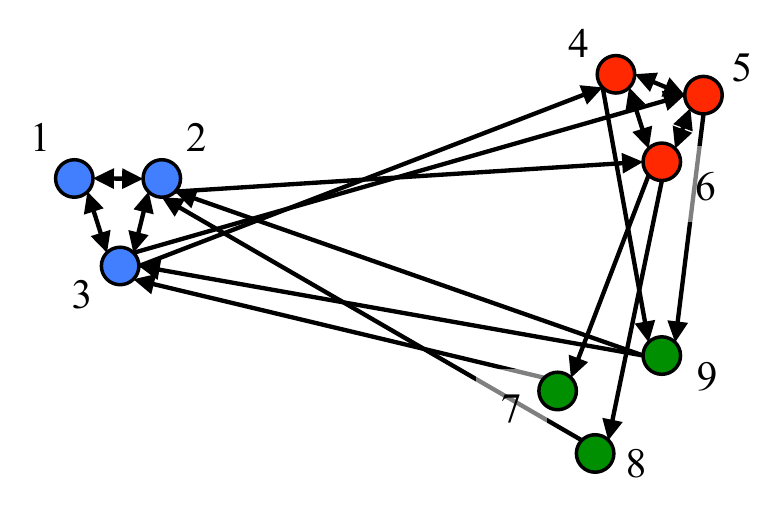} \hfill
  \includegraphics[width=0.45\textwidth]{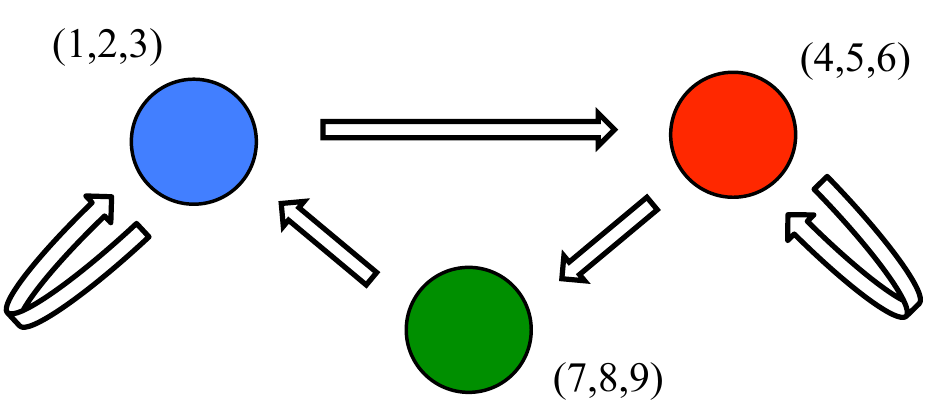}
\caption{Left: An example graph. Right: The corresponding blockmodel,
  where red nodes have been collapsed into the red block and similarly
  for the other colors. Note that this problem is not a typical
  clustering problem, as the green nodes do not share any direct
  connections; each green node, however, has connections directed to
  blue nodes, and connections directed from red nodes. In other words,
  given the partition into monochromatic blocks, the nodes in the
  green block share patterns of connectivity to nodes in other
  blocks.}
\label{fig:example_blockmodel}
\end{figure}

Collapsing nodes into blocks by leveraging the notion of structural
equivalence above is a more general task than clustering. Consider,
for example, the green nodes 7--9 in the left panel of
\autoref{fig:example_blockmodel}. They are structurally equivalent
according to the definition above, as $\mathcal{C}_7\approx
\mathcal{C}_8\approx \mathcal{C}_9$, although there are no direct
connections among the nodes 7--9 themselves. In this sense, nodes 7--9
would not represent a tight cluster according to measures of
similarity based on direct connectivity.
Blocks that would correspond to clusters can be obtained by pre-specifying an identity blockmodel, $B=\mathbb{I}_k$, in which all off-diagonal blocks equal zero and all diagonal blocks equal one.

At the technical level, we need two sets of parameters in order to
instantiate a blockmodel: (i) the blockmodel itself, $B$, is a $K\times K$ matrix, in which the $B(g,h)$ entry specifies, for instance,
the average probability that nodes in block $g$ have connections
directed to nodes in block $h$, and (ii) a mapping between nodes and
blocks, $\vec\pi_{1:N}=\Pi$, where the node-specific array summarizes
some notion of membership. \citet{Airo:Blei:Fien:Xing:2008}, for
instance, specify the mapping in terms of mixed membership arrays, in
which $\pi_n(h)$ specifies the relative frequency of interactions.
Node $n$ participates in 2N-2 interactions in total and instantiates
connectivity patterns that are typical of nodes in its block. These
two sets of parameters, $B$ and $\Pi$, are two latent sources of
variability that compete to explain the observed
connectivity. However, the blockmodel $B$ explains global asymmetric
block connectivity patterns, while the (mixed) membership mapping
$\Pi$ explains node-specific symmetric connectivity patterns. In this
sense, instantiating a blockmodel in terms of $B$ and $\Pi$ does not
introduce any source of non-identifiability beyond the usual
multiplicity of parametric configurations that lead to exactly the
same likelihood---well characterized in this model by
\citet{Nowi:Snij:2001}.
 
As a concrete example, consider the mixed membership stochastic
blockmodel (MMB) introduced by \cite{Airo:Blei:Fien:Xing:2008}; the
data generating process for a graph $G=({\cal N},Y)$ is the following.
\begin{enumerate}
\item[1.] For each node $p \in {\cal N}$:
  \begin{enumerate}
  \item[1.1] Sample mixed membership $\vec{\pi}_p \sim
    \textrm{Dirichlet}_{\,K}\bigm( \vec\alpha \bigm)$.
  \end{enumerate}
\item[2.] For each pair of nodes $(p,q) \in {\cal N} \times {\cal N}$:
  \begin{enumerate}
  \item[2.1] Sample membership indicator, $\vec z_{p
      \rightarrow q} ~ \sim {\rm mult}_{\,K}(\vec\pi_{p})$.
  \item[2.2] Sample membership indicator, $\vec z_{p
      \leftarrow q} ~ \sim {\rm mult}_{\,K}(\vec\pi_{q})$.
  \item[2.3] Sample interaction, $Y(p,q) \sim {\rm
      Bern}~(\vec{z}_{\ptoq}^{~\top} B ~ \vec{z}_{\pfromq})$.
  \end{enumerate}
\end{enumerate}

Note that the group membership of each node is {\em context
  dependent}. That is, each node may assume different membership when
interacting or being interacted with by different peers.  Statistically,
each node is an admixture of group-specific interactions.  The two
sets of latent group indicators are denoted by $\{ \vec{z}_{\ptoq} :
p,q \in \nodeset \} =: Z_\rightarrow$ and $\{ \vec{z}_{\pfromq} : p,q
\in \nodeset \} =: Z_\leftarrow$.  Also note that the pairs of group
memberships that underlie interactions need not be equal; this fact is
useful for characterizing asymmetric interaction networks. Equality
may be enforced when modeling symmetric interactions.

Inference in the blockmodel is challenging, as the integrals that need
to be solved to compute the likelihood cannot be evaluated
analytically. For simplicity, the likelihood is
\[
 \ell(Y\mid\vec\alpha,B) = \int_\Pi \int_Z \Pr (Y\mid Z,B) ~\Pr (Z\mid\Pi) ~ \Pr (\Pi\mid\vec\alpha) ~dZ~d\Pi.
\] 
While the inner integral is easily solvable\footnote{The inner
  integral resolves into a series of sums, each one over the support
  of an individual $\vec z$ variable. The support is the same for all
  such $\vec z$ variables, and it is given by the $N$ vertices of the
  $K$-dimensional unit hypercube. In other words, the inner integral
  is a series of sums, each over the same $N$ elements.}, the outer
  integral is not. Exact inference is thus not an option.  To
  complicate things, the number of observations scales as the square
  of the number of nodes, $O(N^2)$. Sampling algorithms such as Monte
  Carlo Markov chains are typically too slow for real-size problems in
  the natural, social, and computational
  sciences. \citet{Airo:Blei:Fien:Xing:2008} suggest a nested
  variational inference strategy to approximate the posterior
  distribution on the latent variables, $(\Pi,Z)$. (Variational methods
  scale to large problems without loosing much in terms of accuracy
  \citep{Airo:2007a,Brau:McAu:2007,Wain:Jord:2008}.)

Bickel and Chen~\cite{Bick:Chen:2009}, the most recent contribution to
this literature, brings new twists to the model-based approach of
community discovery. They use a blockmodel to formalize a given
network in terms of its community structure. The main result of this
work implies that community detection algorithms based on the
modularity score of Newman and Girvan~\cite{Girv:Newm:2002} are
(asymptotically) biased. It shows that using modularity scores can
lead to the discovery of an incorrect community structure even in the
favorable case of large graphs, where communities are substantial in
size and composed of many individuals. This work also proves that
blockmodels and the corresponding likelihood-based algorithms are
(asymptotically) unbiased and lead to the discovery of the correct
community structure. The proof relies on the exchangeability results
developed in the statistics community~\cite{Aldo:1985,Kall:2005}
applied to paired measurements~\cite{Diac:Jans:2008}.

\section{Latent Space Models}
\label{sec:latent}

The intuition at the core of latent space models is that each node $i
\in \nodeset$ can be represented as a point $\latent_i$ in a ``low
dimensional'' space, say $\mathbb{R}^k$. The existence of an edge in
the adjacency matrix, $Y(i,j)=1$, is determined by the distance among
the corresponding pair of nodes in the low dimensional space,
$d(\latent_i,\latent_j)$, and by the values of a number of covariates
measured on each node individually. The latent space model was first
introduced by \citet{Hoff:Raft:Hand:2002} with applications to social network
analysis, and has been recently extended in a number of directions to
include treatment of transitivity, homophily on node-specific
attributes, clustering, and heterogeneity of nodes
\citep{Hoff:2003,Hand:Raft:Tant:2007,Kriv:Hand:Raft:Hoff:2009}.

The conditional probability model for the adjacency matrix $Y$ is
\[
 \Pr (Y \mid \Latent,X, \Theta)= \prod _{i \neq j} \Pr ( Y(i,j) \mid \Latent_i, \Latent_j , X_{ij} , \Theta),
\]
where $X$ are covariates, $\Theta$ are parameters, and $\Latent$ are
the positions of nodes in the low dimensional latent space. Each
relationship $Y(i,j)$ is sampled from a Bernoulli distribution whose
natural parameter depends on $\Latent_i$, $\Latent_j$, $X_{ij}$ and
$\Theta$. In their model, \citet{Hoff:Raft:Hand:2002} generated the
paired observations $Y(i,j)$ starting from the relevant pair of node
representations, $(\Latent_i, \Latent_j)$, through a distance model,
pair specific covariates $X_{ij}$, and parameters $\Theta =
(\alpha,\beta)$. The log-odds ratio is then:
\[
  \log \frac{\Pr(Y(i,j)=1)}{1-\Pr(Y(i,j)=1)}
  = \alpha + \beta'X_{ij} - |\Latent_i - \Latent_j|
  \equiv \eta_{ij}, 
\]
and the corresponding log likelihood is
\[
 \log \Pr (Y\mid\eta) = \sum_{n\neq m} \bigm( \eta_{ij}\cdot Y_{ij} -\log (1+e^{\eta_{ij}}) \bigm).
\]

One can easily extend the latent space modeling approach to weighted
networks. In the general case, paired observations $Y$ may be modeled
using a generalized linear model that makes use of $\Latent_{1:N}$
$X_{ij}$, and $\Theta$.
Following the formalism in \cite{McCu:Neld:1989}, a generalized
linear model that generates the observed edge weights can be
specified in terms of three quantitative elements:
\begin{itemize}
 \item[i.] the error model $\Pr (Y_{ij})$, i.e., the model for the
   observed edge weights with mean $\mu_{ij} = \mathbb{E} [ y_{ij} ]$;
 \item[ii.] the linear model $\eta_{ij} = \eta_{ij} ( \beta,\Latent_i,\Latent_j )$; %, or rewritten as $\eta_{ij} ( \beta, d(\Latent_i,\Latent_j) )$, for any explicit distance model $d$.
 \item[iii.] the link function $g(\mu_{ij}) = \eta_{ij}$, which maps
   the support of $\mu_{ij}$ to that of $\eta_{ij}$---typically
   $\mathbb{R}$.
\end{itemize}
For example, in the binary graph, the error model is $\Pr (Y_{ij}) =
\Bern(\mu_{ij})$, where $\mu_{ij} \in [0,1]$ for all node pairs $(i,j)
\subseteq \nodeset$; the linear model is $\eta_{ij} = \beta +
d(\Latent_i,\Latent_j)$; the link function is $g(\mu_{ij}) = \log
\bigm( \frac{\mu_{ij}}{1-\mu_{ij}} \bigm)$, with its inverse being
$\mu_{ij} = \frac{1}{1+\exp(-\eta_{ij})}$ \citep{Hoff:Raft:Hand:2002}.
In a graph with non-negative, integer edge weights, we can posit
$\Pr(Y_{ij}) = \Poi~(\mu_{ij})$, where $\mu_{ij} \in \mathbb{R}_+$ for
all node pairs $(i,j) \in \nodeset$; the linear model is $\eta_{ij} =
\beta + d(\Latent_i,\Latent_j)$, the same as in the previous example;
the link function is $g(\mu_{ij}) = \log (\mu_{ij})$, and its inverse
is $\mu_{ij} = e^{\eta_{ij}}$.

In the general case, the generalized linear model for $\eta_{ij}$ may
also include an explicit distance model $d$ in the latent space
$\mathcal{\Latent}$:
\bvq
 \eta_{ij} & = & \eta_{ij} \bigm( \beta, \Latent_i, \Latent_j \bigm) \\
           & = & \eta_{ij} \bigm( \beta, d(\Latent_i, \Latent_j) \bigm).
\evq
Note that it is possible to re-parametrize $\Latent_i = \rho_i \omega_i$ to
separate the position in a latent reference space, $\Omega$,
from its magnitude, $\rho_i$, a scalar.
It is a simple intuition that suggests the use of an explicit distance
model in the latent space.  In a binary graph, for example, edges are
more likely to be generated between pairs of nodes whose
representations in the latent space are close.
A popular choice of distance measures is Euclidean
distance. Estimation can be done via MCMC sampling.

Inference in latent space models has been carried out via Monte Carlo
Markov chain in networks with up to several thousand nodes
\citep{Good:Kitt:Morr:2009}. Scalability issues remain to be addressed
before larger networks can be analyzed.

% T. Brendan Murphy is working on variational inference or latent space models. Let's check with him before going to press for possible addition to references.

\subsection{Comparison with Stochastic Blockmodels}

The latent space model of \citet{Hoff:Raft:Hand:2002} projects nodes
onto a latent Euclidean space by inverting the logistic link. While in
practice there is often interest in identifying groups of similar
nodes, e.g.\ individuals or proteins, there is no explicit clustering
model in the latent space. To identify groups of similar nodes,
clustering methods must be used to analyze the set of latent positions
inferred by the latent space model. To allow joint inference on latent
positions and clusters, \citet{Hand:Raft:Tant:2007} introduce an
explicit clustering model in the latent space in the form of a mixture
of (spherical) Gaussians.
\[
 \left\{ 
 \begin{array}{rcl}
  \Pr (Y \mid \Latent,X, \Theta) &=& \prod _{i \neq j} \Pr ( Y(i,j) \mid \Latent_i, \Latent_j , X_{ij} , \Theta),\\
  \Latent_i &\sim& \sum_k N~(\mu_k,\sigma_k^2\cdot I).  
 \end{array}
 \right.
\]
This model combines the original latent space model \cite{Hoff:Raft:Hand:2002} with a finite mixture of Gaussians approach to clustering \citep{Titt:Smit:Mako:1985,Mari:Meng:Robe:2005}. It posits that the latent positions $Z_i \in \mathbb{R}^d$ come from a $k$-dimensional mixture model.
 
This extension is related to the stochastic blockmodel of
\cite{Airo:Blei:Fien:Xing:2008}, which posits a latent membership
vector for each node.  These vectors can be viewed as cluster
assignment probabilities for each node.  The observed binary
relationships between nodes are mediated by per-pair latent variables,
each drawn conditioned on a node's mixed membership vector.  In its
general form, the blockmodel allows for multiple relations and
covariates.
Similarly, the model in \cite{Hand:Raft:Tant:2007} is also a
hierarchical model, as a Gaussian distribution is placed on the latent
positions $Z_i$.  In contrast, however, each node belongs to a single
cluster and the corresponding partition governs the observed
relationships.  There can be variance in the latent position
variables, but the idea of belonging to two or more groups cannot be
represented.  Posterior uncertainty about cluster membership is
different from having an explicit distribution that controls mixed
membership, which carries with it an additional level of uncertainty.
With that said, the latent space in which nodes are projected in
\cite{Hand:Raft:Tant:2007} is somewhat comparable to the space of
cluster proportions in \cite{Airo:Blei:Fien:Xing:2008}.  The former
maps nodes to a Euclidean space, while the latter maps nodes to the
simplex.

Both models share the same goal: inferring latent structure that
explains the variability of the connectivity in an observed network.
In the mixed membership model, full MCMC for any but the simplest
problems is unreasonably expensive.  \citet{Airo:Blei:Fien:Xing:2008}
appeal to variational methods for a computationally efficient
approximation to the posterior.  These methods can scale to large matrices (e.g., millions of nodes) because of the simplified approximation, but at an unknown
cost to accuracy.  It would be interesting to explore computational
tradeoffs for the latent space cluster model
\citep{Hand:Raft:Tant:2007} as the sample size grows and when large
numbers of covariates are added.
 
\paragraph{Remark.}

\citet{Blei:Fien:2007} argue that a stochastic
blockmodel and node-specific mixed membership vectors are two sets of
parameters that are directly interpretable in terms of notions and
concepts relevant to social scientists, and better suited to assist
these scientists in extracting substantive knowledge from noisy data,
to ultimately inform or support the development of new hypotheses and
theories.

Applying the mixed membership stochastic blockmodel (MMB) to SampsonÕs
data demonstrates both similarities and differences \citep{Airo:2007}.
For instance, BIC suggests the existence of three factions among the
18 monks when fitting the MMB, but the groupings differ from those
found by the latent space cluster model.  
\comment{(This is too
detailed.  There's no such comparison in the rest of the paper -- az)
In the analysis with the former, for instance, Romuald and Victor (two
of SampsonÕs waverers) stand out; and so do Greg and John who were
expelled from the monastery first. The map of hierarchical relations
among factions is specified by a ($3 \times 3$) matrix of Bernoulli
hyper-parameters, $B$, where $b(i,j)$ is the probability that monks in
the $i$th faction relate to those in the $j$th faction. The analysis
with the MMB model centers on the membership of monks in factions. For
instance, }
One major benefit of applying mixed membership model to the
data is the ability to quantitatively identify two out of three of the
novices that Sampson labeled as {\em waverers} in his analysis based
on anthropological observations.  This could lead to the formation of
a social theory of failure in isolated communities, with a possibility
to be confirmed with real longitudinal data \citep{Airo:2007}.\\

In the sociology literature, certain specifications of blockmodels are referred to as {\em latent class} models, and certain specifications of latent space models are referred to as {\em latent distance} models. Hoff~\citep{Hoff:2007} provides a nice comparison, both theoretical and empirical, of these two types of models with the {\em eigenmodel}. The eignemodel is based on a singular value decomposition of the socio-matrix, it can capture more connectivity patterns than the latent class and the latent distance models, for a  {\em given degree} of model complexity, which can be the number of classes, the number of dimensions in the latent space, or the number of eigenvectors. There is a price to pay, however. The eigenmodel is the least amenable to interpretation among the three models, as the inferred patters that capture connectivity are in terms of eigenvectors. The latent space model can be interpreted in terms of distances. The latent class models can be interpreted in terms of blocks of connectivity, or tight micro-communities; this is the easiest model to interpret.
 
% + some other finnish group (PCA) style mixed-membership
% + Andrew McCallum's LDA models
% + Parkinen's interaction components model 

\section*{Appendix: Phase Transition Behavior of the {\ER} Model}
\label{sec:erPhaseTrans}

 A simple way to analyze the phrase transition behavior of {\ER}
 models at $\lambda = 1$ is to study the emergence of the giant
 component as a branching process \citep{Durr:2006}. Intuitively,
 consider branching processes that start at every node: for certain
 values of $\lambda$ all the branching processes will keep growing
 with high probability.  Their supports, i.e., the sets of nodes
 involved in each process, will intersect with high probability,
 leading to the emergence of the giant component, $G$, in which each
 node can be reached from every other node.
 
 The following formal argument comes from lecture notes by
 \citet{Guet:Cons:2007} based on proofs given by
 \citet{Jans:Lucz:Ruci:2000}.  Pick a node $v \in \nodeset$.
 If $v$ is connected to all of the nodes in $G$,
then we say that $v$ is {\it saturated} in $G$. Now work as follows:
pick a node $v$ and place it on the list. Then, identify all its
neighbors in $G$, and add them to the list. Next, take the first
unsaturated node on the list and add to the list all of its neighbors
which are not already in it. The proof is constructed by considering the distribution of the number of nodes an unsaturated node adds to the list and by using Chernoff bounds to bound the size of the connected component each node belongs to. For details on this proof please see \citep{Boll:2001}. 
\comment{We can show that the distribution of the
number of nodes that an unsaturated node adds to the list is
\[
 X_k \sim Binomial~(N-k,p) ~\approx  Binomial~(N,p)
\]
where $k$ are the number of nodes already in the list---typically $N$
is large compared to $k$, and so we can ignore the latter. To prove
the proposition above, consider the probability that node $v$ belongs to a
connected component of size at most $k$,
\[
 \Pr~(X_1+X_2+\ldots+X_k \geq k-1).
\]
We can bound each random variable $X_i$ from above by a random
variable $X_{i}^{+}$ which has the Binomial$(N,p)$ distribution.
The Chernoff bounds for a random variable $X$ with distribution
Binomial$(N,p)$ are
\begin{eqnarray}
\label{eq:cb+}
 \Pr\bigm(X > Np(1+\epsilon)\bigm) &\leq& \exp\left(\frac{Np\epsilon^2}{3}\right) \\
\label{eq:cb-}
 \Pr\bigm(X < Np(1-\epsilon)\bigm) &\leq& \exp\left(\frac{-Np\epsilon^2}{2}\right).
\end{eqnarray}
 
Using \autoref{eq:cb+}, we have 
\begin{eqnarray*}
 \Pr~(X_1+X_2+\ldots+X_k \geq k-1)
  &\leq& \Pr~(\sum_i X_{i}^{+}  \geq k-1) \\
  &\leq& \exp\left(\frac{-(1-\epsilon)^2)k}{2}\right) \\        
\end{eqnarray*}
Using $k=3\log N/(1-\epsilon)^2$ yields the probability that any node
is part of a component of size $k$ or greater as being $\le
N^{-3/2}=k^{+}$.  Along the same lines, we can bound the $X_i$ from
below and apply the other Chernoff bound in
\autoref{eq:cb-}, to get a value $k^{-}$.  Then we can show
that for $k^{-}<k<k^{+}$, with high probability the process either
dies out before time $k^{-}$ or reaches time $k$ with at least
$(\epsilon -1)/2$ nodes.}
\citet{Boll:Jans:Rior:2007} carried out an extensive
 analysis of the phase transition that mathematically characterizes
 emergence of the giant component in inhomogeneous random graphs.

%%%%%%%%%%%%%%%%%%%%%%%%%%%%%%%%%%%%%%%%%%%%%%%%%%%%%%%%%%%%%%%%%%%%%%%%%%%%%%
%%%%%%%%%%%%%%%%%%%%%%%%%%%%%%%%%%%%%%%%%%%%%%%%%%%%%%%%%%%%%%%%%%%%%%%%%%%%%%
%%%%%%%%%%%%%%%%%%%%%%%%%%%%%%%%%%%%%%%%%%%%%%%%%%%%%%%%%%%%%%%%%%%%%%%%%%%%%%
%%%%%%%%%%%%%%%%%%%%%%%%%%%%%%%%%%%%%%%%%%%%%%%%%%%%%%%%%%%%%%%%%%%%%%%%%%%%%%
%%%%%%%%%%%%%%%%%%%%%%%%%%%%%%%%%%%%%%%%%%%%%%%%%%%%%%%%%%%%%%%%%%%%%%%%%%%%%%
%%%%%%%%%%%%%%%%%%%%%%%%%%%%%%%%%%%%%%%%%%%%%%%%%%%%%%%%%%%%%%%%%%%%%%%%%%%%%%
%%%%%%%%%%%%%%%%%%%%%%%%%%%%%%%%%%%%%%%%%%%%%%%%%%%%%%%%%%%%%%%%%%%%%%%%%%%%%%
%%%%%%%%%%%%%%%%%%%%%%%%%%%%%%%%%%%%%%%%%%%%%%%%%%%%%%%%%%%%%%%%%%%%%%%%%%%%%%
%%%%%%%%%%%%%%%%%%%%%%%%%%%%%%%%%%%%%%%%%%%%%%%%%%%%%%%%%%%%%%%%%%%%%%%%%%%%%%
%%%%%%%%%%%%%%%%%%%%%%%%%%%%%%%%%%%%%%%%%%%%%%%%%%%%%%%%%%%%%%%%%%%%%%%%%%%%%%
% WAS: \input{dynamicmodels.tex}

\chapter{Dynamic Models for Longitudinal Data}
\label{chap:dym}

In \autoref{chapter:static} we focused on models for static networks,
that consider a cross-section of a real network at a given point in
time.  However, real networks often contain a dynamic component. In
the language of networks, dynamics can be translated into the birth
and death of edges and nodes.  For example, in a friendship network,
new nodes may be introduced at any time and old nodes may drop out due
to inactivity; links of friendships and alliances may be even more
brittle.  Dynamic network modeling has been a neglected sibling of
static network modeling, partly due to the added complexity and partly
due to a lack of datasets to study.  Sampson's monastery
study~\citep{Samp:1968} produced one of the earliest datasets with
information on the dynamics in the network of the 18 initiates. The
original research, however, focused on the network structure at each
given time point, rather than modeling the underlying dynamics
explicitly. As online communities gain in popularity, we are beginning
to get access to an increasing number of dynamic network datasets of
much larger size and longer time span.  At the same time, advances in
statistical and computational methods for inference and learning have
enabled development of richer models.  Bearing this in mind, in this
chapter we consider three different classes of models. We begin by
revisiting the {\ER} random graph model and its generalizations,
viewing them as models for dynamic processes. Then we turn to
continuous time Markov process models (CMPM) and their discrete time
cousins (such as a dynamic version of ERGM and other recently
proposed models).

\section{Random Graphs and the Preferential Attachment Model}
\label{sec:dym-randomgraph}

Many variations on the classical {\ER} random graph model in
\autoref{sec:er-model} are typically considered to be static models,
in that they model a single, static snapshot of the network, as
opposed to multiple snapshots recorded at different time steps.
However, they also contain processes for link addition and
modification, which is a dynamic process that may have generated the
observed graph, though there is no attempt to fit these dynamic model
properties to observed data.  For this reason, we view them as
``pseudo-dynamic'' models and discuss three examples here: the {\ER}
model, preferential attachment model, and small-world models.

For example, we can view the {\ER} model {$G(N,E)$}, itself as
a dynamic process used to generate a random graph:
\begin{itemize}
\item start from the graph of $\node$ unconnected nodes at time 0;
\item at each subsequent time step, add a different edge to the
  network with probability $p=E/{N\choose 2}$.
\end{itemize}
By convention, we usually fix the number of nodes at $\node$, although
we can extend the process to allow for addition of nodes.  This model
assumes that edges (and nodes) are not removed once they are
added. The degree distribution for $G(N,E)$ is binomial.  But as
$\node$ gets large, $Np$ tends to a constant, so it is approximately
Poisson. \citet{Durr:2006} provides a rich discussion for situating
this dynamic description with the tradition of discrete time random
walks and branching processes. In particular, he uses this
representation to explore the emergence of the giant component
described in \autoref{sec:er-model} (see appendix of
\autoref{chapter:static}).

The {\ER} model is simple and easy to study but does not address many
issues present in real network dynamics. One of the major criticisms
\citep{Bara:Albe:1999} of this model centers on the fact that it does
not produce a scale-free network, i.e., the resulting node degree
distribution does not follow a power law.  The network literature is
replete with claims that many real networks exhibit the power-law
phenomenon, (cf. \citep{Albe:Bara:2002}), and much subsequent research
has focused on how various generalizations of the {\ER} model conform
to the power law degree distribution.  \citet{Moll:Reed:1995} were the
first to describe how to construct graphs with a general degree
distribution and they went on to describe the emergence of the giant
component in that context as well~\citep{Moll:Reed:1998}.

\citet{Bara:Albe:1999} described a dynamic preferential attachment (PA)
model specifically designed to generate scale-free networks.    At time
0, the model starts out with $\node_0$ unconnected nodes.  At each
subsequent time step, a new node is added with $m \leq \node_0$ edges.
The probability that the new node is connected to an existing node is
proportional to the degree of the latter.  In other words, the new
node picks $m$ nodes out of the existing network according to the
multinomial distribution
\[ p_i = \frac{\deg_i}{\sum_j \deg_j}, \]
where $\deg_i$ denotes the (undirected) degree of node $i$.  This
model, which was described much earlier in the statistical literature
by \citet{Yule:1925} and \citet{Simo:1955}, is intended to describe
networks that grow from a small nucleus of nodes and follow a
``rich-get-richer" scheme.  The assumption is that, for instance, a
new web page will more likely link via a URL to a well-known web page
as opposed to a little-known one.  \citet{Mitz:2004} gives a brief
history of generative models for power law distributions.

The preferential attachment model of Barab\'asi and Albert results in
a network with a power law degree distribution whose exponent is
empirically determined to be $\pexp_{BA} = 2.9 \pm 0.1$, whereas the
{\ER} model has a Poisson degree distribution.  Many extensions of the
model have been proposed that allow for flexible power-law exponents,
edge modifications, non-uniform dependence on the node degree
distributions, etc.  For example, \citet{Doro:Mend:2000a} proposed
that creating an edge to node $i$ should be proportional not just to
its degree $k_i$ but also to its age, decaying as $(t-t_i)^{-\nu}$,
where $\nu$ is a tunable parameter. This leads to a power law degree
distribution only if $\nu < 1$.  \citet{Bara:Jeon:Neda:Rava:2002} and
\citet{Durr:2006} provide an account of this and other extensions to
the original model of Albert and Barab\'asi. Alternative graph
generation mechanisms appear every day---R--MAT
\cite{Chak:Zhan:Falo:2004},`winners don't take all'
\cite{Penn:Flak:Lawr:Glov:2002},`forest fire'
\cite{Lesc:Klei:Falo:2007},`butterfly' \cite{McGl:Akog:Falo:2008} and
RTG \cite{Akog:Falo:2009}, to name a few. The latest, RTG model,
proves conformance to $11$ empirical laws observed in real networks.
\comment{ \citet{Durr:2006} also describes the linkage of these models
for scale-free graphs to susceptible-infected-susceptible epidemics. }
The main goal of these random graph models is to describe a process
that could generate networks emulating certain known network
properties.  The generative process could then give an insight into
the dynamics that led to the observed network. But these models are
often applied to network data are gathered at a few points in time
(sometimes only once).  Thus the networks are often examined
statically.
 
\begin{figure}[h!]
  \includegraphics[width=0.50\textwidth]{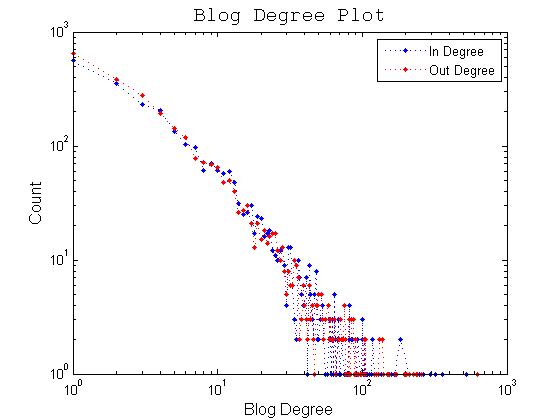} \hfill
  \includegraphics[width=0.50\textwidth]{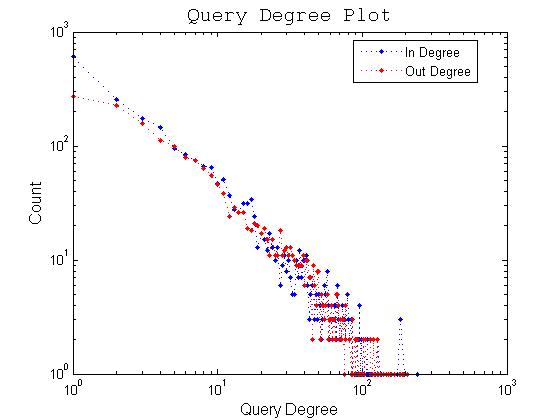}
\caption{Log-log plots of degree distributions for a query data bases
  and a blog data base from a company database.  Left: Blog indegree
  and outdegree distributions. Right: Query indegree and outdegree
  distributions. Source: Data from an unnamed large company, stored in
  iLab, Carnegie Mellon University.}
\label{fig:powerlaws}
\end{figure}

It has been recently pointed out that, contrary to previous claims,
the empirical laws that generative models aim to emulate are not
always supported by real data.  Visual comparison are not sufficient
for determining the goodness of fit of a model. For example, a lot of
attention has recently been paid to the degree
distribution. \autoref{fig:powerlaws} shows indegree and outdegree
distributions for blog and query databases from an unnamed large
company.  They are plotted on a log-log scale and the downward slopes,
if fitted by straight lines, would be visually similar to power law
distributions with exponents less than 2.  A careful examination of
these plots, however, reveals a curvilinear relationship in all cases,
which suggests that there is a different generating process than those
usually used to justify power laws in empirical network data. Data
such as that displayed in \autoref{fig:powerlaws} are often fitted by
ordinary least squares or even by eye; often the claim is that a
degree distribution is scale-free except for a cutoff at very high or
very low degrees, without any adjustment for searching for a cutoff!
There have been a number of recent efforts to assess the fit of degree
distributions, such as those associated with power laws in log-log
plots, with more rigor, e.g., see \citep{Clau:Shal:Newm:2007}.  As in
this example, results from such careful assessments of fit often
contradict the assumption of linearity.

\citet{Li:Alde:Doyl:Will:2006} give a ``structural metric" for
examining simple connected graphs having identical degree
distributions and derive theoretical properties of scale-free
graphs. They provide at least one possible way to assess whether a
graph corresponding to a network is in fact scale-free.  For more
informal discussions related to this theoretical work, see
\citep{Alde:2008,Will:Alde:Doyl:2009}.  Flaxman et
al.~\citep{Flax:Frie:Vera:2006,Flax:Frie:Vera:2007} describe a class
of network models linked to the preferential attachment model that also
yield a power-law degree distribution.

Most descriptions of generative models fall short of studying the full
parameter space and do not propose procedures for fitting the proposed
methods to real data, though there are a few works that suggest maximum likelihood, MCMC and other frameworks for fitting these models to data (for e.g. \citep{Beza:Kala:Sant:2006,Cleg:Land:Hard:Rio:2009,Midd:Ziv:Wigg:2004,Will:Mart:2000}). One of the notable exceptions is work based on
Kronecker graph multiplication. What started as yet another generative
procedure \cite{Lesc:Chak:Klei:Falo:2005} has turned into a well
analyzed methodology \cite{NdDC:Klei:Falo:Ghah:2009} with an
efficient algorithm for model fitting, analysis of the parameter
space, and model selection.  This work goes further in understanding
real network structure and provides a way for principled graph
sampling.

\section{Small-World Models}
\label{sec:dym-smallworld}

\begin{figure}[h!]
 \begin{center}
  \includegraphics[width=0.60\textwidth]{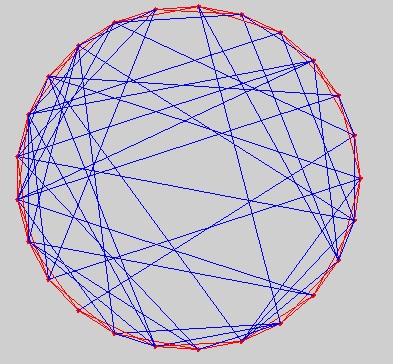}
  \end{center}
\caption{Small-world graph for $N=25$ nodes and 2 rewirings per
  node. The red edges form the ring lattice and the blue edges the
  rewiring.  This graph was generated using the Java applet at \url{
  http://cs.gmu.edu/~astavrou/smallworld.html}}
\label{fig:WattsStrogatz}
\end{figure}

\citet{Watt:Stro:1998} proposed a small-world model which can be
thought of as a ``pseudo-dynamic'' model in the sense we described in
\autoref{sec:dym-randomgraph}.  This one-parameter ``small-world''
model interpolates between an ordered finite-dimensional lattice and
an {\ER} random graph in order to produce local clustering and triadic
closures. \citet{Boll:Chun:1988} had previously noted that adding
random edges to a ring of $N$ nodes drastically reduces the diameter
of the network. The Watts-Strogatz model begins with a ring lattice
with $N$ nodes and $k$ edges per node, and randomly rewires each edge
with probability $p$.  As $p$ goes from $0$ to $1$, the construction
moves toward an {\ER} model. They and others who followed, studied the
behavior of such small-world networks when $0<p<1$.  This model is not
dynamic although it is often used to describe networks that evolve
over time.  \autoref{fig:WattsStrogatz} shows a small-world graph for
$n=25$ nodes and 2 rewirings per node.

\citet{Klei:2001} introduced a variation on the small-world model
where random edges are added to a fixed grid. Starting with an
underlying finite-dimensional grid, he added shortcut edges, where the
probability that two nodes are connected by a long edge depends on the
distance between them in the grid.  More precisely, the probability
that two non-adjacent nodes $x$ and $y$ are connected is proportional
to $d(x, y)^{-\alpha}$.  With $\alpha$ set to the dimension of the
lattice, the greedy routing algorithm can find paths from one node to
another in a polylogarithmic number of expected steps.

Several follow-up works have made adjustments to Kleinberg's rewiring procedure in attempt to improve the understanding and efficiency of the navigability of networks. For example, \citet{Clau:Moor:2003} suggested to rewire a long distance edge from node $x$, if while performing a greedy walk over to $y$, the original topology of the network did not allow to reach $y$ within $T_{\texttt{thresh}}$ steps. The edge was rewired to the place where the search gave up (the node reached after $T_{\texttt{thresh}}$ steps of the walk).They show that through this rewiring procedure the network degree distribution converges to a power law, where $\alpha = \alpha_{\texttt{rewired}}$. Their work also studied finite size effects and showed that $\alpha_{\texttt{opt}}\to d$, as $n\to\infty$ rather slowly.

\citet{Sand:2005,Sand:2008} and \citet{Sand:Clar:2008} introduced a
different rewiring scheme with the end goal to make the network more amenable to statistical analysis. Starting with $N$ nodes on a ring, each with two neighbor links and a long range link, the model of \citet{Sand:2005}
randomly rewires a graph in the following steps:
\begin{itemize}
\item at each time step $j = 1, 2, 3,\dots,$ choose a random starting
  node $x$ and a target node $y$ and perform greedy routing from $x$
  to $y$;
\item independently and with (small) probability $x$, update the
  long-range link of each node on the resulting path to point to $y$.
\end{itemize}
This defines a Markov chain on a collection of labeled graphs.
\citet{Sand:Clar:2008} conjecture that when the chain achieves
stationarity, the distribution of distances spanned by long-range
links is (close to) theoretical optimum for search and the expected
length of searches is polylogarithmic.  They support the conjecture by
a series of simulations.  This methodology has been applied to the
study of peer-to-per (P2P) networks.

\citet{Durr:2006} discusses links between small-world models and
stochastic processes.  Typical usage of small-world models include
empirical analyses involving aggregate summary statistics (see,
e.g.,~\cite{Amar:Scal:Bart:Stan:2000,Newm:Bara:Watt:2006}).  There are
as yet no formal statistical methods for examining the evolution of
small-world network models and for assessing their fit to network data
measured over time.

\section{Duplication-Attachment Models}
\label{sec:da}

Duplication-Attachment models were originally developed in the
computer science theory community to study the world wide web as a
directed graph
\citep{Klei:Kuma:Ragh:Raja:1999,Kuma:Ragh:Raja:Siva:2000}.
These models aim at describing properties of a snapshot of the web
graph at a specific time, that is, a static directed graph. The data
generating process underlying these models, however, is explicitly
dynamic.
The following example demonstrates some basic assumptions behind the
dynamics. Consider a newly added web page $A$, which provides a new
node in the web graph. The creator of web page $A$ will then add {\em
hyper-links} to it, which provide new directed edges in the web
graph. In particular, {\em some} of these hyper-links will point to
other web pages regardless of whether their topical content matches
the topical content of web page $A$, but {\em most} of these
hyper-links will point to web pages with a topical content that
closely matches the topical content of web page $A$.

Technically, there are many possible specifications and variants. The
basic duplication-attachment model proposed and analyzed by Kumar et
al.~\citep{Kuma:Ragh:Raja:Siva:2000} is as follows. Denote the graph
at time $t$ as $G_t = (\nodeset_t,\edgeset_t)$. At each step, say
$t+1$, one new node $\node$ is added to $G_t$. The new node is
connected to a {\em prototype} node $m$, chosen uniformly at random
among those in $\nodeset_t$. Then $d$ out-links are added to node
$\node$. The $i$th out-link is chosen as follows: with probability
$\alpha$ the destination node is chosen uniformly at random among
those in $\nodeset_t$, and with probability $1-\alpha$ the destination
node is taken to be the $i$th out-link of the prototype node
$m$. Note that this is possible since the algorithm generates a
constant degree graph.
Rather than proposing estimation strategies for the two parameters
$(\alpha,d)$ of this particular duplication-attachment model, the goal
of the analysis of Kumar et al.~\citep{Kuma:Ragh:Raja:Siva:2000} is on
deriving results about topological properties of
duplication-attachment graphs, described as functions of the two
parameters $(\alpha,d)$.
Recent extensions of this model include a model where fractions of
both out-links and in-links of the prototype node $m$ are {\em copied}
by the newly added node $\node$ \citep{Lesk:Klei:Falo:2005}. The goal
of the analyses in this line of research, however, remains that of
replicating properties of observed graphs, with a few exceptions. In the biological context, duplication-attachment models have appeared to be useful in modeling protein-protein interaction networks. For example, \citet{Ratm:Jorg:Hink:Stum:2007} proposed a mixture of preferential attachment and duplication divergence with parent-child attachment model to assess evolutionary dynamics of protein interaction networks of {\it H. pylori} and {\it P. falciparum}.  They proposed a likelihood-free MCMC-based routine to estimate posterior of network summary statistics. A more general review of work in modeling dynamics (evolution) on the basis of protein-protein interaction data is available in \cite{Ratm:Wiuf:Pinn:2009}.

\citet{Wiuf:Bram:Hagb:Stum:2006} have developed a recursive construction of the likelihood for duplication-attachment models, effectively enabling principled statistical data analysis, estimation and inference.

\section{Continuous Time Markov Chain Models}
\label{sec:dym-cmpm}

The use of continuous Markov processes to model dynamic networks was
first proposed by \citet{Holl:Lein:1977} and \citet{Wass:1977} and
most recently studied by Snijders and colleagues
\citep{Snij:2005,Snij:2006}.  As shall become clear in this
section, continuous Markov process models (CMPM) are intimately tied
to the ERGM models described in \autoref{sec:ergm}.  Within the CMPM
family, network edges are taken to be binary (either absent or
present, but not weighted), and the evolution occurs one edge at a
time.  Model variants arise due to the many possible specifications of
edge change probability. Some exceptions to this general approach
include the party model of \citet{Maye:1984}, where multiple edges
are allowed to change at the same time, and the work of
\citet{Kosk:Snij:2007}, which deals with Bayesian parameter inference
methods for the case where not all edge modifications are observed.

We begin by providing a quick reminder of continuous Markov processes,
borrowing notation from \cite{Snij:2005}.  Define $\{\X(t) \mid
t\in\cmtimeint\}$ to be a stochastic process, where $\X(t)$ has a
finite outcome space $\cmeventspace$ and $\cmtimeint$ is a continuous
time interval.  Suppose that a Markov condition holds: for any
possible outcome $\tilde{\x}\in\cmeventspace$ and any pair of time
points $\{t_a<t_b \mid t_a,t_b\in\cmtimeint\}$,
\begin{equation}
 \Pr \{\X(t_b)=\tilde{\x} \mid \X(t)=\x(t), \forall t:t \leq t_a\} = \Pr \{\X(t_b)=\tilde{\x}|\X(t_a)=\x(t_a)\} . \label{eqn:cmpm-condprob}
\end{equation}
In other words, supposing that $t_b$ denotes the future and $t_a$ the
present, then conditioning on the past is equivalent to conditioning
on the present when it comes to determining the future.  If the
probability in \autoref{eqn:cmpm-condprob} depends only on $t_b -
t_a$, then one can prove that $\X(t)$ has a stationary transition
distribution, and the transition matrix
\begin{equation}
  \Pr (t_b - t_a ) := \Bigl[ \Pr \{\X(t_b) = \tilde \x \mid \X(t_a) = \x\} \Bigr]_{\x, \tilde \x \in \cmeventspace}
\end{equation}
can be written as a matrix exponential
\begin{equation}
  \Pr (t) = e^{t Q} ,
\end{equation}
where $Q$ is known as the \emph{intensity matrix} with elements
$q(\x,\tilde \x)$.  The elements $q(\x, \tilde \x)$ can be thought of
as the slope (rate of change) of the probability of state change as a
function of time, i.e., $\Pr\{\X(t+\epsilon) = \tilde \x \mid \X(t) =
\x\} \approx \epsilon q(\x,\tilde \x)$.  The diagonal elements
$q(\x,\x)$ are negative and are defined so that the rows of $Q$ sum
to zero.

When modeling a social network, the outcome space $\cmeventspace$ is
taken to be all possible edge configurations of an $\node$-node
network, and an individual configuration $\xv \in \cmeventspace$ is
taken to be a binary vector of length $N\choose 2$.  We use the
shorthand $q_{ij}(\xv)$ to denote the propensity for the edge between
node $i$ and $j$ to flip into its opposite value under configuration
$\xv$.  The function $q_{ij}(\xv)$ completely specifies the dynamics
of the network model.  We now review several variants of CMPM which
differ only in their definition of $q_{ij}(\xv)$.

\paragraph{Independent arc, reciprocity, and popularity models.}
 
The {\it independent arc} model employs the simplest definition of $q_{ij}(\xv)$:
\begin{equation}
  \mbox{Independent arc model:} \quad q_{ij}(\xv) = \lambda_{\x_{ij}},
\end{equation} 
i.e., $\X_{ij}$ changes from $0$ to $1$ at a rate $\lambda_0$, and
from $1$ to $0$ at rate $\lambda_1$.  In this model, modification to
one edge does not depend on the setting of other edges.  The model is
simple enough that the transition probabilities $\Pr(t)$ can be
derived in closed form (see, e.g., \citet{Tayl:Carl:1998} p.\ 362-364).
Maximum likelihood parameter estimation for this model was discussed in
\cite{Snij:VanD:1997}.

In the {\it reciprocity} model, the rate of change in $\x_{ij}$
depends only on the reciprocal edge $\x_{ji}$:
\begin{equation}
  \mbox{Reciprocity model:} \quad q_{ij}(\xv) = \lambda_{\x_{ij}} + \mu_{\x_{ij}} \x_{ji} .
\end{equation}
Thus, if no link currently exists between nodes $i$ and $j$, then the
propensity for adding either directed edge is $\lambda_0$; if one
directed edge exists, then the reciprocal edge is added with
propensity $\lambda_0 + \mu_0$. If one directed edge exists, then it
is deleted with rate $\lambda_1$.  If both edges exist, then the
deletion propensity for either is $\lambda_1 + \mu_1$.  The transition
matrix $\Pr(t)$ can be derived but has a complicated
form~\citep{Leen:1995,Snij:1999}.

Along the same line of development, the {\it popularity} model and the
{\it expansiveness} model \citep{Wass:1977,Wass:1980} define the
change rate for edge $\x_{ij}$ to be dependent on $y_{+j}$, the
in-degree of node $j$, or $y_{i+}$, the out-degree of node $i$:
\begin{eqnarray}
  \mbox{Popularity model:} \quad q_{ij} (\xv) = \lambda_{\x_{ij}} + \pi_{\x_{ij}} \x_{+j} , \\
  \mbox{Expansiveness model:} \quad q_{ij} (\xv) = \lambda_{\x_{ij}} + \pi_{\x_{ij}} \x_{i+} .
\end{eqnarray}
%
%need to think about how to change this - repetative?
%

\paragraph{Edge-oriented dynamics.}

\citet{Snij:2006} outlines two \comment{additional} categories of
transition dynamics: edge-oriented and node-oriented.  In both cases,
the intensity matrix is factored into two components: one controls the
\emph{opportunity} for change, and the other specifies the propensity
of change.  More precisely, the continuous time Markov process is now
split into two sub-processes; the first operating in the continuous
time domain and dictating \emph{when} a change should occur; the
second dealing with the probability of the discrete event of individual
edge flips.  Both edge-oriented and node-oriented dynamics can be
interpreted as stochastic optimizations of a potential function
$f(\xv)$ on the network configuration.  The difference is that, in the
edge-oriented case, $f$ is based on global statistics of the network,
whereas in the node-oriented case, $f$ is defined for each node's
local neighborhood. Moreover, the choice of \emph{which} edge to flip
differs between the two formulations.

Using $\x(i,j,z)$ to denote the configuration where the edge $e_{ij}$
has the value $z \in \{0, 1\}$, edge-oriented dynamics can be written
in the following general form:
\begin{equation}
  q_{ij}(\xv) = \rho p_{ij} (\xv) ,
\end{equation}
where
\begin{equation}
p_{ij}(\xv) = \frac{\exp(f(\x(i,j,1-\x_{ij})))}{\exp(f(\x(i,j,0)))+\exp(f(\x(i,j,1)))}.
\end{equation}
Thus, in edge-oriented dynamics each edge follows an independent
Poisson process, so that the time until the next event has an
exponential distribution with parameter $\rho$.  When an event occurs
for edge $i\rightarrow j$, the edge flips to its opposite value with
probability $p_{ij}(\xv)$.

The potential function $f(\xv)$ is usually defined as a linear
combination of network statistics:
\begin{equation}
  f(\xv) = \sum_k \beta_k s_k(\xv). \label{eqn:cmpm-edge-potential}
\end{equation}
This should start to look familiar.  Indeed the CMPM process with
edge-oriented dynamics is equivalent to the Gibbs sampling process for
ERGMs (where the next edge to be updated is selected randomly).  The
statistics $s_k(\xv)$ for node $k$ take on the usual forms
(see~\autoref{tab:cmpmstats}).

\begin{table}[!h]
\begin{tabular}{ll}
Number of directed arcs: & $s_1(\xv) = \displaystyle{\sum_{ij}} \x_{ij}$ \\
Number of reciprocated arcs: & $s_2(\xv) = \displaystyle{\sum_{ij}} \x_{ij}\x_{ji}$ \\
Number of pairs of arcs with the same target: & $s_3(\xv) = \displaystyle{\sum_{ijk}} \x_{kj}\x_{ji}$ \\
Number of pairs of arcs with the same origin: & $s_4(\xv) = \displaystyle{\sum_{ijk}} \x_{ik}\x_{ij}$ \\
Number of paths of length two: & $s_5(\xv) = \displaystyle{\sum_{ijk}} \x_{ij}\x_{jk}$ \\
Number of transitive triplets: & $s_6(\xv) = \displaystyle{\sum_{ijk}} \x_{ij}\x_{ik}\x_{jk}$ \\
\end{tabular}
\centering{
\caption{The table of network statistics for a directed
    social network.} \label{tab:cmpmstats}
}
\end{table}

The statistics in \autoref{tab:cmpmstats} assume directed graphs,
however it is easy to come up with the corresponding statistics for
undirected graphs.  For example, in the undirected case all the edges
are ``reciprocal'' and thus $s_1$ and $s_2$ are combined into $s'(\xv)
= \sum_{i,j>i\in\natnum} \x_{ij}$.

Due to their close relations to ERGMs, edge-oriented models suffer the
same fate of degeneracy.  For example, if the parameter $\beta$ for
transitive triplets is not too small, then with high probability the
simulated network will be a complete graph.  However, compared to
static networks, degeneracy in the longitudinal case is not as much a
concern, as the complete graph will only emerge at some distant time
in the future.

\paragraph{Node-oriented dynamics.}
 
Fully node-oriented dynamics~\citep{Snij:2005} defines the
intensity matrix as
\begin{equation}
  q_{ij}(\xv) = \rho_i p_{ij}(\xv),
\end{equation}
where
\begin{equation}
  p_{ij}(\xv) = \frac{\exp (f_i(\xv(i,j,1-\x_{ij})))}
  {\sum_{h \neq i} \exp ( f_i(\xv (i,h, 1-\x_{ih})))} .
\end{equation}
Thus the independent Poisson processes for determining edge change
\emph{opportunity} are now defined for each node (with intensity
$\rho_i$) as opposed to each edge.  Given the opportunity for edge
change, each node seeks to optimize its own potential function as
defined by
\begin{equation}
  f_i(\xv) = \sum_k \beta_k s_{ik} (\xv).
\end{equation}
The function $f_i(\xv)$ is similar to the global potential $f(\xv)$
in~\autoref{eqn:cmpm-edge-potential} but only aggregates over the
local neighborhood of node $i$.  Node $i$ favors changing the incident
edge that would lead to the biggest increase in its potential.
 
\paragraph{Edge-node mixed dynamics.}

\citet{Snij:2006} also suggested a form of mixed dynamics where
the opportunity for change is edge-oriented, but the potential
functions are node-oriented:
\begin{equation}
  q_{ij}(\xv) = \rho \frac{\exp (f_i(\xv(i,j,1-\x_{ij})))}
  {\sum_{h \neq i} \exp ( f_i(\xv (i,h, 1-\x_{ih})))} .
\end{equation}
Thus the opportunity to modify each edge $i\rightarrow j$ follows
independent Poisson processes with parameter $\rho$. But given the
opportunity for change, the probability of an actual flip depends on
node $i$'s local network configuration.

\paragraph{Remark.}
Parameter estimation in CPCM models has until recently been done via
method of moments, where the expected values are obtained through MCMC
on simulated networks \citep{Snij:2001}. \citet{Kosk:Snij:2007}
proposed a Bayesian inference method that allows for computation of
the posterior distribution of the parameters and treats missing values
more adequately. For details of the procedure, please refer to
\citet{Kosk:Snij:2007}.

\section{Discrete Time Markov Models}
\label{sec:dym-discrete}

In this section, we outline three recent proposals of dynamic network
models operating in the discrete time domain (see also
\citep{Bank:Carl:1996}). All three models have the Markov property and
represent the likelihood as a sequence of factored conditional
probabilities
\begin{equation}
  \Pr(\X^1, \X^2, \ldots, \X^T) = \Pr(\X^T \mid \X^{T-1}) \Pr(\X^{T-1} \mid \X^{T-2}) \cdots \Pr(\X^2 \mid \X^1)),
\end{equation}
where $\{\X^1, \ldots, \X^T\}$ is a sequence of $T$ observed snapshots
of the network.  \citet{Bank:Carl:1996} discussed the simplest version
of such models.  See also \cite{Robi:Patt:2001}. 

\subsection{Discrete Markov ERGM Model}
\label{subsec:dym-dm-ergm}

\citet{Hann:Xing:2007} proposed a natural extension of the ERGM model
in the discrete Markov domain.  Unlike the set up in the continuous
domain, the potential function in this model involve the statistics of
two consecutive configurations of the network:
\begin{equation}
  \Pr(\xv^t \mid \xv^{t-1}) = \frac{1}{Z} \exp \{\sum_k \beta_k s_k (\xv^t, \xv^{t-1})\}.
\end{equation}
\autoref{tab:pairstats} lists a few examples of network statistics defined on
pairs of network snapshots.
\begin{table}[!h]
\begin{tabular}{ll}
Density of edges: & $s_1(\xv^t, \xv^{t-1}) = \frac{1}{(n-1)} \displaystyle{\sum_{ij}} \x^t_{ij}$ \\
Stability: & $s_2(\xv^t, \xv^{t-1}) = \frac{1}{(n-1)} \displaystyle{\sum_{ij}} [\x^t_{ij} \x^{t-1}_{ij} + (1-\x_{ij}^t) (1-\x_{ij}^{t-1})]$\\
Reciprocity: & $s_3(\xv^t, \xv^{t-1}) = n \displaystyle{\sum_{ij}} \x^t_{ji} \x^{t-1}_{ij} \Big/ \displaystyle{\sum_{ij}} \x_{ij}^{t-1}$ \\
Transitivity: & $s_4(\xv^t, \xv^{t-1}) = n \displaystyle{\sum_{ijk}} \x^t_{ik} \x^{t-1}_{ij} \x^{t-1}_{jk} \Big/ \displaystyle{\sum_{ijk}} \x_{ij}^{t-1} \x_{jk}^{t-1} $ \
\end{tabular}
\caption{The table of network statistics for pairs of network snapshots.}
\label{tab:pairstats}
\end{table}

The basic model may be extended to allow for multiple relations, node
attributes, and K-th order Markov dependencies of the form
\begin{equation}
  \Pr(\X^{K+1}, \X^{K+2}, \ldots, \X^T \mid \X^1, \ldots, \X^K) = \prod_{t = K+1}^T \Pr(\X^t \mid \X^{t-K}, \ldots, \X^{t-1}),
\end{equation}
where
\begin{equation}
  \Pr(\X^t \mid \X^{t-K}, \ldots, \X^{t-1}) = \frac{1}{Z} \exp \{ \sum_k \beta_k s_k (\X^t, \ldots, \X^{t-K}) .
\end{equation}
The joint distribution of the first $K$ network snapshots may be
represented by an ERGM for the first snapshot, and a $(k-1)$-th order
discrete Markov dependency model for $\X_k$.  The paired network
statistics may be extended over $K$ network sequences.

Maximum likelihood parameter estimates may be computed via any
numerical approximation technique such as the Newton-Raphson method.
Computation of the gradient and Hessian requires the mean and
covariance of the sequence network statistics, which are exactly
computable for a pair of networks, but require Gibbs sampling in the
$K$-sequence case \citep{Hann:Xing:2007}. The likelihood of this model is well behaved if the minimum sufficient statistics involve only dyads, however, similar to its static
counterpart, the full dynamic ERGM is prone to likelihood degeneracy.

\subsection{Dynamic Latent Space Model}
\label{subsec:dym-discrete-ls}

\citet{Sark:Moor:2005a} extended the static latent space model of
\citet{Hoff:Raft:Hand:2002} (cf.~\autoref{sec:latent}) in the time
domain.  Recall that in the static latent space model, the log odds
ratio of a link between nodes $i$ and $j$ depends on the distance
between their latent positions $\latent_i$ and $\latent_j$.  The
dynamic latent space model allows the latent positions to change over
time in Gaussian-distributed random steps:
\begin{equation}
  \Latent_t \mid \Latent_{t-1} \sim \Gaussian(\Latent_{t-1}, \sigma^2 I).
\end{equation}
The observation model is a modified version of the original latent
space model\footnote{Note that in this dynamic version of the latent
  space model, links are assumed to be undirected.}:
\begin{equation}
  p^L_{ij} := p^L(y_{ij} = 1) = \frac{1}{1 + \exp(d_{ij} - r_{ij})},
\end{equation}
where $d_{ij}$ is the Euclidean distance between $i$ and $j$ in latent
space, and $r_{ij}$ is a radius of influence defined as $c \times
(\max(\deg_i, \deg_j) + 1)$ ($\deg_i$ and $\deg_j$ being the degrees
of node $i$ and $j$, respectively).  The ``radius of influence'' is based on the assumption that the higher the maximum degree of the two end nodes, the more likely the edge.  This may be true in citation networks where prolific authors are more likely to form new co-authorships. The constant $1$ is added to
ensure that the radius is non-zero, and $c$ is estimated from data by
a line-search (a minimization method in one dimension).

The link probability $p_{ij}$ is defined to be a mixture between the
modified latent space link probability $p^L_{ij}$ and a noise
probability $\rho$.  The idea is that pairs of nodes who are outside
of each other's radius have only a low noise probability of
establishing a link, while nodes within each other's radii follow the
probability $p^L_{ij}$:
\begin{equation}
  p_{ij} = \kappa(d_{ij}) p^L_{ij} K(d_{ij}) + (1-\kappa(d_{ij})) \rho .
\end{equation}
\comment{(too much detail -- az)
The mixture weight $\kappa$ is taken to be a biquadratic kernel
transform of the latent space distance $d_{ij} = | z_i - z_j |$:
\begin{equation}
  \kappa(d_{ij}) = 
  \begin{cases}
    (1- (d_{ij}/r_{ij})^2 )^2 \quad \mbox{when $d_{ij} \leq r_{ij}$}, \\
    0 \quad\quad\quad \mbox{otherwise.}
  \end{cases}
\end{equation}
The biquadratic kernel has the property of being continuous and
differentiable at the boundary $d_{ij} = r_{ij}$.  Hence the above
expression is easy to optimize.
}

The full observation model is then
\begin{equation}
  \Pr(\X^t \mid \Latent^t) = \prod_{i \sim j} p_{ij} \prod_{i \nsim j} (1-p_{ij}),
\end{equation}
where $i \sim j$ denotes the presence of an edge from $i$ to $j$. The
latent space positions $\Latent^t$ are estimated in sequence for $t =
1 \ldots T$ by maximizing the likelihood of the observed $\X^t$:
\begin{equation}
  \Latent^t = \argmax_{\Latent} \Pr(\X^t \mid \Latent) \Pr(\Latent \mid \Latent^{t-1}).
\end{equation}
The authors propose conjugate gradient optimization starting from an
initial estimate of the latent positions based on a multidimensional
scaling (MDS) transform of the observed pairwise distances.  To
eliminate rotational ambiguity, a Procrustean (rotationally invariant) transform is applied to
the MDS transform so that $\Latent^t$ is aligned with $\Latent^{t-1}$.

Applying the model to the NIPS paper co-authorship dataset
(cf.~\autoref{sec:nips-data}), the authors gave anecdotal evidence of
the validity of the changing embeddings of several well known machine
learning researchers over time. The dynamics of the researchers' latent
positions allowed for an insight into the evolution of the machine
learning community.

\citet{Sark:Sidd:Gord:2007} also proposed a richer model based on
\citep{Glob:Chec:Tish:Weis:2007}, which improved upon previous
work in two ways. One of the differentiating features of this work was
the ability to simultaneously embed words and authors into the latent
space, which allowed for representation of a two-mode network. The
major advantage, however, was the inference method---the authors
proposed a Kalman-filter like dynamic procedure, which allowed for
estimation of the posterior distributions over the positions of the
authors in the latent space. Proposed procedure was applied to a
simulated NIPS dataset.
 
The impact of this line of work is dichotomous: first, it offers an
explanation of the network at every time step, and second, it
enables an accurate and efficient prediction of the state of the
network at a time step in the future. The proposed inference
procedures made it possible for network modeling to scale to large
dynamic collections of data.  The drawback of this approach is the
lack of an explicit mechanism that could explain the dynamics behind
the real networks.

Another latent model for citation networks was developed in the
physics community. \citet{Leic:Clar:Shed:Newm:2007} proposed to use
latent variables to capture the grouping of papers that have similar
citation profiles over time.  The network in this case is a directed
acyclic graph and the nodes are papers rather than authors. Using as
example a set of opinions from the US Supreme Court and their
citations between the years of 1789 and 2007, the authors showed how a
simple latent model was able to recover, in a completely unsupervised
manner, the different eras in US Supreme court opinion references. The
parameters of the model, except for the number of latent classes, were
estimated using an EM algorithm. Different numbers of latent classes
were tested and each revealed something new about the underlying
data. The authors also compared the latent method to a clustering
based on network {\it modularity} \citep{Newm:2006}. Even with the
information about time (directionality in the graph) removed, the
latent variable model was still able to discover the same split
between two groups of opinions that happened around 1937.  The
network modularity clustering in a way validated the outcome of the
latent model. 

In a separate experiment, \citet{Leic:Clar:Shed:Newm:2007} showed that
deterministic approaches such as ``hubs and authorities'' and
eigenvector centrality \citep{Klei:1999} discovered interesting
network properties that were not revealed by the statistical models.
The deterministic analyses showed several significant drops in the age
of authorities sited, meaning that once in a while, the younger set of
opinions became the new authorities and that the process happened in a
``decisive'' manner, rather than gradually. In this way, deterministic
network analysis approaches complement statistical models.

\subsection{Dynamic Contextual Friendship Model (DCFM)}
\label{subsec:dym-discrete-dcfm}

The dynamic contextual friendship model (DCFM) of
\citet{Gold:Zhen:2007} represents an attempt to capture several
aspects of the complexity of the evolution of real social networks
over time.  In a real-life friendship network, people may meet and
interact with each other under different contexts (e.g., school, work
projects, social outings, etc.), and the strength of interpersonal
relationships change over time based on these interactions.  DCFM
offers such a mechanism for network evolution, where edges have
weights that indicate the strength of the relationship, and each node
is given a distribution over social interaction spheres (contexts).
Context is defined to be any activity where people may interact with
each other. At each given time step, each node chooses a random
context according to the node's distribution over contexts. Nodes that
appear in the same context update the weights of the links between
them. The probability of a weight increase (or decrease) depends on
whether the pair had a chance to meet (a coin toss in a model) and the
``friendliness'' parameter of the individuals involved. The
possibility of both positive and negative weight updates allows for
edge birth and death over time.  An extension of the model also allows
for addition and deletion of nodes.

The underlying dynamics is captured by a first-order Markov chain
model. Letting $W^t$ denote the weighted adjacency matrix at time $t$,
the basic generative process at time $t$ can be formalized as follows:

\begin{itemize}
  \item[1.] For each node $i$, sample context $C_i \sim \mult
    (\theta_i)$, where $\theta_i$ denotes the context distribution
    parameters.
  \item[2.] For each pair of nodes $i$ and $j$ in the same context,
    sample meeting variable $M_{ij} \sim \Bern(\nu_i \nu_j)$, where
    $\nu_i$ and $\nu_j$ represent the ``friendliness'' of nodes $i$
    and $j$;
  \item[3.] 
    \[
W^t_{ij} = \begin{cases}
    \Poi(\lambda_h (W^{t-1}_{ij} + 1)) \quad \mbox{if }M_{ij} = 1, \\
    \Poi(\lambda_\ell (W^{t-1}_{ij})) \quad \mbox{otherwise},
  \end{cases}
\]
    where $\lambda_h$ and $\lambda_\ell$ are hyperparameters indicating the
    rates of growth and decay, respectively. The idea is that a
    meeting should increase the edge weight with high probability,
    otherwise the weight decays.
\end{itemize}
The parameters $\theta_i, \nu_i, \lambda_h, \lambda_\ell$ all have
conjugate priors and are estimated through Gibbs sampling
\citep{Zhen:Gold:2006}.

\begin{figure}[!h]
  \centering
  \includegraphics[width=3.5in]{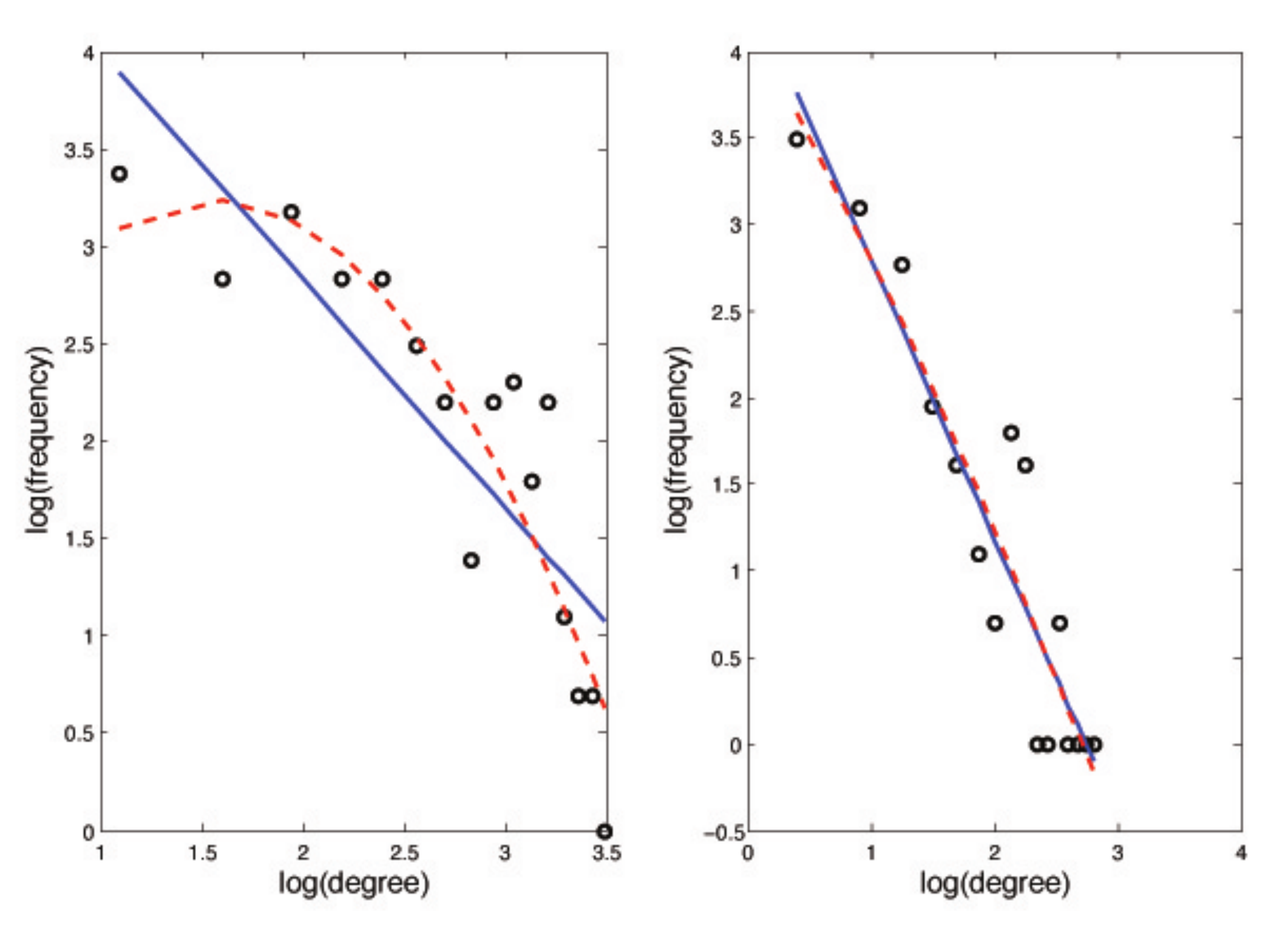}
  \caption{Log-log plot of the degree distributions of a network with $200$ 
    people.  $\nu_i$ is drawn from $\Bet(1,3)$ for the plot on the left,
    and from $\Bet(1,8)$ for the right hand side. Solid lines represent a 
    linear fit and dashed lines quadratic fit to the data.  Contexts are drawn 
    every $50$-th timesteps. }
  \label{fig:degdist}
\end{figure}

\begin{figure}[!htbp]
  \centering
  \includegraphics[width=4in]{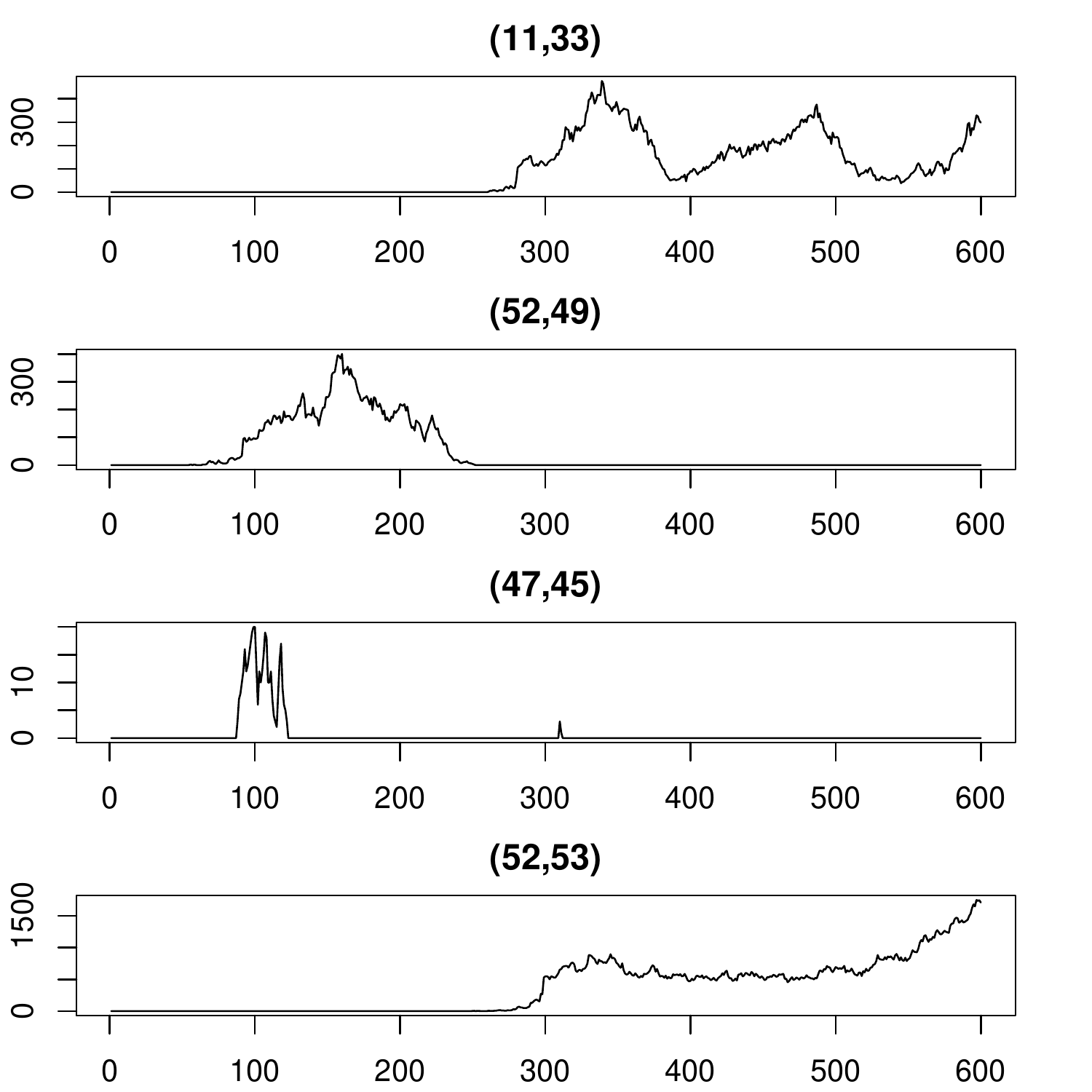}
  \caption{Weight dynamics for 4 different pairs in a DCFM simulated network of $600$ people 
    over $600$ time steps. Contexts switches occur every $50$-th timestep
    and $b = 3$.}
  \label{fig:weights}
\end{figure}

The model can generate networks with a number of different properties.
For example, \autoref{fig:degdist} shows various degree distributions
generated by DCFM, while Figure \ref{fig:weights} demonstrates
possible relation dynamics. Pair $(47,45)$ shows a brief resuming of
the relationship, which dissolves again in the next moment.  While
DCFM is capable of emulating such long-term memory of past
relationships, it does so at the cost of added model complexity.

Few datasets contain weighted relationships.  The Enron dataset
(cf.~\autoref{sec:enron}) contains email exchanges that can be
aggregated on a weekly basis to simulate strength of relationships.
In the NIPS dataset (cf.~\autoref{sec:nips-data}), the number of joint
publications per year can represent the strength of the coauthorship.
In these cases, the DCFM contexts can be taken to be the topics of
emails or articles, and the friendliness parameters can be estimated
using the method of moments.

One drawback of DCFM is its lack of identifiability; it is impossible
to tell without additional knowledge whether an individual formed many
friendships because he frequently changes contexts and is very
friendly or because the contexts themselves tend to be large. Also,
weighted network data are hard to come by and thus pseudo-weights
often have to be used.

The DCFM model is important in its own right: the life-mimicking, rich
generative mechanism is a step towards realistic complex models that
ultimately can be used to explain the intricacies of observed data,
especially if additional information about contexts and individuals'
friendliness is available.

%%%%%%%%%%%%%%%%%%%%%%%%%%%%%%%%%%%%%%%%%%%%%%%%%%%%%%%%%%%%%%%%%%%%%%%%%%%%%%
%%%%%%%%%%%%%%%%%%%%%%%%%%%%%%%%%%%%%%%%%%%%%%%%%%%%%%%%%%%%%%%%%%%%%%%%%%%%%%
%%%%%%%%%%%%%%%%%%%%%%%%%%%%%%%%%%%%%%%%%%%%%%%%%%%%%%%%%%%%%%%%%%%%%%%%%%%%%%
%%%%%%%%%%%%%%%%%%%%%%%%%%%%%%%%%%%%%%%%%%%%%%%%%%%%%%%%%%%%%%%%%%%%%%%%%%%%%%
%%%%%%%%%%%%%%%%%%%%%%%%%%%%%%%%%%%%%%%%%%%%%%%%%%%%%%%%%%%%%%%%%%%%%%%%%%%%%%
%%%%%%%%%%%%%%%%%%%%%%%%%%%%%%%%%%%%%%%%%%%%%%%%%%%%%%%%%%%%%%%%%%%%%%%%%%%%%%
%%%%%%%%%%%%%%%%%%%%%%%%%%%%%%%%%%%%%%%%%%%%%%%%%%%%%%%%%%%%%%%%%%%%%%%%%%%%%%
%%%%%%%%%%%%%%%%%%%%%%%%%%%%%%%%%%%%%%%%%%%%%%%%%%%%%%%%%%%%%%%%%%%%%%%%%%%%%%
%%%%%%%%%%%%%%%%%%%%%%%%%%%%%%%%%%%%%%%%%%%%%%%%%%%%%%%%%%%%%%%%%%%%%%%%%%%%%%
%%%%%%%%%%%%%%%%%%%%%%%%%%%%%%%%%%%%%%%%%%%%%%%%%%%%%%%%%%%%%%%%%%%%%%%%%%%%%%
% WAS: \input{issues.tex}

\chapter{Issues in Network Modeling}
\label{ch:issues}

There are a number of major statistical modeling and inferential
challenges in the analysis of network data that go well beyond those
described in previous sections of this article.  These relate to both
the quality and the ease of statistical inference and we mention a few
of them here:

\paragraph{Network Visualization.}
With the rise of online social networks and network modeling, we have
seen a proliferation of visualization tools, especially those based on
variations of constraint-based spring model algorithms, e.g., see the
discussion and references in \citet{Shne:Aris:2006}.  The automated
algorithms often use node degrees or some form of distance metric
between nodes to arrange their placement.  For example, {\it
SoNIA}\footnote{\url{http://sonia.stanford.edu/}} is a popular package for
visualizing dynamic or longitudinal network data; it can be used as a
platform for the development, testing, and comparison of various
static and dynamic layout techniques.  However, little is known about
how to effectively combine visualization with the kinds of statistical
models we review here, especially if one wants to use the
visualization as another tool in the analysis of network data.

\paragraph{Computability.}  

Can we do statistical estimation computations and model fitting
exactly for large networks, e.g., by full MCMC methods for mixed
membership and exponential random graph models, or do we need to
resort to approximations such as those involved in the variational
approximation employed by
\cite{Airo:Blei:Fien:Xing:2007,Airo:Blei:Fien:Xing:2008}? 
 
For ERGM models a newly updated suite of programs and documentation is
now
available~\cite{Hand:Hunt:Butt:Good:2008,Hunt:Hand:Butt:Good:2008,Morr:Hand:Hunt:2008,Good:Hand:Hunt:Butt:2008}. The
SIENA package\footnote{\url{http://stat.gamma.rug.nl/siena.html}}
developed by Snijders and colleagues contains a complementary suite of
programs that are particularly useful for longitudinal network
analyses (though \citet{Rina:Fien:Zhou:2009} speak words of
caution). The packages are capable of learning networks of size up to
a few thousand nodes.

The truth is that it is unrealistic to expect that really large
networks with millions of nodes can be estimated using exact
methods. Even variational approximations, which have their own
drawbacks such as sensitivity to the starting point, are not
realizable for networks on a really large scale. The key to network
modeling and parameter estimation is to take into account the sparsity
that comes with size. The methods that are good on small or
medium-sized but relatively dense networks, might be computationally
infeasible or contain invalid assumptions for larger networks. As we
gear up to model very large networks, it is important to focus not
only on the disadvantages that size brings but also on its advantages.

\paragraph{Asymptotics and Assessing Goodness of Fit.}  

There is no standard large sample asymptotics for networks (e.g., as
$N$ goes to infinity) that can be used to assess the goodness-of-fit
of models.  Thus we may have serious problems with variance estimates
for parameters and with confidence or posterior interval estimates.
While a few models with a small number of fixed parameters have
well-behaved asymptotics, the problems here tend to be the inherent
dependence of network data and the growth in the number of parameters
to be estimated as $N$ increases. \citet{Habe:1981} comments briefly
on asymptotics in his discussion of the {\pone} model, and notes the
similarity to issues for the Rasch model from item response theory.
The lack of asymptotics means that we may have problems of consistency
of estimators, but it also means that there is no standard basis for
model comparison and assessing goodness of fit.  Most other authors
have addressed these issues either empirically, e.g.,
\citet{Hunt:Good:Hand:2008}, or not at all.
 
There are two alternative approaches.  We can consider assessing fit
or comparing models using exact distributions given the minimal
sufficient statistics (MSSs).  This works for simple models but not obviously
for the general class of ERGMs or most dynamic models in the
literature.  Further, for many of the models, especially those
involving latent variables, the MSSs are the data
themselves. Alternatively, we could think in terms of some form of
cross-validation for model selection and assessment.  The problem with
cross-validation is the boundary effects associated with subsets of
nodes.  This is directly related to the problem of sampling in
networks.

\citet{Bick:Chen:2009} address the problem of asymptotics in the
context of blockmodeling or community discovery, and the methods they
exploit may be useful in a broader context when the number of
parameters to be estimated grows as $N$ increases.

\paragraph{Sampling.}

Do our data represent the entire network or are they based on only a
subnetwork or subgraph?  When the data come from a subgraph, even one
selected at random, we need to worry about the effects at the boundary\footnote{The boundary is the collection of observed nodes which have links to the unobserved nodes. The boundary can potentially include all observed nodes. Only nodes for which the set of known links are certain to be complete are not included in the boundary {--} the condition that is hard to satisfy in real world networks.}   and the
attendant biases they bring to parameter estimates, cf.\ the negative
result in Stumpf et al. for scale-free models in which they show the
extent and nature of the bias \citep{Stum:Wiuf:May:2005}.  Most of the
early results on sampling for network data focused on random subgraphs
and exploited the traditional statistical theory of design-based
sampling, in which the properties of the network are assumed to be
fixed, and we evaluate sample quantities by considering their
distribution under all possible similarly selected subgraphs.  For
details, see the many papers by Ove
Frank~\cite{Fran:2005,Thom:Fran:2000} and
others~\cite{Goel:Salg:2007,Hand:Gile:2010,Salg:Heck:2004}. \citet{Wiuf:Stum:2006}
and \citet{Stum:Thor:2006a} recently adopted a related but different
approach focusing on properties such as degree distributions using
binomial random sample sizes from ``large" graphs. Others such
as~\citet{Lesk:Falo:2006} examine aspects of the question in an
empirical but ad hoc fashion.  The relevance of sampling for
model-based network inference was first addressed by
\citet{Thom:Fran:2000}, and further developed by
\citet{Hand:Gile:2010}, who adapt MCMC algorithms for exponential
random graph models to account for sampling designs.  To date, these
are the only works to seriously explore this important topic.
\citet{Airo:Carl:2005} quantify the sensitivity of alternative
sampling algorithms to generate graphs that share similar topological
properties, as well as the divergence of topological properties of
algorithms for sampling popular network models.
 
We expect the issue of sampling to be of relevance to virtually all of
the models and we need to explore their consequences. This will be
especially true when we try to update model parameter estimates based
on extracts of data in a dynamic fashion.

 \paragraph{Missing data.} 
 
 Along with sampling arises a question of the treatment of missing data in statistical networks.  Usually, the non-respondents to surveys are excluded from the analysis and the modeling considers only individuals for which all data is available. A few works deal with missing data directly. The empirical impact of nonrespondents in a survey to analysis is considered in \cite{Stor:Rich:1992}, the modeling implications and inference for non-respondents in ERGM can be found in \citep{Robi:Patt:Wool:2004,Gile:Hand:2006,Kosk:Robi:Patt:2008}. Missing data in longitudinal studies is the subject of  \citep{Huis:Steg:2008}. This work makes assumptions about sampling strategies to justify the estimation of missing edges using a Missing at Random assumption.  Because this is not in general a correct assumption we have an interesting set of open problems. \citet{Koss:2008} considers three missing data mechanisms: network boundary specification (non-inclusion of actors or affiliations), survey non-response, and censoring by vertex degree (fixed choice design), and examines their effect on a study of a scientific collaboration network.  One type of missing data - links or relations - can be treated as a prediction task by treating links between nodes in a given network as probabilistic quantities and using statistical models based on the available data to estimate the likelihood of those edges being there. The problem of prediction is often addressed in the machine learning community and we discuss it next.

\paragraph{Prediction.}

In our review of the literature on networks across many disciplines we
have found limited methodological work focussing on evaluating and
comparing the predictive ability of various models, static or
dynamic. There are papers on link prediction in the relational network
model literature
(e.g. \cite{OMad:Smyt:Adam:2005}). \citet{Libe:Klei:2003} develop
approaches to link prediction based on measures for analyzing the
``proximity" of nodes in a network, e.g., the WWW.  In biological
literature, a number of papers examine the problem of predicting
missing links in biological networks
(e.g. \cite{Wong:Zhan:Tong:Li:2003} is one of the earlier
works). However, these papers focus on how to cleverly combine
heterogeneous data in order to discover new links. The evaluation is
usually limited to cross-validation on the known links---information
that is incomplete and available only for a few organisms. In the
sociological literature on organizations, there is often interest in
distinguishing among organizations on the basis of their network
structure, so there would clearly be interest in utilizing methodology
for prediction based on network structure.  Because making predictions
of various sorts from dynamic network models fits well within the
machine learning paradigm, we expect to see many more papers on the
topic in the not too distant future.

\paragraph{Embeddability.} 
 
Underlying most dynamic network models is a continuous time stochastic
process even though the data used to study the models and their
implications may come in the form of repeated snapshots at discrete
time points (epochs)---a form of time sampling as opposed to node
sampling referred to above---or cumulative network links.  In such
circumstances we need to take special care in how we represent and
estimate the continuous-time parameters in the actual data
realizations used to fit models.  This is known in the statistical
literature as the {\bf embeddability} problem and was studied for
Markov processes in the 1970s
by~\citet{Sing:Spil:1974,Sing:Spil:1976} for social
processes, and more recently by~\citet{Hans:Sche:1995} in the context of
econometric models and by others in the computational finance
literature. \citet{Wass:1980} and various papers by Snijders and his collaborators
illustrate how to address embedding in some simple dynamic models.

\paragraph{Identifiability.}

Identifiability of model parameters is a technical issue in statistics
that refers to the fact that multiple solutions may exist (in the
parametric space) that lead to exactly the same likelihood. In this
sense, no inference procedure can distinguish between these
solutions. For instance, in a mixture model we can permute the
assignments of points to mixture components to obtain an equivalent
solution. There are a number of papers that describe the issue in
various models (e.g., \citep{Step:2000,Grun:Leis:2008}) and from
different perspectives (e.g., \citep{Buot:Rich:2006,Buot:Rich:2006a}
from the algebraic perspective).  A few solutions to address this
issue have been proposed recently.  Some consider inference on
equivalence classes in a blockmodel for network data
\citep{Nowi:Snij:2001}.  Others pre-process the data to identify a
reference solution that drives the inference
\citep{Hand:Raft:Tant:2007}.

\paragraph{Combining links with their attributes.}  

In many network data sets, especially those arising in machine
learning contexts, there are attributes associated with the network
links.  For example in e-mail and blog databases, the attributes may
be taken to be the contents of the messages or postings.  There is an
emerging literature focused on cascades of such links but few papers
are situated in a full network model setting and few authors attempt
to combine the models for links with models for message or posting
texts.  This is a natural extension to models described here,
especially the mixed membership stochastic blockmodels of
\autoref{sec:mmsb}, since the text could naturally be modeled by
mixed-membership topic models. \citet{McCa:Wang:Moha:2007} and
\citet{Chan:Blei:2008} suggest different ways to approach this kind of
combination model.  Dynamic models that combine evolving block and
topic structures would be of special interest for such applications.

%%%%%%%%%%%%%%%%%%%%%%%%%%%%%%%%%%%%%%%%%%%%%%%%%%%%%%%%%%%%%%%%%%%%%%%%%%%%%%
%%%%%%%%%%%%%%%%%%%%%%%%%%%%%%%%%%%%%%%%%%%%%%%%%%%%%%%%%%%%%%%%%%%%%%%%%%%%%%
%%%%%%%%%%%%%%%%%%%%%%%%%%%%%%%%%%%%%%%%%%%%%%%%%%%%%%%%%%%%%%%%%%%%%%%%%%%%%%
%%%%%%%%%%%%%%%%%%%%%%%%%%%%%%%%%%%%%%%%%%%%%%%%%%%%%%%%%%%%%%%%%%%%%%%%%%%%%%
%%%%%%%%%%%%%%%%%%%%%%%%%%%%%%%%%%%%%%%%%%%%%%%%%%%%%%%%%%%%%%%%%%%%%%%%%%%%%%
%%%%%%%%%%%%%%%%%%%%%%%%%%%%%%%%%%%%%%%%%%%%%%%%%%%%%%%%%%%%%%%%%%%%%%%%%%%%%%
%%%%%%%%%%%%%%%%%%%%%%%%%%%%%%%%%%%%%%%%%%%%%%%%%%%%%%%%%%%%%%%%%%%%%%%%%%%%%%
%%%%%%%%%%%%%%%%%%%%%%%%%%%%%%%%%%%%%%%%%%%%%%%%%%%%%%%%%%%%%%%%%%%%%%%%%%%%%%
%%%%%%%%%%%%%%%%%%%%%%%%%%%%%%%%%%%%%%%%%%%%%%%%%%%%%%%%%%%%%%%%%%%%%%%%%%%%%%
%%%%%%%%%%%%%%%%%%%%%%%%%%%%%%%%%%%%%%%%%%%%%%%%%%%%%%%%%%%%%%%%%%%%%%%%%%%%%%

\chapter{Summary}

\begin{figure}[t]
\begin{center}
 \includegraphics[width=\textwidth]{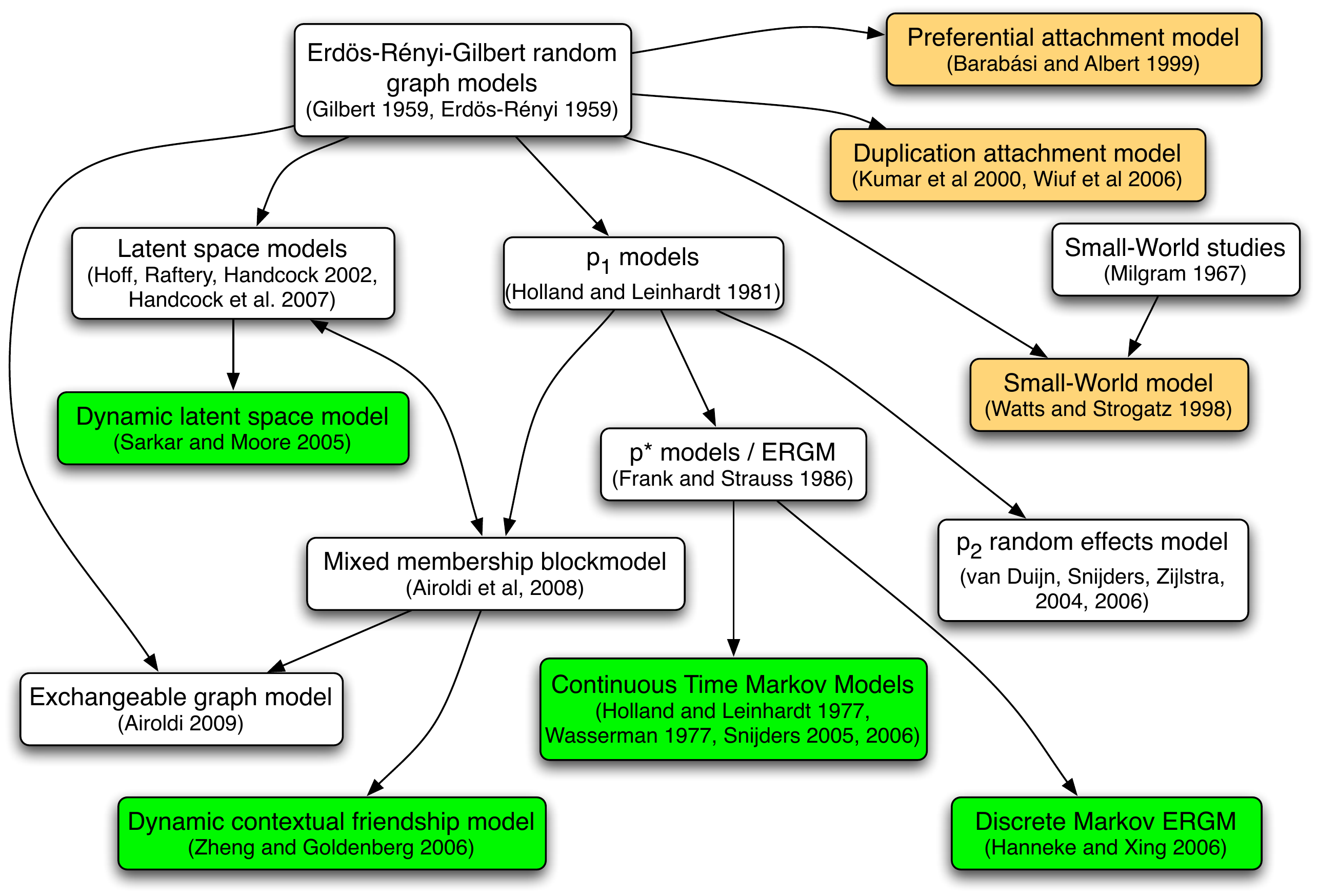}
\end{center}
\caption{Network summarizing the relations between models discussed in
  our review. White nodes denote static models, yellow nodes {--} ``pseudo-dynamic'' and green {--}
  dynamic models. Arrows indicate inspiration or influence of the model
  at the source on the model at the target.}
\label{fig:famtree}
\end{figure}

The ubiquity of networks in areas as diverse as the social sciences,
biology, computer science, physics, and economics, has spawned
extensive literature on the subject. In this review, we discussed in
detail a few main trends in the \emph{statistical} network modeling
literature, focussing on models that have historically inspired many
others as well as a few recent proposals.  By charting the evolution
of statistical network modeling approaches, we pointed out explicit
connections between the discussed models.  Figure~\ref{fig:famtree}
provides a visual diagram of model influence; an arrow pointing from A
to B means either that the development of model A influenced the
subsequent development of model B, or that B can be viewed as a
generalization of A.

The literature on network modeling may be divided along different
lines of motivation.  Models primarily introduced in the physics
literature are motivated by asymptotic properties of networks, whereas
the literature stemming from statistics and statistical social science
is concerned with the inference step in addition. Thus, the main criticism
of the random graph models primarily developed in statistical physics
is the lack of the assessment of the fit of the models to the
data. The main drawback in the statistical literature is the lack of
the comprehensive asymptotic analysis. Though degeneracy found in the
limiting case of the earlier versions of the ERGM has been addressed,
a more broad analysis is still missing.

In this work we made a distinction between static and dynamic
models. Descriptive models such as {\pone}, {\ptwo}, and ERGM are
clearly static as they infer a set of sufficient statistics from a
single snapshot of an existing network.  The families of continuous
and discrete time Markov models, on the other hand, are clearly
dynamic as they seek to model multiple snapshots of an evolving
network.  The {\ER}, preferential attachment, and
small-world models, while ultimately aim to model a single time point snapshot of a network, are usually described via generative processes, where edges are added one at a time. These models can thus be considered as either static, with respect to what they model, or dynamic, with respect to how they're represented. In this work we refer to them as pseudo-dynamic.

Within the category of static models we discussed two main directions:
models that take networks as given (see \autoref{sec:pone},
\autoref{sec:ptwo}, and \autoref{sec:ergm}) and models that assume and
estimate latent structures (\autoref{sec:mmsb} and
\autoref{sec:latent}). Latent structure models have to make certain
assumptions about the data. Stochastic blockmodels assume structural
equivalence of the nodes, whereas latent space models assume the
existence of an embedding of the network in a low dimensional
space. These models allow for better understanding of the data in
cases where it is believed to contain hidden structure.

We divided the category of dynamic models into continuous time Markov
models and discrete time Markov models.  CMPM (\autoref{sec:dym-cmpm})
assumes that the adjacency matrix evolves according to a continuous
Markov chain whose intensity matrix can depend on various edge and
node dynamics.  Discrete time Markov network models deal with a set of network snapshots observed at various time points. Examples of discrete time Markov network models
include dynamic extensions of ERGM (\autoref{subsec:dym-dm-ergm}) and
the latent space model (\autoref{subsec:dym-discrete-ls}), the
duplication-attachment model, as well as a generative dynamic model
for friendship networks (\autoref{subsec:dym-discrete-dcfm}).

Despite the many advances in network modeling over the last decade,
there remains a host of unresolved issues. We listed some of the issues
in \autoref{ch:issues}.  We feel that, from a statistics or machine
learning perspective, the biggest breakthroughs are to be made in the
areas of inference and dynamic modeling. Creating a model or perhaps
fixing an existing one in such a way that provides realistic
generative and inference mechanisms which can identifiably infer
parameters of a large real world network would make a great
contribution to the statistical network modeling community.

\chapter*{Acknowledgments}
%\addcontentsline{toc}{unnumchapter}{Acknowledgments}%
%\markboth{Acknowledgments}{Acknowledgments}

This research was partly supported by United States 
National Institute of General Medical Sciences Center of Excellence grant 
 P50 GM071508, by 
National Science Foundation grants
 DBI-0546275, 
 IIS-0513552,
by
National Institutes of Health grant
 R01 GM071966
to Princeton University,
by National Science Foundation grant DMS-0907009 to Harvard
University, and by National Science Foundation grant DMS-0631589 and
partial support from U.S. Army Research Office Contract W911NFo910360
to the Department of Statistics, Carnegie Mellon University.  Edoardo
M. Airoldi was a postdoctoral fellow in the Department of Computer
Science and the Lewis-Sigler Institute for Integrative Genomics at
Princeton University when a large portion of this work was carried
out.  We thank three anonymous reviewers for their valuable comments, as well as
their helpful additions and corrections to our citation list.  We
thank Joseph Blitzstein and Pavel Krivitsky for a careful reading and the correction of a
number of infelicities. We finally wish to thank L\'aszl\'o Barab\'asi and Z\'oltan Oltvai; Peter Bearman, James Moody, and Katherine Stovel; James Fowler and Nicholas Christakis; Purnamrita Sarkar and Andrew Moore for giving permission to re-print figures from their original papers \citep{Bara:Oltv:2004,Bear:Mood:Stov:2004,Chri:Fowl:2007,Sark:Moor:2005}.

%%%%%%%%%%%%%%%%%%%%%%%%%%%%%%%%%%%%%%%%%%%%%%%%%%%%%%%%%%%%%%%%%%%%%%%%%%%%%%
%%%%%%%%%%%%%%%%%%%%%%%%%%%%%%%%%%%%%%%%%%%%%%%%%%%%%%%%%%%%%%%%%%%%%%%%%%%%%%
%%%%%%%%%%%%%%%%%%%%%%%%%%%%%%%%%%%%%%%%%%%%%%%%%%%%%%%%%%%%%%%%%%%%%%%%%%%%%%
%%%%%%%%%%%%%%%%%%%%%%%%%%%%%%%%%%%%%%%%%%%%%%%%%%%%%%%%%%%%%%%%%%%%%%%%%%%%%%
%%%%%%%%%%%%%%%%%%%%%%%%%%%%%%%%%%%%%%%%%%%%%%%%%%%%%%%%%%%%%%%%%%%%%%%%%%%%%%
%%%%%%%%%%%%%%%%%%%%%%%%%%%%%%%%%%%%%%%%%%%%%%%%%%%%%%%%%%%%%%%%%%%%%%%%%%%%%%
%%%%%%%%%%%%%%%%%%%%%%%%%%%%%%%%%%%%%%%%%%%%%%%%%%%%%%%%%%%%%%%%%%%%%%%%%%%%%%
%%%%%%%%%%%%%%%%%%%%%%%%%%%%%%%%%%%%%%%%%%%%%%%%%%%%%%%%%%%%%%%%%%%%%%%%%%%%%%
%%%%%%%%%%%%%%%%%%%%%%%%%%%%%%%%%%%%%%%%%%%%%%%%%%%%%%%%%%%%%%%%%%%%%%%%%%%%%%
%%%%%%%%%%%%%%%%%%%%%%%%%%%%%%%%%%%%%%%%%%%%%%%%%%%%%%%%%%%%%%%%%%%%%%%%%%%%%%

\addcontentsline{toc}{chapter}{\bibname}%
\markboth{\bibname}{\bibname}
\bibliography{newAllRefs9}

\end{document}